\documentclass[twocolumn,tighten]{aastex61}
\usepackage[]{txfonts}
\usepackage{natbib, twoopt}
\usepackage{float}
\usepackage{graphicx}
\usepackage{epstopdf}
\usepackage{enumitem}
\usepackage[caption=false]{subfig}
\usepackage{color}\definecolor{AliceBlue}{rgb}{0.94,0.97,1.00}
\definecolor{AntiqueWhite1}{rgb}{1.00,0.94,0.86}
\definecolor{AntiqueWhite2}{rgb}{0.93,0.87,0.80}
\definecolor{AntiqueWhite3}{rgb}{0.80,0.75,0.69}
\definecolor{AntiqueWhite4}{rgb}{0.55,0.51,0.47}
\definecolor{AntiqueWhite}{rgb}{0.98,0.92,0.84}
\definecolor{BlanchedAlmond}{rgb}{1.00,0.92,0.80}
\definecolor{BlueViolet}{rgb}{0.54,0.17,0.89}
\definecolor{CadetBlue1}{rgb}{0.60,0.96,1.00}
\definecolor{CadetBlue2}{rgb}{0.56,0.90,0.93}
\definecolor{CadetBlue3}{rgb}{0.48,0.77,0.80}
\definecolor{CadetBlue4}{rgb}{0.33,0.53,0.55}
\definecolor{CadetBlue}{rgb}{0.37,0.62,0.63}
\definecolor{CornflowerBlue}{rgb}{0.39,0.58,0.93}
\definecolor{DarkBlue}{rgb}{0.00,0.00,0.55}
\definecolor{DarkCyan}{rgb}{0.00,0.55,0.55}
\definecolor{DarkGoldenrod1}{rgb}{1.00,0.73,0.06}
\definecolor{DarkGoldenrod2}{rgb}{0.93,0.68,0.05}
\definecolor{DarkGoldenrod3}{rgb}{0.80,0.58,0.05}
\definecolor{DarkGoldenrod4}{rgb}{0.55,0.40,0.03}
\definecolor{DarkGoldenrod}{rgb}{0.72,0.53,0.04}
\definecolor{DarkGray}{rgb}{0.66,0.66,0.66}
\definecolor{DarkGreen}{rgb}{0.00,0.39,0.00}
\definecolor{DarkGrey}{rgb}{0.66,0.66,0.66}
\definecolor{DarkKhaki}{rgb}{0.74,0.72,0.42}
\definecolor{DarkMagenta}{rgb}{0.55,0.00,0.55}
\definecolor{DarkOliveGreen1}{rgb}{0.79,1.00,0.44}
\definecolor{DarkOliveGreen2}{rgb}{0.74,0.93,0.41}
\definecolor{DarkOliveGreen3}{rgb}{0.64,0.80,0.35}
\definecolor{DarkOliveGreen4}{rgb}{0.43,0.55,0.24}
\definecolor{DarkOliveGreen}{rgb}{0.33,0.42,0.18}
\definecolor{DarkOrange1}{rgb}{1.00,0.50,0.00}
\definecolor{DarkOrange2}{rgb}{0.93,0.46,0.00}
\definecolor{DarkOrange3}{rgb}{0.80,0.40,0.00}
\definecolor{DarkOrange4}{rgb}{0.55,0.27,0.00}
\definecolor{DarkOrange}{rgb}{1.00,0.55,0.00}
\definecolor{DarkOrchid1}{rgb}{0.75,0.24,1.00}
\definecolor{DarkOrchid2}{rgb}{0.70,0.23,0.93}
\definecolor{DarkOrchid3}{rgb}{0.60,0.20,0.80}
\definecolor{DarkOrchid4}{rgb}{0.41,0.13,0.55}
\definecolor{DarkOrchid}{rgb}{0.60,0.20,0.80}
\definecolor{DarkRed}{rgb}{0.55,0.00,0.00}
\definecolor{DarkSalmon}{rgb}{0.91,0.59,0.48}
\definecolor{DarkSeaGreen1}{rgb}{0.76,1.00,0.76}
\definecolor{DarkSeaGreen2}{rgb}{0.71,0.93,0.71}
\definecolor{DarkSeaGreen3}{rgb}{0.61,0.80,0.61}
\definecolor{DarkSeaGreen4}{rgb}{0.41,0.55,0.41}
\definecolor{DarkSeaGreen}{rgb}{0.56,0.74,0.56}
\definecolor{DarkSlateBlue}{rgb}{0.28,0.24,0.55}
\definecolor{DarkSlateGray1}{rgb}{0.59,1.00,1.00}
\definecolor{DarkSlateGray2}{rgb}{0.55,0.93,0.93}
\definecolor{DarkSlateGray3}{rgb}{0.47,0.80,0.80}
\definecolor{DarkSlateGray4}{rgb}{0.32,0.55,0.55}
\definecolor{DarkSlateGray}{rgb}{0.18,0.31,0.31}
\definecolor{DarkSlateGrey}{rgb}{0.18,0.31,0.31}
\definecolor{DarkTurquoise}{rgb}{0.00,0.81,0.82}
\definecolor{DarkViolet}{rgb}{0.58,0.00,0.83}
\definecolor{DeepPink1}{rgb}{1.00,0.08,0.58}
\definecolor{DeepPink2}{rgb}{0.93,0.07,0.54}
\definecolor{DeepPink3}{rgb}{0.80,0.06,0.46}
\definecolor{DeepPink4}{rgb}{0.55,0.04,0.31}
\definecolor{DeepPink}{rgb}{1.00,0.08,0.58}
\definecolor{DeepSkyBlue1}{rgb}{0.00,0.75,1.00}
\definecolor{DeepSkyBlue2}{rgb}{0.00,0.70,0.93}
\definecolor{DeepSkyBlue3}{rgb}{0.00,0.60,0.80}
\definecolor{DeepSkyBlue4}{rgb}{0.00,0.41,0.55}
\definecolor{DeepSkyBlue}{rgb}{0.00,0.75,1.00}
\definecolor{DimGray}{rgb}{0.41,0.41,0.41}
\definecolor{DimGrey}{rgb}{0.41,0.41,0.41}
\definecolor{DodgerBlue1}{rgb}{0.12,0.56,1.00}
\definecolor{DodgerBlue2}{rgb}{0.11,0.53,0.93}
\definecolor{DodgerBlue3}{rgb}{0.09,0.45,0.80}
\definecolor{DodgerBlue4}{rgb}{0.06,0.31,0.55}
\definecolor{DodgerBlue}{rgb}{0.12,0.56,1.00}
\definecolor{FloralWhite}{rgb}{1.00,0.98,0.94}
\definecolor{ForestGreen}{rgb}{0.13,0.55,0.13}
\definecolor{GhostWhite}{rgb}{0.97,0.97,1.00}
\definecolor{GreenYellow}{rgb}{0.68,1.00,0.18}
\definecolor{HotPink1}{rgb}{1.00,0.43,0.71}
\definecolor{HotPink2}{rgb}{0.93,0.42,0.65}
\definecolor{HotPink3}{rgb}{0.80,0.38,0.56}
\definecolor{HotPink4}{rgb}{0.55,0.23,0.38}
\definecolor{HotPink}{rgb}{1.00,0.41,0.71}
\definecolor{IndianRed1}{rgb}{1.00,0.42,0.42}
\definecolor{IndianRed2}{rgb}{0.93,0.39,0.39}
\definecolor{IndianRed3}{rgb}{0.80,0.33,0.33}
\definecolor{IndianRed4}{rgb}{0.55,0.23,0.23}
\definecolor{IndianRed}{rgb}{0.80,0.36,0.36}
\definecolor{LavenderBlush1}{rgb}{1.00,0.94,0.96}
\definecolor{LavenderBlush2}{rgb}{0.93,0.88,0.90}
\definecolor{LavenderBlush3}{rgb}{0.80,0.76,0.77}
\definecolor{LavenderBlush4}{rgb}{0.55,0.51,0.53}
\definecolor{LavenderBlush}{rgb}{1.00,0.94,0.96}
\definecolor{LawnGreen}{rgb}{0.49,0.99,0.00}
\definecolor{LemonChiffon1}{rgb}{1.00,0.98,0.80}
\definecolor{LemonChiffon2}{rgb}{0.93,0.91,0.75}
\definecolor{LemonChiffon3}{rgb}{0.80,0.79,0.65}
\definecolor{LemonChiffon4}{rgb}{0.55,0.54,0.44}
\definecolor{LemonChiffon}{rgb}{1.00,0.98,0.80}
\definecolor{LightBlue1}{rgb}{0.75,0.94,1.00}
\definecolor{LightBlue2}{rgb}{0.70,0.87,0.93}
\definecolor{LightBlue3}{rgb}{0.60,0.75,0.80}
\definecolor{LightBlue4}{rgb}{0.41,0.51,0.55}
\definecolor{LightBlue}{rgb}{0.68,0.85,0.90}
\definecolor{LightCoral}{rgb}{0.94,0.50,0.50}
\definecolor{LightCyan1}{rgb}{0.88,1.00,1.00}
\definecolor{LightCyan2}{rgb}{0.82,0.93,0.93}
\definecolor{LightCyan3}{rgb}{0.71,0.80,0.80}
\definecolor{LightCyan4}{rgb}{0.48,0.55,0.55}
\definecolor{LightCyan}{rgb}{0.88,1.00,1.00}
\definecolor{LightGoldenrod1}{rgb}{1.00,0.93,0.55}
\definecolor{LightGoldenrod2}{rgb}{0.93,0.86,0.51}
\definecolor{LightGoldenrod3}{rgb}{0.80,0.75,0.44}
\definecolor{LightGoldenrod4}{rgb}{0.55,0.51,0.30}
\definecolor{LightGoldenrodYellow}{rgb}{0.98,0.98,0.82}
\definecolor{LightGoldenrod}{rgb}{0.93,0.87,0.51}
\definecolor{LightGray}{rgb}{0.83,0.83,0.83}
\definecolor{LightGreen}{rgb}{0.56,0.93,0.56}
\definecolor{LightGrey}{rgb}{0.83,0.83,0.83}
\definecolor{LightPink1}{rgb}{1.00,0.68,0.73}
\definecolor{LightPink2}{rgb}{0.93,0.64,0.68}
\definecolor{LightPink3}{rgb}{0.80,0.55,0.58}
\definecolor{LightPink4}{rgb}{0.55,0.37,0.40}
\definecolor{LightPink}{rgb}{1.00,0.71,0.76}
\definecolor{LightSalmon1}{rgb}{1.00,0.63,0.48}
\definecolor{LightSalmon2}{rgb}{0.93,0.58,0.45}
\definecolor{LightSalmon3}{rgb}{0.80,0.51,0.38}
\definecolor{LightSalmon4}{rgb}{0.55,0.34,0.26}
\definecolor{LightSalmon}{rgb}{1.00,0.63,0.48}
\definecolor{LightSeaGreen}{rgb}{0.13,0.70,0.67}
\definecolor{LightSkyBlue1}{rgb}{0.69,0.89,1.00}
\definecolor{LightSkyBlue2}{rgb}{0.64,0.83,0.93}
\definecolor{LightSkyBlue3}{rgb}{0.55,0.71,0.80}
\definecolor{LightSkyBlue4}{rgb}{0.38,0.48,0.55}
\definecolor{LightSkyBlue}{rgb}{0.53,0.81,0.98}
\definecolor{LightSlateBlue}{rgb}{0.52,0.44,1.00}
\definecolor{LightSlateGray}{rgb}{0.47,0.53,0.60}
\definecolor{LightSlateGrey}{rgb}{0.47,0.53,0.60}
\definecolor{LightSteelBlue1}{rgb}{0.79,0.88,1.00}
\definecolor{LightSteelBlue2}{rgb}{0.74,0.82,0.93}
\definecolor{LightSteelBlue3}{rgb}{0.64,0.71,0.80}
\definecolor{LightSteelBlue4}{rgb}{0.43,0.48,0.55}
\definecolor{LightSteelBlue}{rgb}{0.69,0.77,0.87}
\definecolor{LightYellow1}{rgb}{1.00,1.00,0.88}
\definecolor{LightYellow2}{rgb}{0.93,0.93,0.82}
\definecolor{LightYellow3}{rgb}{0.80,0.80,0.71}
\definecolor{LightYellow4}{rgb}{0.55,0.55,0.48}
\definecolor{LightYellow}{rgb}{1.00,1.00,0.88}
\definecolor{LimeGreen}{rgb}{0.20,0.80,0.20}
\definecolor{MediumAquamarine}{rgb}{0.40,0.80,0.67}
\definecolor{MediumBlue}{rgb}{0.00,0.00,0.80}
\definecolor{MediumOrchid1}{rgb}{0.88,0.40,1.00}
\definecolor{MediumOrchid2}{rgb}{0.82,0.37,0.93}
\definecolor{MediumOrchid3}{rgb}{0.71,0.32,0.80}
\definecolor{MediumOrchid4}{rgb}{0.48,0.22,0.55}
\definecolor{MediumOrchid}{rgb}{0.73,0.33,0.83}
\definecolor{MediumPurple1}{rgb}{0.67,0.51,1.00}
\definecolor{MediumPurple2}{rgb}{0.62,0.47,0.93}
\definecolor{MediumPurple3}{rgb}{0.54,0.41,0.80}
\definecolor{MediumPurple4}{rgb}{0.36,0.28,0.55}
\definecolor{MediumPurple}{rgb}{0.58,0.44,0.86}
\definecolor{MediumSeaGreen}{rgb}{0.24,0.70,0.44}
\definecolor{MediumSlateBlue}{rgb}{0.48,0.41,0.93}
\definecolor{MediumSpringGreen}{rgb}{0.00,0.98,0.60}
\definecolor{MediumTurquoise}{rgb}{0.28,0.82,0.80}
\definecolor{MediumVioletRed}{rgb}{0.78,0.08,0.52}
\definecolor{MidnightBlue}{rgb}{0.10,0.10,0.44}
\definecolor{MintCream}{rgb}{0.96,1.00,0.98}
\definecolor{MistyRose1}{rgb}{1.00,0.89,0.88}
\definecolor{MistyRose2}{rgb}{0.93,0.84,0.82}
\definecolor{MistyRose3}{rgb}{0.80,0.72,0.71}
\definecolor{MistyRose4}{rgb}{0.55,0.49,0.48}
\definecolor{MistyRose}{rgb}{1.00,0.89,0.88}
\definecolor{NavajoWhite1}{rgb}{1.00,0.87,0.68}
\definecolor{NavajoWhite2}{rgb}{0.93,0.81,0.63}
\definecolor{NavajoWhite3}{rgb}{0.80,0.70,0.55}
\definecolor{NavajoWhite4}{rgb}{0.55,0.47,0.37}
\definecolor{NavajoWhite}{rgb}{1.00,0.87,0.68}
\definecolor{NavyBlue}{rgb}{0.00,0.00,0.50}
\definecolor{OldLace}{rgb}{0.99,0.96,0.90}
\definecolor{OliveDrab1}{rgb}{0.75,1.00,0.24}
\definecolor{OliveDrab2}{rgb}{0.70,0.93,0.23}
\definecolor{OliveDrab3}{rgb}{0.60,0.80,0.20}
\definecolor{OliveDrab4}{rgb}{0.41,0.55,0.13}
\definecolor{OliveDrab}{rgb}{0.42,0.56,0.14}
\definecolor{OrangeRed1}{rgb}{1.00,0.27,0.00}
\definecolor{OrangeRed2}{rgb}{0.93,0.25,0.00}
\definecolor{OrangeRed3}{rgb}{0.80,0.22,0.00}
\definecolor{OrangeRed4}{rgb}{0.55,0.15,0.00}
\definecolor{OrangeRed}{rgb}{1.00,0.27,0.00}
\definecolor{PaleGoldenrod}{rgb}{0.93,0.91,0.67}
\definecolor{PaleGreen1}{rgb}{0.60,1.00,0.60}
\definecolor{PaleGreen2}{rgb}{0.56,0.93,0.56}
\definecolor{PaleGreen3}{rgb}{0.49,0.80,0.49}
\definecolor{PaleGreen4}{rgb}{0.33,0.55,0.33}
\definecolor{PaleGreen}{rgb}{0.60,0.98,0.60}
\definecolor{PaleTurquoise1}{rgb}{0.73,1.00,1.00}
\definecolor{PaleTurquoise2}{rgb}{0.68,0.93,0.93}
\definecolor{PaleTurquoise3}{rgb}{0.59,0.80,0.80}
\definecolor{PaleTurquoise4}{rgb}{0.40,0.55,0.55}
\definecolor{PaleTurquoise}{rgb}{0.69,0.93,0.93}
\definecolor{PaleVioletRed1}{rgb}{1.00,0.51,0.67}
\definecolor{PaleVioletRed2}{rgb}{0.93,0.47,0.62}
\definecolor{PaleVioletRed3}{rgb}{0.80,0.41,0.54}
\definecolor{PaleVioletRed4}{rgb}{0.55,0.28,0.36}
\definecolor{PaleVioletRed}{rgb}{0.86,0.44,0.58}
\definecolor{PapayaWhip}{rgb}{1.00,0.94,0.84}
\definecolor{PeachPuff1}{rgb}{1.00,0.85,0.73}
\definecolor{PeachPuff2}{rgb}{0.93,0.80,0.68}
\definecolor{PeachPuff3}{rgb}{0.80,0.69,0.58}
\definecolor{PeachPuff4}{rgb}{0.55,0.47,0.40}
\definecolor{PeachPuff}{rgb}{1.00,0.85,0.73}
\definecolor{PowderBlue}{rgb}{0.69,0.88,0.90}
\definecolor{RosyBrown1}{rgb}{1.00,0.76,0.76}
\definecolor{RosyBrown2}{rgb}{0.93,0.71,0.71}
\definecolor{RosyBrown3}{rgb}{0.80,0.61,0.61}
\definecolor{RosyBrown4}{rgb}{0.55,0.41,0.41}
\definecolor{RosyBrown}{rgb}{0.74,0.56,0.56}
\definecolor{RoyalBlue1}{rgb}{0.28,0.46,1.00}
\definecolor{RoyalBlue2}{rgb}{0.26,0.43,0.93}
\definecolor{RoyalBlue3}{rgb}{0.23,0.37,0.80}
\definecolor{RoyalBlue4}{rgb}{0.15,0.25,0.55}
\definecolor{RoyalBlue}{rgb}{0.25,0.41,0.88}
\definecolor{SaddleBrown}{rgb}{0.55,0.27,0.07}
\definecolor{SandyBrown}{rgb}{0.96,0.64,0.38}
\definecolor{SeaGreen1}{rgb}{0.33,1.00,0.62}
\definecolor{SeaGreen2}{rgb}{0.31,0.93,0.58}
\definecolor{SeaGreen3}{rgb}{0.26,0.80,0.50}
\definecolor{SeaGreen4}{rgb}{0.18,0.55,0.34}
\definecolor{SeaGreen}{rgb}{0.18,0.55,0.34}
\definecolor{SkyBlue1}{rgb}{0.53,0.81,1.00}
\definecolor{SkyBlue2}{rgb}{0.49,0.75,0.93}
\definecolor{SkyBlue3}{rgb}{0.42,0.65,0.80}
\definecolor{SkyBlue4}{rgb}{0.29,0.44,0.55}
\definecolor{SkyBlue}{rgb}{0.53,0.81,0.92}
\definecolor{SlateBlue1}{rgb}{0.51,0.44,1.00}
\definecolor{SlateBlue2}{rgb}{0.48,0.40,0.93}
\definecolor{SlateBlue3}{rgb}{0.41,0.35,0.80}
\definecolor{SlateBlue4}{rgb}{0.28,0.24,0.55}
\definecolor{SlateBlue}{rgb}{0.42,0.35,0.80}
\definecolor{SlateGray1}{rgb}{0.78,0.89,1.00}
\definecolor{SlateGray2}{rgb}{0.73,0.83,0.93}
\definecolor{SlateGray3}{rgb}{0.62,0.71,0.80}
\definecolor{SlateGray4}{rgb}{0.42,0.48,0.55}
\definecolor{SlateGray}{rgb}{0.44,0.50,0.56}
\definecolor{SlateGrey}{rgb}{0.44,0.50,0.56}
\definecolor{SpringGreen1}{rgb}{0.00,1.00,0.50}
\definecolor{SpringGreen2}{rgb}{0.00,0.93,0.46}
\definecolor{SpringGreen3}{rgb}{0.00,0.80,0.40}
\definecolor{SpringGreen4}{rgb}{0.00,0.55,0.27}
\definecolor{SpringGreen}{rgb}{0.00,1.00,0.50}
\definecolor{SteelBlue1}{rgb}{0.39,0.72,1.00}
\definecolor{SteelBlue2}{rgb}{0.36,0.67,0.93}
\definecolor{SteelBlue3}{rgb}{0.31,0.58,0.80}
\definecolor{SteelBlue4}{rgb}{0.21,0.39,0.55}
\definecolor{SteelBlue}{rgb}{0.27,0.51,0.71}
\definecolor{VioletRed1}{rgb}{1.00,0.24,0.59}
\definecolor{VioletRed2}{rgb}{0.93,0.23,0.55}
\definecolor{VioletRed3}{rgb}{0.80,0.20,0.47}
\definecolor{VioletRed4}{rgb}{0.55,0.13,0.32}
\definecolor{VioletRed}{rgb}{0.82,0.13,0.56}
\definecolor{WhiteSmoke}{rgb}{0.96,0.96,0.96}
\definecolor{YellowGreen}{rgb}{0.60,0.80,0.20}
\definecolor{aliceblue}{rgb}{0.94,0.97,1.00}
\definecolor{antiquewhite}{rgb}{0.98,0.92,0.84}
\definecolor{aquamarine1}{rgb}{0.50,1.00,0.83}
\definecolor{aquamarine2}{rgb}{0.46,0.93,0.78}
\definecolor{aquamarine3}{rgb}{0.40,0.80,0.67}
\definecolor{aquamarine4}{rgb}{0.27,0.55,0.45}
\definecolor{aquamarine}{rgb}{0.50,1.00,0.83}
\definecolor{azure1}{rgb}{0.94,1.00,1.00}
\definecolor{azure2}{rgb}{0.88,0.93,0.93}
\definecolor{azure3}{rgb}{0.76,0.80,0.80}
\definecolor{azure4}{rgb}{0.51,0.55,0.55}
\definecolor{azure}{rgb}{0.94,1.00,1.00}
\definecolor{beige}{rgb}{0.96,0.96,0.86}
\definecolor{bisque1}{rgb}{1.00,0.89,0.77}
\definecolor{bisque2}{rgb}{0.93,0.84,0.72}
\definecolor{bisque3}{rgb}{0.80,0.72,0.62}
\definecolor{bisque4}{rgb}{0.55,0.49,0.42}
\definecolor{bisque}{rgb}{1.00,0.89,0.77}
\definecolor{black}{rgb}{0.00,0.00,0.00}
\definecolor{blanchedalmond}{rgb}{1.00,0.92,0.80}
\definecolor{blue1}{rgb}{0.00,0.00,1.00}
\definecolor{blue2}{rgb}{0.00,0.00,0.93}
\definecolor{blue3}{rgb}{0.00,0.00,0.80}
\definecolor{blue4}{rgb}{0.00,0.00,0.55}
\definecolor{blueviolet}{rgb}{0.54,0.17,0.89}
\definecolor{blue}{rgb}{0.00,0.00,1.00}
\definecolor{brown1}{rgb}{1.00,0.25,0.25}
\definecolor{brown2}{rgb}{0.93,0.23,0.23}
\definecolor{brown3}{rgb}{0.80,0.20,0.20}
\definecolor{brown4}{rgb}{0.55,0.14,0.14}
\definecolor{brown}{rgb}{0.65,0.16,0.16}
\definecolor{burlywood1}{rgb}{1.00,0.83,0.61}
\definecolor{burlywood2}{rgb}{0.93,0.77,0.57}
\definecolor{burlywood3}{rgb}{0.80,0.67,0.49}
\definecolor{burlywood4}{rgb}{0.55,0.45,0.33}
\definecolor{burlywood}{rgb}{0.87,0.72,0.53}
\definecolor{cadetblue}{rgb}{0.37,0.62,0.63}
\definecolor{chartreuse1}{rgb}{0.50,1.00,0.00}
\definecolor{chartreuse2}{rgb}{0.46,0.93,0.00}
\definecolor{chartreuse3}{rgb}{0.40,0.80,0.00}
\definecolor{chartreuse4}{rgb}{0.27,0.55,0.00}
\definecolor{chartreuse}{rgb}{0.50,1.00,0.00}
\definecolor{chocolate1}{rgb}{1.00,0.50,0.14}
\definecolor{chocolate2}{rgb}{0.93,0.46,0.13}
\definecolor{chocolate3}{rgb}{0.80,0.40,0.11}
\definecolor{chocolate4}{rgb}{0.55,0.27,0.07}
\definecolor{chocolate}{rgb}{0.82,0.41,0.12}
\definecolor{coral1}{rgb}{1.00,0.45,0.34}
\definecolor{coral2}{rgb}{0.93,0.42,0.31}
\definecolor{coral3}{rgb}{0.80,0.36,0.27}
\definecolor{coral4}{rgb}{0.55,0.24,0.18}
\definecolor{coral}{rgb}{1.00,0.50,0.31}
\definecolor{cornflowerblue}{rgb}{0.39,0.58,0.93}
\definecolor{cornsilk1}{rgb}{1.00,0.97,0.86}
\definecolor{cornsilk2}{rgb}{0.93,0.91,0.80}
\definecolor{cornsilk3}{rgb}{0.80,0.78,0.69}
\definecolor{cornsilk4}{rgb}{0.55,0.53,0.47}
\definecolor{cornsilk}{rgb}{1.00,0.97,0.86}
\definecolor{cyan1}{rgb}{0.00,1.00,1.00}
\definecolor{cyan2}{rgb}{0.00,0.93,0.93}
\definecolor{cyan3}{rgb}{0.00,0.80,0.80}
\definecolor{cyan4}{rgb}{0.00,0.55,0.55}
\definecolor{cyan}{rgb}{0.00,1.00,1.00}
\definecolor{darkblue}{rgb}{0.00,0.00,0.55}
\definecolor{darkcyan}{rgb}{0.00,0.55,0.55}
\definecolor{darkgoldenrod}{rgb}{0.72,0.53,0.04}
\definecolor{darkgray}{rgb}{0.66,0.66,0.66}
\definecolor{darkgreen}{rgb}{0.00,0.39,0.00}
\definecolor{darkgrey}{rgb}{0.66,0.66,0.66}
\definecolor{darkkhaki}{rgb}{0.74,0.72,0.42}
\definecolor{darkmagenta}{rgb}{0.55,0.00,0.55}
\definecolor{darkolive}{rgb}{0.33,0.42,0.18}
\definecolor{darkorange}{rgb}{1.00,0.55,0.00}
\definecolor{darkorchid}{rgb}{0.60,0.20,0.80}
\definecolor{darkred}{rgb}{0.55,0.00,0.00}
\definecolor{darksalmon}{rgb}{0.91,0.59,0.48}
\definecolor{darksea}{rgb}{0.56,0.74,0.56}
\definecolor{darkslate}{rgb}{0.18,0.31,0.31}
\definecolor{darkslate}{rgb}{0.18,0.31,0.31}
\definecolor{darkslate}{rgb}{0.28,0.24,0.55}
\definecolor{darkturquoise}{rgb}{0.00,0.81,0.82}
\definecolor{darkviolet}{rgb}{0.58,0.00,0.83}
\definecolor{deeppink}{rgb}{1.00,0.08,0.58}
\definecolor{deepsky}{rgb}{0.00,0.75,1.00}
\definecolor{dimgray}{rgb}{0.41,0.41,0.41}
\definecolor{dimgrey}{rgb}{0.41,0.41,0.41}
\definecolor{dodgerblue}{rgb}{0.12,0.56,1.00}
\definecolor{firebrick1}{rgb}{1.00,0.19,0.19}
\definecolor{firebrick2}{rgb}{0.93,0.17,0.17}
\definecolor{firebrick3}{rgb}{0.80,0.15,0.15}
\definecolor{firebrick4}{rgb}{0.55,0.10,0.10}
\definecolor{firebrick}{rgb}{0.70,0.13,0.13}
\definecolor{floralwhite}{rgb}{1.00,0.98,0.94}
\definecolor{forestgreen}{rgb}{0.13,0.55,0.13}
\definecolor{gainsboro}{rgb}{0.86,0.86,0.86}
\definecolor{ghostwhite}{rgb}{0.97,0.97,1.00}
\definecolor{gold1}{rgb}{1.00,0.84,0.00}
\definecolor{gold2}{rgb}{0.93,0.79,0.00}
\definecolor{gold3}{rgb}{0.80,0.68,0.00}
\definecolor{gold4}{rgb}{0.55,0.46,0.00}
\definecolor{goldenrod1}{rgb}{1.00,0.76,0.15}
\definecolor{goldenrod2}{rgb}{0.93,0.71,0.13}
\definecolor{goldenrod3}{rgb}{0.80,0.61,0.11}
\definecolor{goldenrod4}{rgb}{0.55,0.41,0.08}
\definecolor{goldenrod}{rgb}{0.85,0.65,0.13}
\definecolor{gold}{rgb}{1.00,0.84,0.00}
\definecolor{gray0}{rgb}{0.00,0.00,0.00}
\definecolor{gray100}{rgb}{1.00,1.00,1.00}
\definecolor{gray10}{rgb}{0.10,0.10,0.10}
\definecolor{gray11}{rgb}{0.11,0.11,0.11}
\definecolor{gray12}{rgb}{0.12,0.12,0.12}
\definecolor{gray13}{rgb}{0.13,0.13,0.13}
\definecolor{gray14}{rgb}{0.14,0.14,0.14}
\definecolor{gray15}{rgb}{0.15,0.15,0.15}
\definecolor{gray16}{rgb}{0.16,0.16,0.16}
\definecolor{gray17}{rgb}{0.17,0.17,0.17}
\definecolor{gray18}{rgb}{0.18,0.18,0.18}
\definecolor{gray19}{rgb}{0.19,0.19,0.19}
\definecolor{gray1}{rgb}{0.01,0.01,0.01}
\definecolor{gray20}{rgb}{0.20,0.20,0.20}
\definecolor{gray21}{rgb}{0.21,0.21,0.21}
\definecolor{gray22}{rgb}{0.22,0.22,0.22}
\definecolor{gray23}{rgb}{0.23,0.23,0.23}
\definecolor{gray24}{rgb}{0.24,0.24,0.24}
\definecolor{gray25}{rgb}{0.25,0.25,0.25}
\definecolor{gray26}{rgb}{0.26,0.26,0.26}
\definecolor{gray27}{rgb}{0.27,0.27,0.27}
\definecolor{gray28}{rgb}{0.28,0.28,0.28}
\definecolor{gray29}{rgb}{0.29,0.29,0.29}
\definecolor{gray2}{rgb}{0.02,0.02,0.02}
\definecolor{gray30}{rgb}{0.30,0.30,0.30}
\definecolor{gray31}{rgb}{0.31,0.31,0.31}
\definecolor{gray32}{rgb}{0.32,0.32,0.32}
\definecolor{gray33}{rgb}{0.33,0.33,0.33}
\definecolor{gray34}{rgb}{0.34,0.34,0.34}
\definecolor{gray35}{rgb}{0.35,0.35,0.35}
\definecolor{gray36}{rgb}{0.36,0.36,0.36}
\definecolor{gray37}{rgb}{0.37,0.37,0.37}
\definecolor{gray38}{rgb}{0.38,0.38,0.38}
\definecolor{gray39}{rgb}{0.39,0.39,0.39}
\definecolor{gray3}{rgb}{0.03,0.03,0.03}
\definecolor{gray40}{rgb}{0.40,0.40,0.40}
\definecolor{gray41}{rgb}{0.41,0.41,0.41}
\definecolor{gray42}{rgb}{0.42,0.42,0.42}
\definecolor{gray43}{rgb}{0.43,0.43,0.43}
\definecolor{gray44}{rgb}{0.44,0.44,0.44}
\definecolor{gray45}{rgb}{0.45,0.45,0.45}
\definecolor{gray46}{rgb}{0.46,0.46,0.46}
\definecolor{gray47}{rgb}{0.47,0.47,0.47}
\definecolor{gray48}{rgb}{0.48,0.48,0.48}
\definecolor{gray49}{rgb}{0.49,0.49,0.49}
\definecolor{gray4}{rgb}{0.04,0.04,0.04}
\definecolor{gray50}{rgb}{0.50,0.50,0.50}
\definecolor{gray51}{rgb}{0.51,0.51,0.51}
\definecolor{gray52}{rgb}{0.52,0.52,0.52}
\definecolor{gray53}{rgb}{0.53,0.53,0.53}
\definecolor{gray54}{rgb}{0.54,0.54,0.54}
\definecolor{gray55}{rgb}{0.55,0.55,0.55}
\definecolor{gray56}{rgb}{0.56,0.56,0.56}
\definecolor{gray57}{rgb}{0.57,0.57,0.57}
\definecolor{gray58}{rgb}{0.58,0.58,0.58}
\definecolor{gray59}{rgb}{0.59,0.59,0.59}
\definecolor{gray5}{rgb}{0.05,0.05,0.05}
\definecolor{gray60}{rgb}{0.60,0.60,0.60}
\definecolor{gray61}{rgb}{0.61,0.61,0.61}
\definecolor{gray62}{rgb}{0.62,0.62,0.62}
\definecolor{gray63}{rgb}{0.63,0.63,0.63}
\definecolor{gray64}{rgb}{0.64,0.64,0.64}
\definecolor{gray65}{rgb}{0.65,0.65,0.65}
\definecolor{gray66}{rgb}{0.66,0.66,0.66}
\definecolor{gray67}{rgb}{0.67,0.67,0.67}
\definecolor{gray68}{rgb}{0.68,0.68,0.68}
\definecolor{gray69}{rgb}{0.69,0.69,0.69}
\definecolor{gray6}{rgb}{0.06,0.06,0.06}
\definecolor{gray70}{rgb}{0.70,0.70,0.70}
\definecolor{gray71}{rgb}{0.71,0.71,0.71}
\definecolor{gray72}{rgb}{0.72,0.72,0.72}
\definecolor{gray73}{rgb}{0.73,0.73,0.73}
\definecolor{gray74}{rgb}{0.74,0.74,0.74}
\definecolor{gray75}{rgb}{0.75,0.75,0.75}
\definecolor{gray76}{rgb}{0.76,0.76,0.76}
\definecolor{gray77}{rgb}{0.77,0.77,0.77}
\definecolor{gray78}{rgb}{0.78,0.78,0.78}
\definecolor{gray79}{rgb}{0.79,0.79,0.79}
\definecolor{gray7}{rgb}{0.07,0.07,0.07}
\definecolor{gray80}{rgb}{0.80,0.80,0.80}
\definecolor{gray81}{rgb}{0.81,0.81,0.81}
\definecolor{gray82}{rgb}{0.82,0.82,0.82}
\definecolor{gray83}{rgb}{0.83,0.83,0.83}
\definecolor{gray84}{rgb}{0.84,0.84,0.84}
\definecolor{gray85}{rgb}{0.85,0.85,0.85}
\definecolor{gray86}{rgb}{0.86,0.86,0.86}
\definecolor{gray87}{rgb}{0.87,0.87,0.87}
\definecolor{gray88}{rgb}{0.88,0.88,0.88}
\definecolor{gray89}{rgb}{0.89,0.89,0.89}
\definecolor{gray8}{rgb}{0.08,0.08,0.08}
\definecolor{gray90}{rgb}{0.90,0.90,0.90}
\definecolor{gray91}{rgb}{0.91,0.91,0.91}
\definecolor{gray92}{rgb}{0.92,0.92,0.92}
\definecolor{gray93}{rgb}{0.93,0.93,0.93}
\definecolor{gray94}{rgb}{0.94,0.94,0.94}
\definecolor{gray95}{rgb}{0.95,0.95,0.95}
\definecolor{gray96}{rgb}{0.96,0.96,0.96}
\definecolor{gray97}{rgb}{0.97,0.97,0.97}
\definecolor{gray98}{rgb}{0.98,0.98,0.98}
\definecolor{gray99}{rgb}{0.99,0.99,0.99}
\definecolor{gray9}{rgb}{0.09,0.09,0.09}
\definecolor{gray}{rgb}{0.75,0.75,0.75}
\definecolor{green1}{rgb}{0.00,1.00,0.00}
\definecolor{green2}{rgb}{0.00,0.93,0.00}
\definecolor{green3}{rgb}{0.00,0.80,0.00}
\definecolor{green4}{rgb}{0.00,0.55,0.00}
\definecolor{greenyellow}{rgb}{0.68,1.00,0.18}
\definecolor{green}{rgb}{0.00,1.00,0.00}
\definecolor{grey0}{rgb}{0.00,0.00,0.00}
\definecolor{grey100}{rgb}{1.00,1.00,1.00}
\definecolor{grey10}{rgb}{0.10,0.10,0.10}
\definecolor{grey11}{rgb}{0.11,0.11,0.11}
\definecolor{grey12}{rgb}{0.12,0.12,0.12}
\definecolor{grey13}{rgb}{0.13,0.13,0.13}
\definecolor{grey14}{rgb}{0.14,0.14,0.14}
\definecolor{grey15}{rgb}{0.15,0.15,0.15}
\definecolor{grey16}{rgb}{0.16,0.16,0.16}
\definecolor{grey17}{rgb}{0.17,0.17,0.17}
\definecolor{grey18}{rgb}{0.18,0.18,0.18}
\definecolor{grey19}{rgb}{0.19,0.19,0.19}
\definecolor{grey1}{rgb}{0.01,0.01,0.01}
\definecolor{grey20}{rgb}{0.20,0.20,0.20}
\definecolor{grey21}{rgb}{0.21,0.21,0.21}
\definecolor{grey22}{rgb}{0.22,0.22,0.22}
\definecolor{grey23}{rgb}{0.23,0.23,0.23}
\definecolor{grey24}{rgb}{0.24,0.24,0.24}
\definecolor{grey25}{rgb}{0.25,0.25,0.25}
\definecolor{grey26}{rgb}{0.26,0.26,0.26}
\definecolor{grey27}{rgb}{0.27,0.27,0.27}
\definecolor{grey28}{rgb}{0.28,0.28,0.28}
\definecolor{grey29}{rgb}{0.29,0.29,0.29}
\definecolor{grey2}{rgb}{0.02,0.02,0.02}
\definecolor{grey30}{rgb}{0.30,0.30,0.30}
\definecolor{grey31}{rgb}{0.31,0.31,0.31}
\definecolor{grey32}{rgb}{0.32,0.32,0.32}
\definecolor{grey33}{rgb}{0.33,0.33,0.33}
\definecolor{grey34}{rgb}{0.34,0.34,0.34}
\definecolor{grey35}{rgb}{0.35,0.35,0.35}
\definecolor{grey36}{rgb}{0.36,0.36,0.36}
\definecolor{grey37}{rgb}{0.37,0.37,0.37}
\definecolor{grey38}{rgb}{0.38,0.38,0.38}
\definecolor{grey39}{rgb}{0.39,0.39,0.39}
\definecolor{grey3}{rgb}{0.03,0.03,0.03}
\definecolor{grey40}{rgb}{0.40,0.40,0.40}
\definecolor{grey41}{rgb}{0.41,0.41,0.41}
\definecolor{grey42}{rgb}{0.42,0.42,0.42}
\definecolor{grey43}{rgb}{0.43,0.43,0.43}
\definecolor{grey44}{rgb}{0.44,0.44,0.44}
\definecolor{grey45}{rgb}{0.45,0.45,0.45}
\definecolor{grey46}{rgb}{0.46,0.46,0.46}
\definecolor{grey47}{rgb}{0.47,0.47,0.47}
\definecolor{grey48}{rgb}{0.48,0.48,0.48}
\definecolor{grey49}{rgb}{0.49,0.49,0.49}
\definecolor{grey4}{rgb}{0.04,0.04,0.04}
\definecolor{grey50}{rgb}{0.50,0.50,0.50}
\definecolor{grey51}{rgb}{0.51,0.51,0.51}
\definecolor{grey52}{rgb}{0.52,0.52,0.52}
\definecolor{grey53}{rgb}{0.53,0.53,0.53}
\definecolor{grey54}{rgb}{0.54,0.54,0.54}
\definecolor{grey55}{rgb}{0.55,0.55,0.55}
\definecolor{grey56}{rgb}{0.56,0.56,0.56}
\definecolor{grey57}{rgb}{0.57,0.57,0.57}
\definecolor{grey58}{rgb}{0.58,0.58,0.58}
\definecolor{grey59}{rgb}{0.59,0.59,0.59}
\definecolor{grey5}{rgb}{0.05,0.05,0.05}
\definecolor{grey60}{rgb}{0.60,0.60,0.60}
\definecolor{grey61}{rgb}{0.61,0.61,0.61}
\definecolor{grey62}{rgb}{0.62,0.62,0.62}
\definecolor{grey63}{rgb}{0.63,0.63,0.63}
\definecolor{grey64}{rgb}{0.64,0.64,0.64}
\definecolor{grey65}{rgb}{0.65,0.65,0.65}
\definecolor{grey66}{rgb}{0.66,0.66,0.66}
\definecolor{grey67}{rgb}{0.67,0.67,0.67}
\definecolor{grey68}{rgb}{0.68,0.68,0.68}
\definecolor{grey69}{rgb}{0.69,0.69,0.69}
\definecolor{grey6}{rgb}{0.06,0.06,0.06}
\definecolor{grey70}{rgb}{0.70,0.70,0.70}
\definecolor{grey71}{rgb}{0.71,0.71,0.71}
\definecolor{grey72}{rgb}{0.72,0.72,0.72}
\definecolor{grey73}{rgb}{0.73,0.73,0.73}
\definecolor{grey74}{rgb}{0.74,0.74,0.74}
\definecolor{grey75}{rgb}{0.75,0.75,0.75}
\definecolor{grey76}{rgb}{0.76,0.76,0.76}
\definecolor{grey77}{rgb}{0.77,0.77,0.77}
\definecolor{grey78}{rgb}{0.78,0.78,0.78}
\definecolor{grey79}{rgb}{0.79,0.79,0.79}
\definecolor{grey7}{rgb}{0.07,0.07,0.07}
\definecolor{grey80}{rgb}{0.80,0.80,0.80}
\definecolor{grey81}{rgb}{0.81,0.81,0.81}
\definecolor{grey82}{rgb}{0.82,0.82,0.82}
\definecolor{grey83}{rgb}{0.83,0.83,0.83}
\definecolor{grey84}{rgb}{0.84,0.84,0.84}
\definecolor{grey85}{rgb}{0.85,0.85,0.85}
\definecolor{grey86}{rgb}{0.86,0.86,0.86}
\definecolor{grey87}{rgb}{0.87,0.87,0.87}
\definecolor{grey88}{rgb}{0.88,0.88,0.88}
\definecolor{grey89}{rgb}{0.89,0.89,0.89}
\definecolor{grey8}{rgb}{0.08,0.08,0.08}
\definecolor{grey90}{rgb}{0.90,0.90,0.90}
\definecolor{grey91}{rgb}{0.91,0.91,0.91}
\definecolor{grey92}{rgb}{0.92,0.92,0.92}
\definecolor{grey93}{rgb}{0.93,0.93,0.93}
\definecolor{grey94}{rgb}{0.94,0.94,0.94}
\definecolor{grey95}{rgb}{0.95,0.95,0.95}
\definecolor{grey96}{rgb}{0.96,0.96,0.96}
\definecolor{grey97}{rgb}{0.97,0.97,0.97}
\definecolor{grey98}{rgb}{0.98,0.98,0.98}
\definecolor{grey99}{rgb}{0.99,0.99,0.99}
\definecolor{grey9}{rgb}{0.09,0.09,0.09}
\definecolor{grey}{rgb}{0.75,0.75,0.75}
\definecolor{honeydew1}{rgb}{0.94,1.00,0.94}
\definecolor{honeydew2}{rgb}{0.88,0.93,0.88}
\definecolor{honeydew3}{rgb}{0.76,0.80,0.76}
\definecolor{honeydew4}{rgb}{0.51,0.55,0.51}
\definecolor{honeydew}{rgb}{0.94,1.00,0.94}
\definecolor{hotpink}{rgb}{1.00,0.41,0.71}
\definecolor{indianred}{rgb}{0.80,0.36,0.36}
\definecolor{ivory1}{rgb}{1.00,1.00,0.94}
\definecolor{ivory2}{rgb}{0.93,0.93,0.88}
\definecolor{ivory3}{rgb}{0.80,0.80,0.76}
\definecolor{ivory4}{rgb}{0.55,0.55,0.51}
\definecolor{ivory}{rgb}{1.00,1.00,0.94}
\definecolor{khaki1}{rgb}{1.00,0.96,0.56}
\definecolor{khaki2}{rgb}{0.93,0.90,0.52}
\definecolor{khaki3}{rgb}{0.80,0.78,0.45}
\definecolor{khaki4}{rgb}{0.55,0.53,0.31}
\definecolor{khaki}{rgb}{0.94,0.90,0.55}
\definecolor{lavenderblush}{rgb}{1.00,0.94,0.96}
\definecolor{lavender}{rgb}{0.90,0.90,0.98}
\definecolor{lawngreen}{rgb}{0.49,0.99,0.00}
\definecolor{lemonchiffon}{rgb}{1.00,0.98,0.80}
\definecolor{lightblue}{rgb}{0.68,0.85,0.90}
\definecolor{lightcoral}{rgb}{0.94,0.50,0.50}
\definecolor{lightcyan}{rgb}{0.88,1.00,1.00}
\definecolor{lightgoldenrod}{rgb}{0.93,0.87,0.51}
\definecolor{lightgoldenrod}{rgb}{0.98,0.98,0.82}
\definecolor{lightgray}{rgb}{0.83,0.83,0.83}
\definecolor{lightgreen}{rgb}{0.56,0.93,0.56}
\definecolor{lightgrey}{rgb}{0.83,0.83,0.83}
\definecolor{lightpink}{rgb}{1.00,0.71,0.76}
\definecolor{lightsalmon}{rgb}{1.00,0.63,0.48}
\definecolor{lightsea}{rgb}{0.13,0.70,0.67}
\definecolor{lightsky}{rgb}{0.53,0.81,0.98}
\definecolor{lightslate}{rgb}{0.47,0.53,0.60}
\definecolor{lightslate}{rgb}{0.47,0.53,0.60}
\definecolor{lightslate}{rgb}{0.52,0.44,1.00}
\definecolor{lightsteel}{rgb}{0.69,0.77,0.87}
\definecolor{lightyellow}{rgb}{1.00,1.00,0.88}
\definecolor{limegreen}{rgb}{0.20,0.80,0.20}
\definecolor{linen}{rgb}{0.98,0.94,0.90}
\definecolor{magenta1}{rgb}{1.00,0.00,1.00}
\definecolor{magenta2}{rgb}{0.93,0.00,0.93}
\definecolor{magenta3}{rgb}{0.80,0.00,0.80}
\definecolor{magenta4}{rgb}{0.55,0.00,0.55}
\definecolor{magenta}{rgb}{1.00,0.00,1.00}
\definecolor{maroon1}{rgb}{1.00,0.20,0.70}
\definecolor{maroon2}{rgb}{0.93,0.19,0.65}
\definecolor{maroon3}{rgb}{0.80,0.16,0.56}
\definecolor{maroon4}{rgb}{0.55,0.11,0.38}
\definecolor{maroon}{rgb}{0.69,0.19,0.38}
\definecolor{mediumaquamarine}{rgb}{0.40,0.80,0.67}
\definecolor{mediumblue}{rgb}{0.00,0.00,0.80}
\definecolor{mediumorchid}{rgb}{0.73,0.33,0.83}
\definecolor{mediumpurple}{rgb}{0.58,0.44,0.86}
\definecolor{mediumsea}{rgb}{0.24,0.70,0.44}
\definecolor{mediumslate}{rgb}{0.48,0.41,0.93}
\definecolor{mediumspring}{rgb}{0.00,0.98,0.60}
\definecolor{mediumturquoise}{rgb}{0.28,0.82,0.80}
\definecolor{mediumviolet}{rgb}{0.78,0.08,0.52}
\definecolor{midnightblue}{rgb}{0.10,0.10,0.44}
\definecolor{mintcream}{rgb}{0.96,1.00,0.98}
\definecolor{mistyrose}{rgb}{1.00,0.89,0.88}
\definecolor{moccasin}{rgb}{1.00,0.89,0.71}
\definecolor{navajowhite}{rgb}{1.00,0.87,0.68}
\definecolor{navyblue}{rgb}{0.00,0.00,0.50}
\definecolor{navy}{rgb}{0.00,0.00,0.50}
\definecolor{oldlace}{rgb}{0.99,0.96,0.90}
\definecolor{olivedrab}{rgb}{0.42,0.56,0.14}
\definecolor{orange1}{rgb}{1.00,0.65,0.00}
\definecolor{orange2}{rgb}{0.93,0.60,0.00}
\definecolor{orange3}{rgb}{0.80,0.52,0.00}
\definecolor{orange4}{rgb}{0.55,0.35,0.00}
\definecolor{orangered}{rgb}{1.00,0.27,0.00}
\definecolor{orange}{rgb}{1.00,0.65,0.00}
\definecolor{orchid1}{rgb}{1.00,0.51,0.98}
\definecolor{orchid2}{rgb}{0.93,0.48,0.91}
\definecolor{orchid3}{rgb}{0.80,0.41,0.79}
\definecolor{orchid4}{rgb}{0.55,0.28,0.54}
\definecolor{orchid}{rgb}{0.85,0.44,0.84}
\definecolor{palegoldenrod}{rgb}{0.93,0.91,0.67}
\definecolor{palegreen}{rgb}{0.60,0.98,0.60}
\definecolor{paleturquoise}{rgb}{0.69,0.93,0.93}
\definecolor{paleviolet}{rgb}{0.86,0.44,0.58}
\definecolor{papayawhip}{rgb}{1.00,0.94,0.84}
\definecolor{peachpuff}{rgb}{1.00,0.85,0.73}
\definecolor{peru}{rgb}{0.80,0.52,0.25}
\definecolor{pink1}{rgb}{1.00,0.71,0.77}
\definecolor{pink2}{rgb}{0.93,0.66,0.72}
\definecolor{pink3}{rgb}{0.80,0.57,0.62}
\definecolor{pink4}{rgb}{0.55,0.39,0.42}
\definecolor{pink}{rgb}{1.00,0.75,0.80}
\definecolor{plum1}{rgb}{1.00,0.73,1.00}
\definecolor{plum2}{rgb}{0.93,0.68,0.93}
\definecolor{plum3}{rgb}{0.80,0.59,0.80}
\definecolor{plum4}{rgb}{0.55,0.40,0.55}
\definecolor{plum}{rgb}{0.87,0.63,0.87}
\definecolor{powderblue}{rgb}{0.69,0.88,0.90}
\definecolor{purple1}{rgb}{0.61,0.19,1.00}
\definecolor{purple2}{rgb}{0.57,0.17,0.93}
\definecolor{purple3}{rgb}{0.49,0.15,0.80}
\definecolor{purple4}{rgb}{0.33,0.10,0.55}
\definecolor{purple}{rgb}{0.63,0.13,0.94}
\definecolor{red1}{rgb}{1.00,0.00,0.00}
\definecolor{red2}{rgb}{0.93,0.00,0.00}
\definecolor{red3}{rgb}{0.80,0.00,0.00}
\definecolor{red4}{rgb}{0.55,0.00,0.00}
\definecolor{red}{rgb}{1.00,0.00,0.00}
\definecolor{rosybrown}{rgb}{0.74,0.56,0.56}
\definecolor{royalblue}{rgb}{0.25,0.41,0.88}
\definecolor{saddlebrown}{rgb}{0.55,0.27,0.07}
\definecolor{salmon1}{rgb}{1.00,0.55,0.41}
\definecolor{salmon2}{rgb}{0.93,0.51,0.38}
\definecolor{salmon3}{rgb}{0.80,0.44,0.33}
\definecolor{salmon4}{rgb}{0.55,0.30,0.22}
\definecolor{salmon}{rgb}{0.98,0.50,0.45}
\definecolor{sandybrown}{rgb}{0.96,0.64,0.38}
\definecolor{seagreen}{rgb}{0.18,0.55,0.34}
\definecolor{seashell1}{rgb}{1.00,0.96,0.93}
\definecolor{seashell2}{rgb}{0.93,0.90,0.87}
\definecolor{seashell3}{rgb}{0.80,0.77,0.75}
\definecolor{seashell4}{rgb}{0.55,0.53,0.51}
\definecolor{seashell}{rgb}{1.00,0.96,0.93}
\definecolor{sienna1}{rgb}{1.00,0.51,0.28}
\definecolor{sienna2}{rgb}{0.93,0.47,0.26}
\definecolor{sienna3}{rgb}{0.80,0.41,0.22}
\definecolor{sienna4}{rgb}{0.55,0.28,0.15}
\definecolor{sienna}{rgb}{0.63,0.32,0.18}
\definecolor{skyblue}{rgb}{0.53,0.81,0.92}
\definecolor{slateblue}{rgb}{0.42,0.35,0.80}
\definecolor{slategray}{rgb}{0.44,0.50,0.56}
\definecolor{slategrey}{rgb}{0.44,0.50,0.56}
\definecolor{snow1}{rgb}{1.00,0.98,0.98}
\definecolor{snow2}{rgb}{0.93,0.91,0.91}
\definecolor{snow3}{rgb}{0.80,0.79,0.79}
\definecolor{snow4}{rgb}{0.55,0.54,0.54}
\definecolor{snow}{rgb}{1.00,0.98,0.98}
\definecolor{springgreen}{rgb}{0.00,1.00,0.50}
\definecolor{steelblue}{rgb}{0.27,0.51,0.71}
\definecolor{tan1}{rgb}{1.00,0.65,0.31}
\definecolor{tan2}{rgb}{0.93,0.60,0.29}
\definecolor{tan3}{rgb}{0.80,0.52,0.25}
\definecolor{tan4}{rgb}{0.55,0.35,0.17}
\definecolor{tan}{rgb}{0.82,0.71,0.55}
\definecolor{thistle1}{rgb}{1.00,0.88,1.00}
\definecolor{thistle2}{rgb}{0.93,0.82,0.93}
\definecolor{thistle3}{rgb}{0.80,0.71,0.80}
\definecolor{thistle4}{rgb}{0.55,0.48,0.55}
\definecolor{thistle}{rgb}{0.85,0.75,0.85}
\definecolor{tomato1}{rgb}{1.00,0.39,0.28}
\definecolor{tomato2}{rgb}{0.93,0.36,0.26}
\definecolor{tomato3}{rgb}{0.80,0.31,0.22}
\definecolor{tomato4}{rgb}{0.55,0.21,0.15}
\definecolor{tomato}{rgb}{1.00,0.39,0.28}
\definecolor{turquoise1}{rgb}{0.00,0.96,1.00}
\definecolor{turquoise2}{rgb}{0.00,0.90,0.93}
\definecolor{turquoise3}{rgb}{0.00,0.77,0.80}
\definecolor{turquoise4}{rgb}{0.00,0.53,0.55}
\definecolor{turquoise}{rgb}{0.25,0.88,0.82}
\definecolor{violetred}{rgb}{0.82,0.13,0.56}
\definecolor{violet}{rgb}{0.93,0.51,0.93}
\definecolor{wheat1}{rgb}{1.00,0.91,0.73}
\definecolor{wheat2}{rgb}{0.93,0.85,0.68}
\definecolor{wheat3}{rgb}{0.80,0.73,0.59}
\definecolor{wheat4}{rgb}{0.55,0.49,0.40}
\definecolor{wheat}{rgb}{0.96,0.87,0.70}
\definecolor{whitesmoke}{rgb}{0.96,0.96,0.96}
\definecolor{white}{rgb}{1.00,1.00,1.00}
\definecolor{yellow1}{rgb}{1.00,1.00,0.00}
\definecolor{yellow2}{rgb}{0.93,0.93,0.00}
\definecolor{yellow3}{rgb}{0.80,0.80,0.00}
\definecolor{yellow4}{rgb}{0.55,0.55,0.00}
\definecolor{yellowgreen}{rgb}{0.60,0.80,0.20}
\definecolor{yellow}{rgb}{1.00,1.00,0.00}
\usepackage{amsfonts}
\usepackage{bm}
\usepackage{longtable}
\usepackage{pifont}
\usepackage{booktabs} 
 
\newcommand{\whofont}{
  \fontfamily{pcr}
  \bfseries 
  \color{royalblue}
}

\newcommand{\gskfont}{
  \bfseries 
  \color{red}
}
\newcommand{\jcafont}{
  \color{orange}
}
\newcommand{\mcfont}{
  \color{red}
}

\DeclareTextFontCommand{\who}{\whofont}
\DeclareTextFontCommand{\gsk}{\gskfont}
\DeclareTextFontCommand{\jca}{\jcafont}
\DeclareTextFontCommand{\mc}{\mcfont}

\shorttitle{Mg~\textsc{ii} PRD, Opacity, Irradiation}
\shortauthors{Kerr, Allred, Carlsson}

\begin{document}


	\title{Modelling Mg~\textsc{ii} During Solar Flares, I: Partial Frequency Redistribution, Opacity, and Coronal Irradiation}
	
	\author{Graham~S.~Kerr}
	\email{graham.s.kerr@nasa.gov}
	\altaffiliation{NPP Fellow, administered by USRA}
	\affil{NASA Goddard Space Flight Center, Heliophysics Sciences Division, Code 671, 8800 Greenbelt Rd., Greenbelt, MD 20771, USA}

	 \author{Joel~C.~Allred}
	 \affil{NASA Goddard Space Flight Center, Heliophysics Sciences Division, Code 671, 8800 Greenbelt Rd., Greenbelt, MD 20771, USA}
	  
	 \author{Mats Carlsson}
	 \affil{Rosseland Centre for Solar Physics, University of Oslo, P.O. Box 1029, Blindern, N-0315 Oslo, Norway}
	 \affil{Institute of Theoretical Astrophysics, University of Oslo, P.O. Box 1029, Blindern, N-0315 Oslo, Norway}
	
	\date{Received / Accepted}
	
	
	\begin{abstract}	
 	The Interface Region Imaging Spectrograph (IRIS) has routinely observed the flaring Mg \textsc{ii} NUV spectrum, offering excellent diagnostic potential and a window into the location of energy deposition. A number of studies have forward modelled both the general properties of these lines and specific flare observations. Generally these have forward modelled radiation via post-processing of snapshots from hydrodynamic flare simulations through radiation transfer codes. There has, however, not been a study of how the physics included in these radiation transport codes affects the solution. A baseline setup for forward modelling Mg \textsc{ii} in flares is presented and contrasted with approaches that add or remove complexity. It is shown for Mg \textsc{ii}: (1) PRD is still required during flare simulations despite the increased densities, (2) using full angle-dependent PRD affects the solution but takes significantly longer to process a snapshot, (3) including Mg \textsc{i} in NLTE results in negligible differences to the Mg \textsc{ii} lines but does affect the NUV quasi-continuum, (4) only hydrogen and Mg \textsc{ii} need to be included in NLTE, (5) ideally the non-equilibrium hydrogen populations, with non-thermal collisional rates, should be used rather than the statistical equilibrium populations, (6) an atom consisting of only the ground state, h \& k upper levels, and continuum level is insufficient to model the resonance lines, and (7) irradiation from a hot, dense flaring transition region can affect the formation of Mg \textsc{ii}. We discuss modifications to the RH code allowing straightforward inclusion of transition region and coronal irradiation in flares.
		\end{abstract}

\section{Introduction}\label{sec:intro}
Solar flares are transient yet dramatic events in the solar atmosphere that release tremendous amounts of energy ($>10^{32}$~ergs per event) and drive space weather. It is thought that energy released in the corona during magnetic reconnection is transported to the transition region and chromosphere via directed beams of non-thermal particles (typically electrons), which lose energy via Coulomb interactions. This results in intense plasma heating and ionisation, mass flows, and consequently the broadband enhancement of the solar radiative output \citep[][and references therein]{1971SoPh...18..489B,2011SSRv..159..107H,2011SSRv..159...19F}. Recently it has also been suggested that alternative forms of energy transport can potentially play a role in transporting energy to the lower atmosphere, such as high frequency Alfv\'enic waves \citep{2008ApJ...675.1645F,2018ApJ...853..101R,2016ApJ...827..101K}. 
 
The particle acceleration mechanism is still uncertain, but the properties of the accelerated electrons can be inferred from non-thermal X-ray bremsstrahlung. Observations from the Reuven Ramaty High Energy Solar Spectroscopic Imager \citep[RHESSI;][]{2002SoPh..210....3L} have been extensively studied, and can be used to drive models of solar flares. Simulations of solar flares using non-thermal electron beams to transport energy can then be compared to observations to determine how consistent this model is to reality.
 
Spectroscopic observations of solar flares are essential if we wish to extract information about the state of the flaring plasma, which can be compared to advanced flare simulations. Both the spectral lines themselves, and the diagnostics they provide, are observables with which we can critically attack models of flare energy transport. Modelling the spectral lines under different physical conditions can also help us build diagnostics. The combination of observations and numerical modelling is powerful as a means to shed light on the physical processes at play in flares.   

Spectral lines that originate from the solar chromosphere are of particular interest, as the chromosphere is the location of energy deposition, and the origin of the bulk of the radiative output in flares \citep{2011SSRv..159...19F,2014ApJ...793...70M}. Since it's launch in 2013 the Interface Region Imaging Spectrograph \citep[IRIS;][]{2014SoPh..289.2733D} has provided high spectral- and spatial- resolution observations of the solar chromosphere and transition region. The strongest chromospheric lines observed by IRIS are the Mg~\textsc{ii} h \& k resonance lines and the Mg~\textsc{ii} subordinate triplet which are emitted in the near-ultraviolet (NUV). Indeed, these lines are amongst the strongest lines in the solar spectrum. These lines offer the potential to study a large swathe of the chromosphere, but are optically thick and require forward modelling with radiation transport codes to fully appreciate the complex formation properties, and, ultimately, to extract the information they carry. 

The h \& k lines  are transitions $3p\,^2\!P^{\rm o}_{1/2,3/2} - 3s\,^2\!S_{1/2}$ ($\lambda = 2802.70$~\AA\ \& $\lambda = 2795.52$~\AA), and the subordinate triplet transitions are $3d\,^2\!D_{3/2} - 3p\,^2\!P^{\rm o}_{1/2,3/2}$, and $3d\,^2\!D_{5/2} - 3p\,^2\!P^{\rm o}_{3/2}$ ($\lambda=2790.78$~\AA, $\lambda=2797.92$~\AA\ \& $\lambda=2798.00$~\AA ; the latter two are blended). 

Early observations using balloon-borne and rocket experiments revealed that these lines are strongly affected by opacity, with central reversal features almost  ubiquitous among the quiescent Sun profiles, though with single peaked profiles in sunspots and shallower reversals in plage and active regions, and intensity ratios that indicate optically thick formation \citep[]{1973A&A....22...61L,1976ApJ...205..599K}. Variations of the line width, intensity, emission peak separation and other features were compared to source types \citep{1995A&A...295..517S}, but only one flare observation existed prior to the IRIS era \citep{1984SoPh...90...63L}. The subordinate lines were relatively little studied but were known to be in absorption on disk and in emission off-limb \citep{1977ApJ...212L.147F}.

Modelling efforts used Mg~\textsc{ii} lines to help build semi-empirical atmospheres and to compare models of temperature stratification to observations \citep[e.g.][]{1973A&A....22...61L}. It was demonstrated that partial frequency distribution (PRD) is required to model the formation of the resonance lines \citep[e.g.][]{1974ApJ...192..769M}, rather than the computationally more tractable problem of complete frequency distribution (CRD). In outline, PRD can be described as follows. Atomic species that are strongly scattering and which form in a relatively low density environment may not experience a sufficient number of elastic collisions to destroy coherency between an incident photon and the scattered or re-emitted photon. Photons absorbed in the line wings are re-emitted in the wing in PRD, where it is easier to escape, whereas in CRD they can be redistributed to the core. In PRD the absorption profile is \textsl{not} equal to the emission profile, and the source function is dependent on frequency. See \cite{2001ApJ...557..389U,2002ApJ...565.1312U} and \cite{1982JQSRT..27..593H} for clear discussions of PRD and the form of redistribution functions 

In the IRIS era, the Mg~\textsc{ii} resonance and subordinate lines have been studied in many solar features, including in hundreds of flares or similar transient heating events. In flares, it has been noted that the profiles mostly appear single peaked in flare ribbons, are significantly enhanced and broadened over the quiet Sun, and show various Doppler shifts of a few $\times 10$~km~s$^{-1}$, and asymmetries, that have been attributed to mass flows \citep[e.g.][]{2015A&A...582A..50K,2015SoPh..290.3525L,2015ApJ...807L..22G,2018PASJ...70..100T,2018ApJ...856...34T}. Despite appearing single peaked the lines remain optically thick, with a line intensity ratio of  $R\sim1.2$ ($R=2$ in the optically thin scenario) \citep{2015A&A...582A..50K}. The subordinate lines undergo similar changes, and are almost exclusively in emission during flares. \cite{2018ApJ...861...62P} found that certain profiles (extremely wide with a blueshifted central reversal) appear at the leading front of flare ribbons, and related the line ratio to opacity to infer a higher opacity at flare peak. 

Detailed and accurate modelling is required to understand the behaviour of these lines in flares, to confirm the origin of observational findings, and to develop diagnostics. Recent modelling has largely focussed on the quiet Sun \citep[e.g.][]{2013ApJ...772...89L, 2013ApJ...772...90L,2015ApJ...806...14P,2015ApJ...809L..30C}, which has related atmospheric properties to line features. Some flare modelling has been performed. \cite{2016ApJ...827..101K} noted differences in the Mg~\textsc{ii} line profiles that originated from atmospheres heated by different energy transport mechanisms, due to the resulting mass motions, similar to observations but much too narrow. In one of those cases the mass motions shifted the absorption profile so that the line appeared single peaked and asymmetric.  \cite{2015SoPh..290.3525L} found that a smaller electron beam spectral index (i.e. deeper heating) resulted in more intense line wings, and that the line core was affected by coronal pressure, with larger pressure acting to fill in the central reversal. A parametric study was performed by \cite{2017ApJ...842...82R}, who noted the effect of varying temperature, electron density, bulk velocity, and microturbulence on the emergent profiles, though not in a self-consistent manner. They noted that increasing the electron density would result in a single peaked profile, which was also noted by \cite{2019arXiv190412285Z}, who modelled a strong flare that produced a very high electron density. \cite{2019arXiv190412285Z} also used more up to date Stark broadening data to explain the narrower-than-observed synthetic spectra, though were unable to fully reach the observed line widths.

When modelling Mg~\textsc{ii} and other species that either require advanced radiation transport (e.g. PRD) or are not included in radiation hydrodynamic simulations, which generally only consider a select number of transitions important for energy balance, post-processing is required. The procedure has been to combine snapshots of dynamic atmospheres (from either a hydrodynamic or radiation hydrodynamic simulation) with static radiation transport codes. This means that statistical equilibrium is imposed on the solution, though non-equilibrium effects may be important \citep{2002ApJ...572..626C}. Some studies using semi-empirical flare atmospheres have also been performed. Clearly a dynamic flare simulation capable of non-equilibrium radiation transport that considers blends and PRD is the ideal, but such a resource is not currently available, in large part due to computational demands.

There can be numerous features of these radiation transport codes that can be variously used, such as employing CRD or PRD, or the number of additional atomic species to solve alongside the one of interest. Given the importance of the Mg~\textsc{ii} as a window on the flaring chromosphere we investigate how these features may or may not affect the solution in flares to assess if the typical approaches used to model Mg~\textsc{ii} spectra in flares is appropriate and sufficient.

Specifically, we look at if PRD is required, if the treatment of PRD affects the results, if a small model atom can be used, if Mg~\textsc{i} must be included in the solution, and the impact of re-solving the hydrogen populations in statistical equilibrium. We also investigate transition region and coronal irradiation. All Mg~\textsc{ii} modelling discussed here employs statistical equilibrium to obtain the atomic level populations. Paper 2 in this series will discuss non-equilibrium effects \citep{NEQ_inprep}.


\section{Numerical Resources and Flare Simulations}\label{sec:codes}
\subsection{RADYN}\label{sec:radyn}
\texttt{RADYN} \citep{1992ApJ...397L..59C,1997ApJ...481..500C} is a radiation hydrodynamics numerical code that models the solar atmosphere, with a particular focus on the chromosphere, by considering the coupled non-linear equations of hydrodynamics, radiation transport, and non-equilibrium atomic level populations (for certain species). It has the facility to simulate solar flares by the injection of energy via a directed beam of non-thermal particles \citep{1999ApJ...521..906A,2005ApJ...630..573A,2015ApJ...809..104A} or by downward propagating Alfv\'enic waves \citep{2016ApJ...827..101K}. We focus on electron beams for the remainder of this study. For a recent description of the \texttt{RADYN} code we point the reader to \cite{2019ApJ...871...23K}, and to \cite{FP_inprep}, and references therein. We note  important details below. 

The non local thermodynamic equilibrium (NLTE)  and non-equilibrium (NEQ) radiation transfer problem is solved in detail for species important for chromospheric energy (H, He~\textsc{i}, He~\textsc{ii}, Ca~\textsc{ii}). The bound-bound and bound-free transitions from these species act as sources and sinks of energy, with non-local effects included. Additional radiative cooling is included by summing the emissivities from all transitions in the CHIANTI atomic database \citep{1997A&AS..125..149D,2013ApJ...763...86L}, excluding the bound-bound transitions solved in detail. Backwarming and photoionisation/photoexcitation by X-ray, extreme-ultraviolet, and ultraviolet emission from the flare-heated corona is included by injecting a donward-directed incident radiation from the loop apex. Thermal conduction is a modified form of Spitzer conductivity, saturating at the free-streaming limit \citep{1980ApJ...238.1126S}. Radiation transport assumes complete frequency redistribution, but we truncate the Lyman lines at 10 Doppler widths in an effort to avoid overestimating radiative losses, which can happen if CRD is employed when PRD should really be used \citep[e.g.][]{2002ApJ...565.1312U}.  

Flares are simulated by injecting a non-thermal particle distribution at the loop apex, with a power-law energy spectrum defined by the instantaneous energy flux $F$ [erg cm$^{-2}$ s$^{-1}$], above a low-energy cutoff $E_{\mathrm{c}}$ [keV], and with a spectral index $\delta$. The energy losses and propagation of the beam of particles are modelled using the Fokker-Planck equations, which include the important scattering terms. The Fokker-Planck solver has recently been updated from the implementation described in \cite{2015ApJ...809..104A}, and will be described more fully in \cite{FP_inprep}. Non-thermal collisional ionisation and excitation from the ground state of hydrogen is included using the methods of \cite{1993A&A...274..917F}.

\subsection{Flare Simulations}
\begin{figure*}
	\centering 
	{\includegraphics[width = .95\textwidth, clip = true, trim = 0.cm 0.cm 0.cm 0.cm]{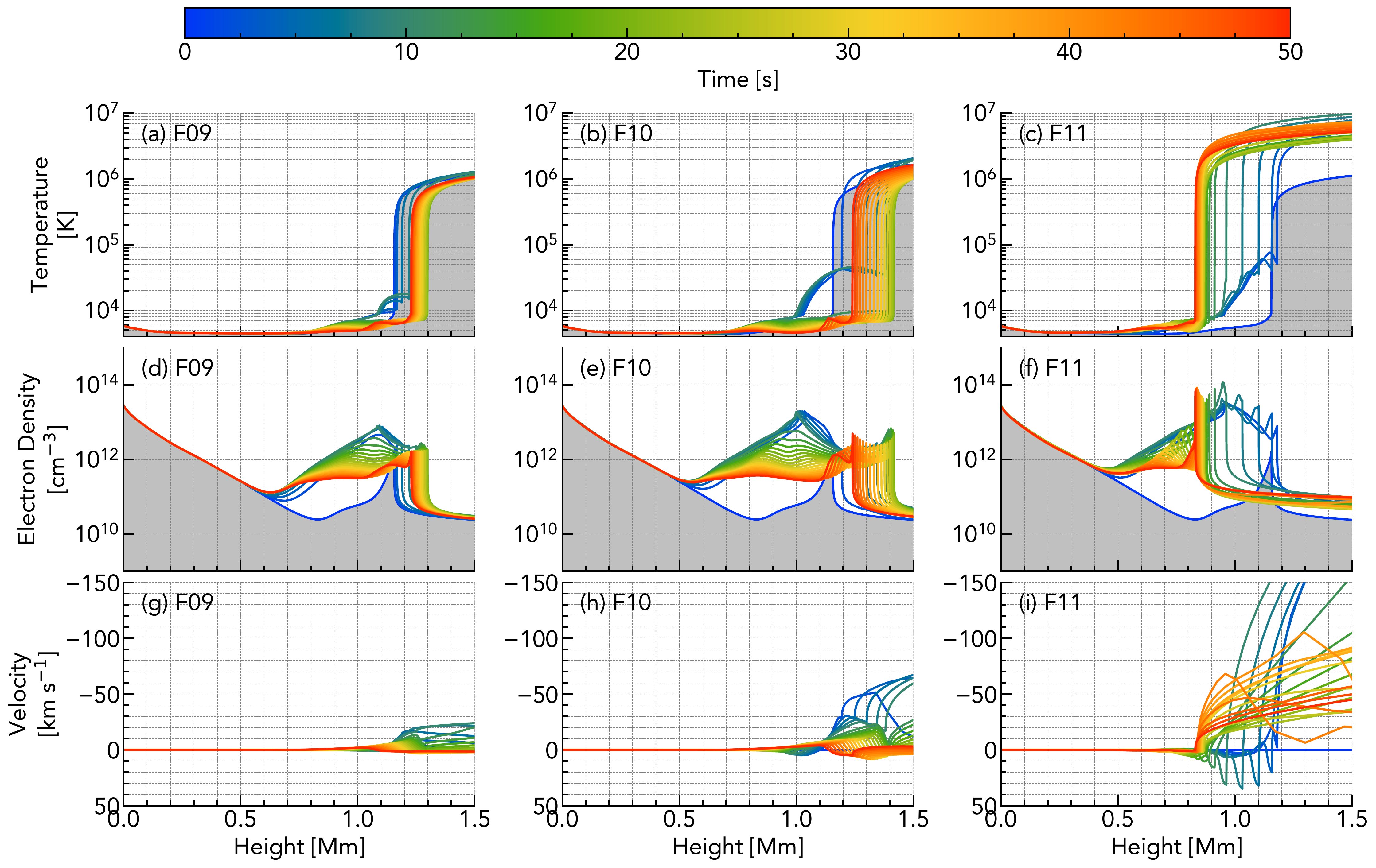}}	
	\caption{\textsl{The stratification of temperature (a,b,c), electron density (d,e,f), and macroscopic velocity (g,h,i; uplows are negative) in the three flare simulations. The first column shows the F9 simulation, second column shows the F10 simulation, and third column shows the F11 simulation. Colour represents time. Recall that heating ceased at $t=10$~s}}
	\label{fig:flare_atmos}
\end{figure*}

Three flares of different strengths were simulated using \texttt{RADYN}, driven by non-thermal electron beams that were injected into a pre-flare atmosphere shown as the shaded portion in Figure~\ref{fig:flare_atmos}. The $z=10$~Mm pre-flare atmosphere spanned the sub-photosphere through to the corona, with an apex temperature of $T_{\mathrm{cor}}=3.2$~MK and apex hydrogen density $n_{\mathrm{H,cor}}=6.6\times10^{9}$~cm$^{-3}$. The photosphere ($z=0$ where $\tau_{5000}=1$) had a temperature $T=5800$~K. These atmospheres were initially in radiative equilibrium, with non-radiative heating applied to maintain the corona and photosphere (column masses $c_{\mathrm{mass}}< 1\times10^{-6}$~g~cm$^{-2}$ and $c_{\mathrm{mass}}> 7.6$~g~cm$^{-2}$). A reflecting boundary condition was employed at the loop apex to mimic incoming disturbances from the other leg of the flare loop, and a transmitting boundary condition was used at the base of the loop. The electron beam fluxes were $F=[1\times10^{9}, 1\times10^{10}, 1\times10^{11}]$~erg~cm$^{-2}$~s$^{-1}$ (hereafter F9, F10, \& F11) which were modelled as power law distributions with spectral index $\delta = 5$ and a low-energy cutoff $E_{c} = 20$~keV. Energy was injected at a constant rate for $t=10$~s, and the simulations allowed to continue to evolve until $t=50$~s. 

Figure~\ref{fig:flare_atmos} shows the evolution of the atmospheres as a function of time in each simulation, where it is clear that the atmosphere is disturbed more strongly and rapidly in the highest flux simulation. Electron densities and macroscopic flows were notably larger in the F11 simulation.

\subsection{RH}\label{sec:RH}
\texttt{RH} \citep{2001ApJ...557..389U} is a NLTE radiation transfer code that solves the coupled equations of statistical equilibrium (SE) and radiation transfer, using the MALI (Multi-level Approximate Lambda Iteration) formalism of \cite{1991A&A...245..171R,1992A&A...262..209R}.  Overlapping wavelengths (blends), and the effects of partial frequency redistribution are included where appropriate. A model atmosphere is provided to \texttt{RH}, with the temperature, electron density, macroscopic velocity, and microturbulent velocity defined on a given depth scale (in our case a 1D plane-parallel geometry, though other geometries are available). Atomic level populations are solved assuming statistical equilibrium for species  specified by the user, with input atomic models containing details of bound-bound and bound-free transitions, collisional rates, photoionisation cross-sections, recombination rates and other relevant data. Multiple species are solved simultaneously. As well as opacities and emissivities from species selected to be treated in NLTE, background sources are included in LTE. These include those atomic species specified by the user, H$^{-}$ bound-bound \& bound-free processes, Thomson and Rayleigh scattering, H free-free processes, and bound-free processes in OH and CH molecules. 

When forward modelling radiation from dynamic atmospheres it is necessary to solve each time as an independent atmosphere. Unfortunately this means that any NEQ effects are not accounted for. In essence we neglect the `history' of the simulation that led to that instantaneous snapshot, which can lead to an erroneous ionisation fraction \citep[e.g.][ discuss this in the case of hydrogen]{2002ApJ...572..626C}.  This is somewhat mitigated by using the non-equilibrium electron density provided in the model atmosphere fixed in the \texttt{RH} solution. Non-equilibrium effects are necessarily sacrificed in order to utilise the more advanced radiation transfer than is typically included in dynamic simulations.

To include hydrogen level populations with non-equilibrium effects and non-thermal ionisations/excitations when computing opacity sources in the flaring atmospheres, the populations from \texttt{RADYN} can be provided using the `OLD\_POPULATIONS' starting condition and setting hydrogen to `PASSIVE' (`PASSIVE' species are treated only as a background opacity source, whereas the NLTE radiation transfer is solved for `ACTIVE' species).  This method allows the user provided populations to be used to compute the background opacity, but the NLTE lines and continua will not be computed by \texttt{RH}. We have instead modified \texttt{RH} to use user provided populations (in our case obtained from \texttt{RADYN}) for an `ACTIVE' species, whilst skipping the level population recalculation step in the iterative loop so that populations remained fixed. \texttt{RH} will therefore use the provided populations directly to solve the NLTE radiation transfer, and for processes such as charge exchange that affect other species. This was implemented by introducing a new starting condition for the initial guess of atomic level populations before the iterations begin. During the iterative loop this `guess' is unchanged. Other species not using this keyword are solved in the usual way, where the populations are recalculated in the iterative loop until convergence is achieved. Our modifications allow us to use NEQ populations from \texttt{RADYN} with the more advanced radiation transfer offered by \texttt{RH}, such as PRD and overlapping transitions. Note that we are forced for the time being to use the SE populations for Mg~\textsc{ii}, as this species was not included in the \texttt{RADYN} flare simulations. A follow up work will investigate those effects, and for now we restrict this study to the other physics involved in the modelling of the Mg~\textsc{ii} spectra. While the focus of this work is Mg~\textsc{ii}  and not hydrogen, we introduce our modifications here because the Mg~\textsc{ii} results presented here were computed using this version of the code. Future research will use the hydrogen lines computed alongside the Mg~\textsc{ii} profiles, and will employ this mechanism for other species.

Snapshots of our \texttt{RADYN} flares atmospheres were post-processed through \texttt{RH}, with a cadence of $0.5$~s in the heating and initial cooling phase ($t = 0 - 15$~s), and with a cadence of $1$~s for the remainder of the simulation ($t = 15-50$~s), giving 65 snapshots in total per simulation. 

Our `standard' setup, which serves as the benchmark, is as follows. As well as a 10-level-plus-continuum singly ionised Mg atom (we use the model atom from \citealt{2013ApJ...772...89L}), the NLTE atomic level populations were solved for: a 23-level-plus-continuum Silicon atom (that included transitions of Si~\textsc{i} \& Si~\textsc{ii}, with the continuum level being the ground state of Si~\textsc{iii}); an eight-level-plus-continuum carbon atom (that included transitions of C~\textsc{i} \& C~\textsc{ii}, with the continuum level being the ground state of C~\textsc{iii}); a thirteen-level-with-continuum oxygen atom (that included transitions of O~\textsc{i} with the continuum level being O~\textsc{ii}).  Additionally we provided the NLTE, non-equilibrium atomic level populations of a five-level-plus-continuum hydrogen atom, and a five-level-plus-continuum singly ionised calcium atom, obtained from the \texttt{RADYN} simulations. The hydrogen populations that we provided also included non-thermal collisional excitations and ionisations from the ground state of hydrogen according to \cite{1993A&A...274..917F}.  

For the lines in which coherent scattering was modelled we used the fast approximation of angle-dependent PRD treatment developed by \cite{2012A&A...543A.109L}, also known as `hybrid' PRD (H-PRD). This scheme approximates full angle-dependent PRD (AD-PRD) for atmospheres with macroscopic flows, with substantially lower computational expense. It works by transforming from the observer's rest frame to the rest frame of the moving gas parcel, allowing the assumption of radiation field isotropy to be used. In our simulations we employ H-PRD for the h \& k lines and subordinate triplet. Since two of the subordinate triplet lines share an upper level ($3d\,^2\!D_{3/2}$) we also modelled cross redistribution effects (XRD), also known as \textsl{Raman} scattering. A comparison to AD-PRD is given in Section~\ref{sec:adprd}. 

\subsection{Note on H in CRD During Flares}\label{sec:neqHpopnote}
While \texttt{RADYN} allows the NEQ H populations to be obtained (with non-thermal effects also included), it uses CRD when modelling the hydrogen lines, potentially leading to inaccuracies in the hydrogen ionisation fraction and electron density. The hydrogen ionisation fraction stratification, and therefore electron density, is affected by the choice of CRD or PRD \citep[e.g.][]{1995ApJ...455..376H}. This will affect the ionisation fraction of other species. In addition to the impact of the electron density, other processes such as charge exchange with neutral hydrogen or protons, and optical pumping, can be important for certain species. For example, charge exchange is important for setting the O~\textsc{i}/O~\textsc{ii} fraction \citep{1993ApJ...402..344C}. 

This is somewhat mitigated by truncating the Lyman lines at 10 Doppler widths, limiting the radiation losses from the wings. Furthermore, in flares densities are significantly enhanced, reducing the impact of PRD on the Lyman lines. 

We forward modelled hydrogen using \texttt{RH}, with \texttt{RADYN} flare atmospheres as input, where Ly$\alpha$ and Ly$\beta$ were treated using both CRD and PRD in SE (with the caveat that the electron density was fixed). It can be concluded from our simulations that the CRD assumption employed by \texttt{RADYN} is, in fact, not so incorrect for moderate-larger flares (the F10 and F11 simulations). Using the NEQ \texttt{RADYN} hydrogen populations computed in CRD will not lead to overly spurious results when forward modelling other species. In fact the inclusion of NEQ and non-thermal effects seems to have a larger impact than the choice of CRD or PRD.  The SE populations in PRD and CRD, the NEQ populations in CRD, and the resulting Ly~$\alpha$ line profiles are are shown in Appendix~\ref{sec:HydPRD}. Further discussion of the treatment of the hydrogen populations is found in Section~\ref{sec:otherspecies}.


\section{Partial or Complete Frequency Redistribution}\label{sec:prd_or_crd}
\subsection{Emergent Intensities and Profiles}\label{sec:intensitycomps}
\begin{figure}
	\centering 
	{\includegraphics[width = .5\textwidth, clip = true, trim = 0.cm 0.cm 0.cm 0.cm]{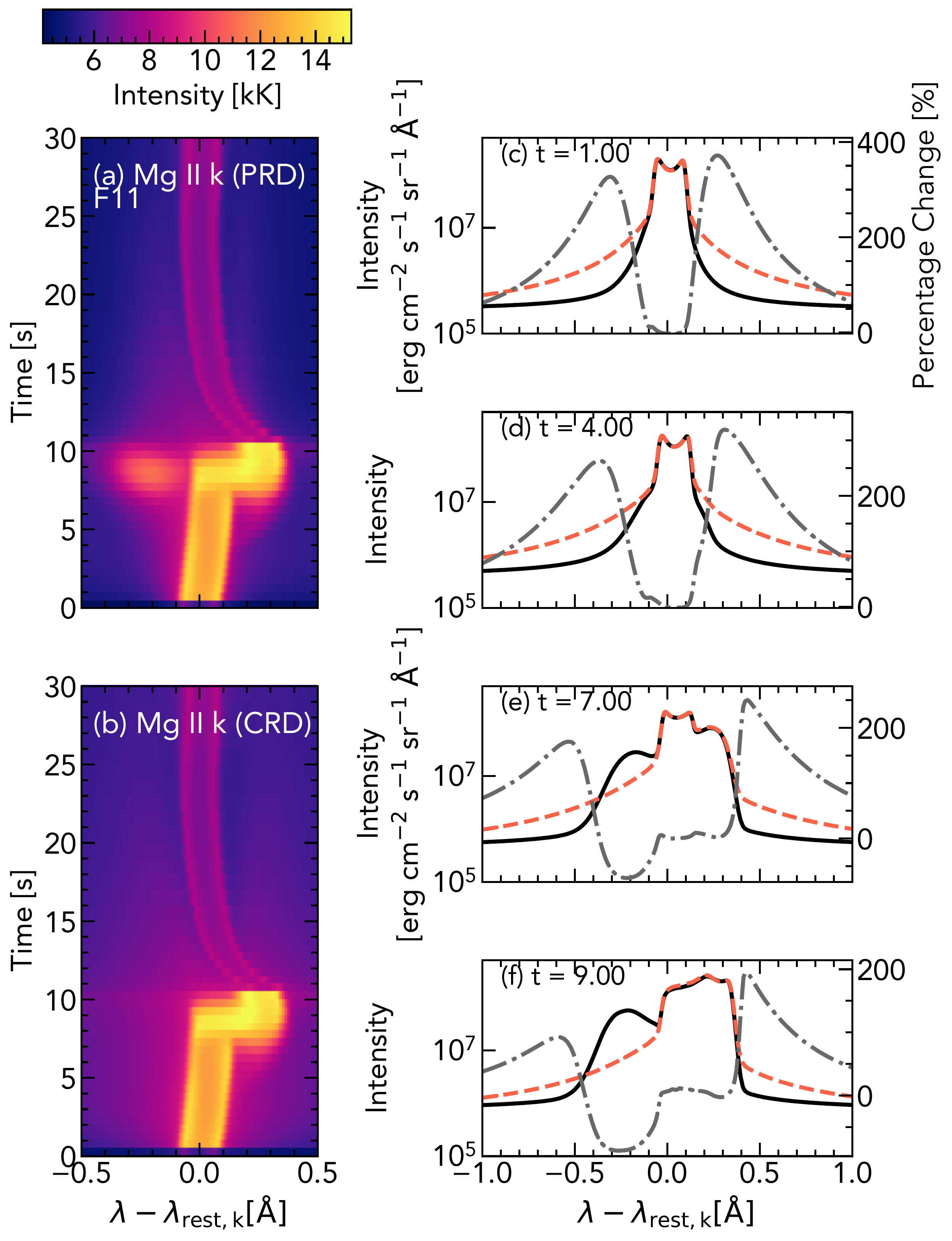}}	
	\caption{\textsl{Emergent intensity of the Mg~\textsc{ii} k line in the F11 simulation. Panels (a) and (b) show stackplots of intensity (radiation temperature units), where the y-axis shows simulation time. Panel (a) is the PRD solution, and panel (b) is the CRD solution. Panels (c-f) show line profiles at different times in the simulation, black solid lines are PRD, and red dashed lines are CRD. The grey dot-dashed lines are the percentage change between the solutions.}}
	\label{fig:stackplot_F11}
\end{figure}
\begin{figure}
	\centering 
	{\includegraphics[width = .5\textwidth, clip = true, trim = 0.cm 0.cm 0.cm 0.cm]{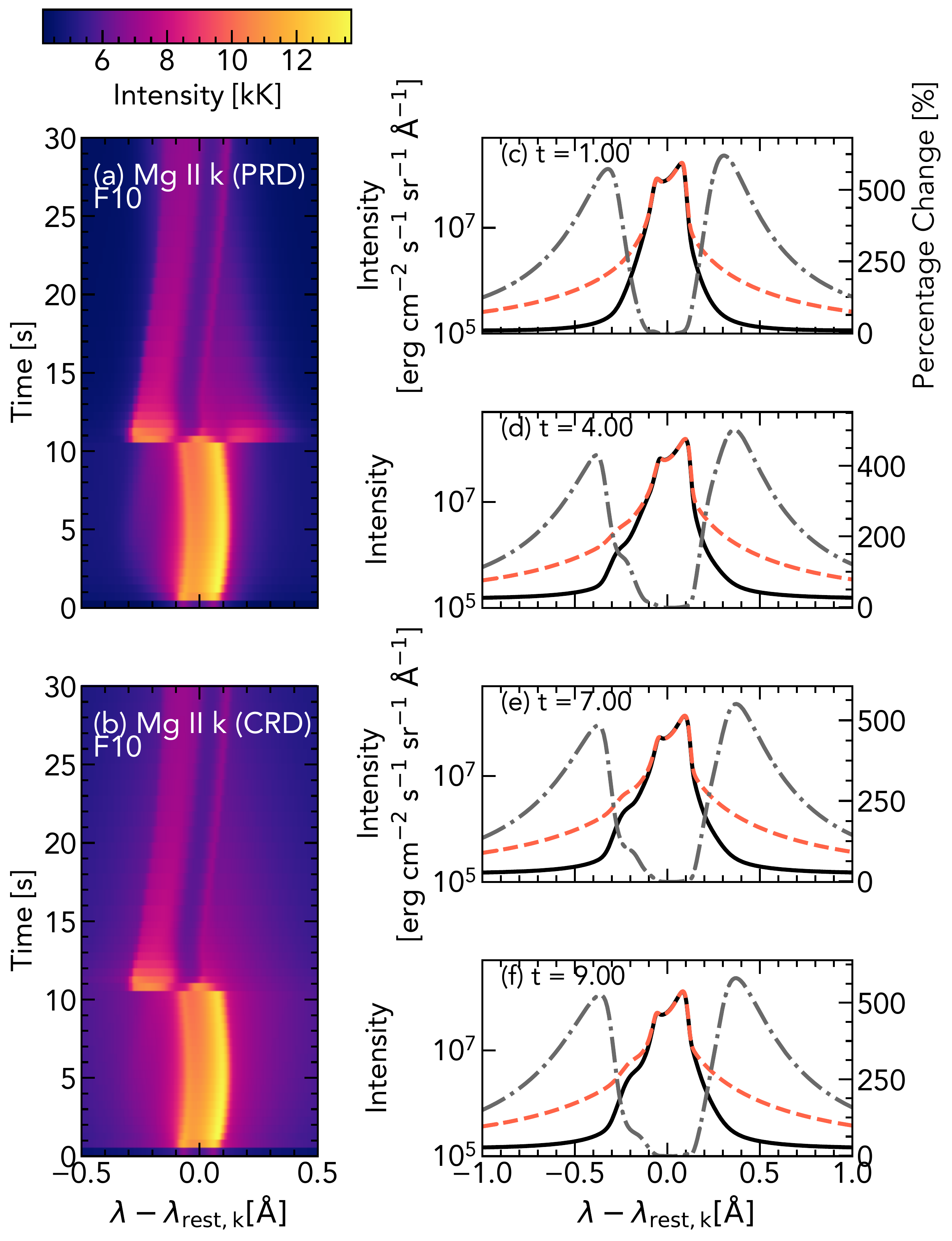}}	
	\caption{\textsl{Same as Figure~\ref{fig:stackplot_F11} but for the F10 simulation.}}
	\label{fig:stackplot_F10}
\end{figure}
\begin{figure}
	\centering 
	{\includegraphics[width = .5\textwidth, clip = true, trim = 0.cm 0.cm 0.cm 0.cm]{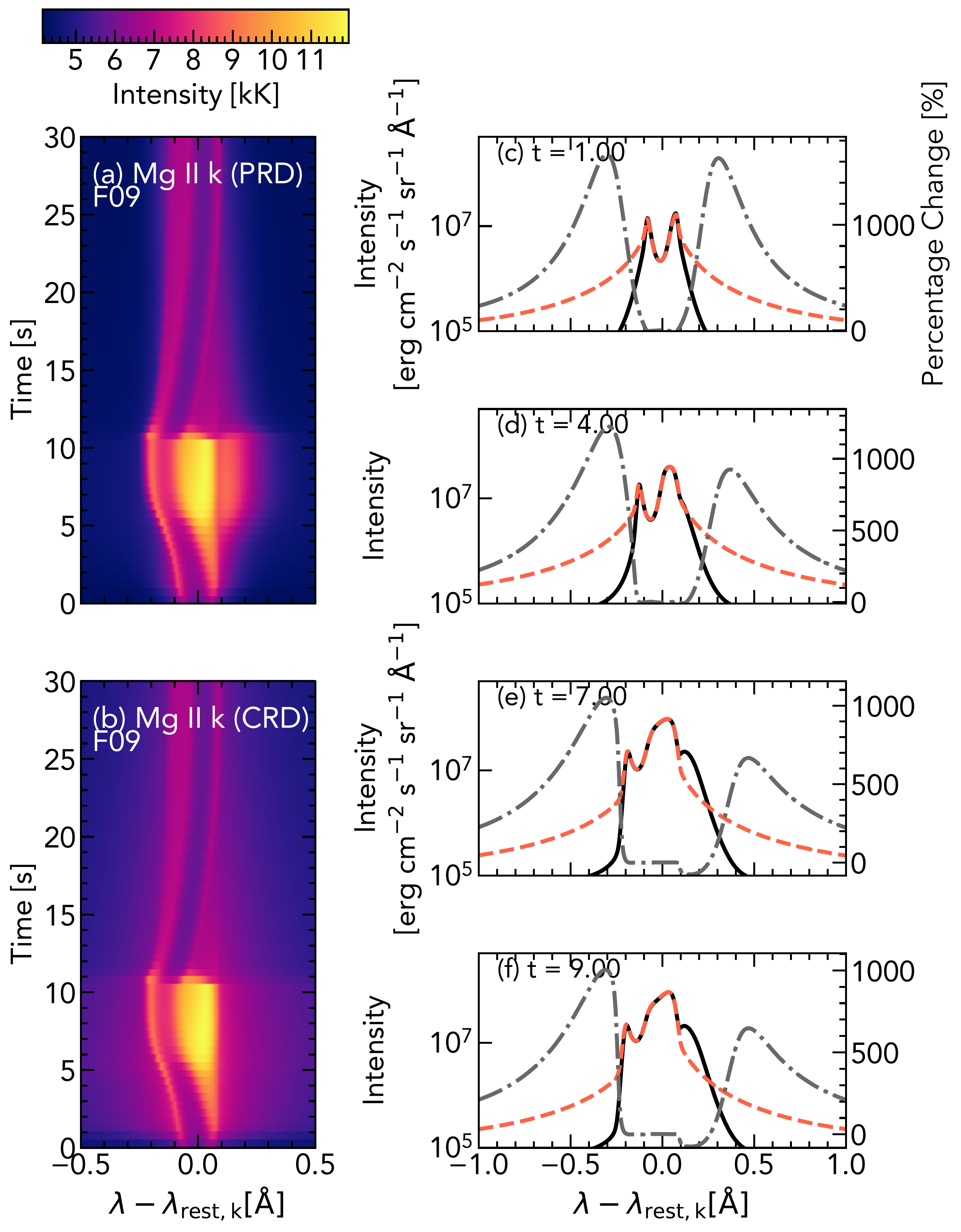}}	
	\caption{\textsl{Same as Figure~\ref{fig:stackplot_F11} but for the F9 simulation.}}
	\label{fig:stackplot_F09}
\end{figure}
\begin{figure}
	\centering 
	{\includegraphics[width = .5\textwidth, clip = true, trim = 0.cm 0.cm 0.cm 0.cm]{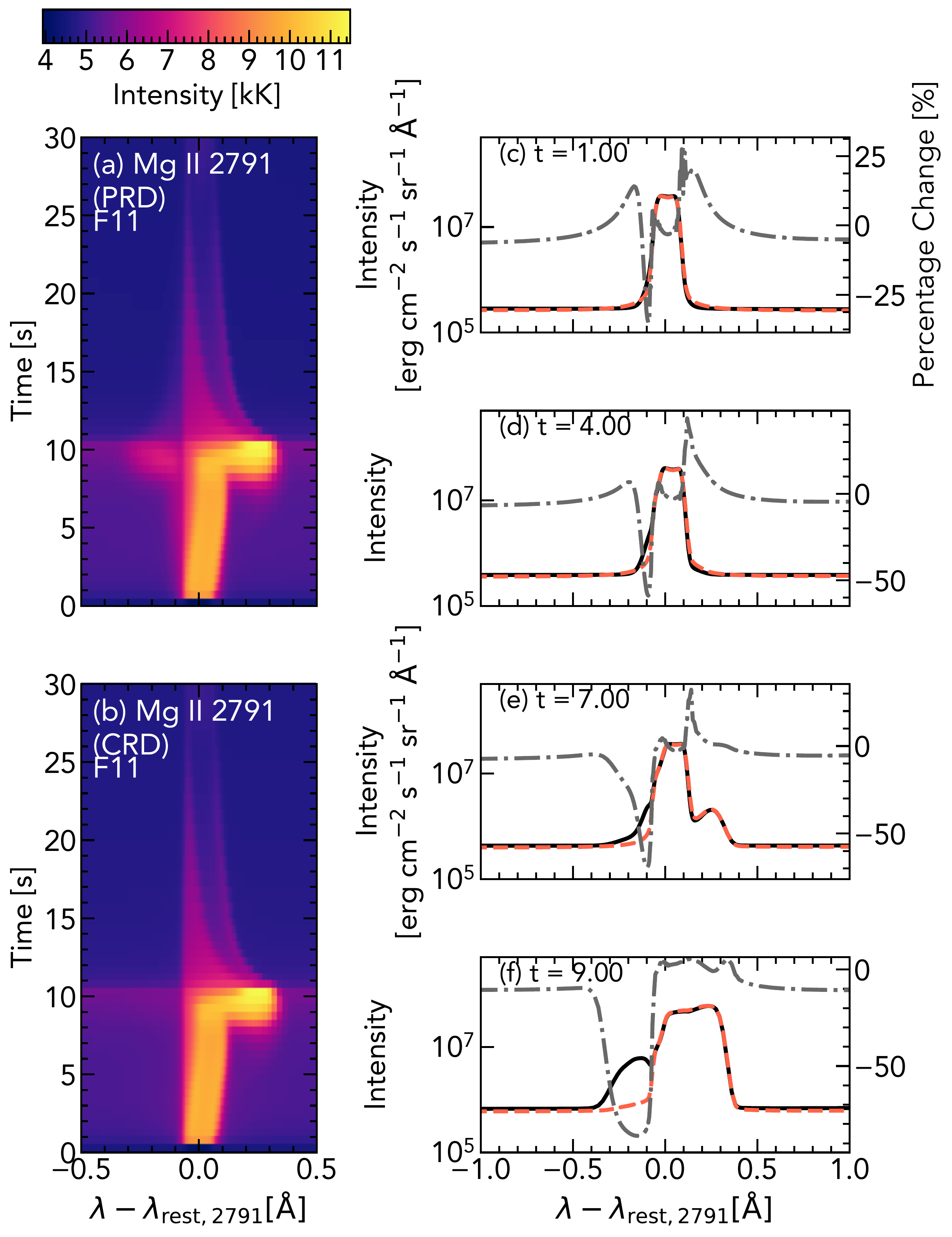}}	
	\caption{\textsl{Same as Figure~\ref{fig:stackplot_F11} but for the Mg~\textsc{ii} 2791~\AA\ subordinate line.}}
	\label{fig:stackplot_F11_sub}
\end{figure}

It is well known that PRD effects are important for the formation of Mg~\textsc{ii} in the quiet Sun, but in flares the density is significantly enhanced, increasing collisions and potentially driving the formation closer to CRD (c.f. Section~\ref{sec:neqHpopnote}). If it transpired that CRD could be used for Mg~\textsc{ii} in larger flares, in which the densities are higher, then the computational problem becomes more tractable. 

The three flare simulations were processed through \texttt{RH} using the standard setup, and again with the Mg~\textsc{ii} lines computed in CRD.  Figures~\ref{fig:stackplot_F11},\ref{fig:stackplot_F10},\ref{fig:stackplot_F09} \& \ref{fig:stackplot_F11_sub} show the profiles of the Mg~\textsc{ii} k line in the F11, F10, \& F9 simulations, and the Mg~\textsc{ii} 2791~\AA\ line in the F11 simulation, respectively. In each of those figures the PRD and CRD solutions are compared. Panels (a,b) show the profiles as functions of time, where the colour represents radiation temperature. Panels (c-f) show individual snapshots, and the percentage change between the two methods of solution, $(I_{\mathrm{CRD}} - I_{\mathrm{PRD}})/I_{\mathrm{PRD}} \times 100$. 

It is immediately clear that unlike the Lyman lines, PRD is still required even in the strongest flare simulation, despite the increased densities. As flare strength (and consequently the electron densities) increases, PRD effects become less important, with the F9 simulation showing up to $\sim1000$~\% intensity changes compared to $\sim200$~\% in the F11 simulation. The resonance line cores and emission peaks were far less affected, if at all, while the inner wings showed the largest differences. While some features in the line wings were preserved in the CRD solution, for example the shoulder in the red wing in the F11 simulation at $t>7$~s (Figure~\ref{fig:stackplot_F11}(e,f)), not all were. In the F11 simulation a feature appears in the blue wing that is only present in the PRD solution, similarly in the F9 simulation a red wing feature forms only in PRD. We will comment on these asymmetries in Section~\ref{sec:formcomps}. Given that both line intensities and, perhaps more importantly, profile shapes are not fully preserved in CRD we recommend that PRD is always used to model the Mg~\textsc{ii} resonance lines even in large flares.

The subordinate lines are typically modelled in CRD, but our results suggest that PRD (with XRD) is required for these lines also. Figure~\ref{fig:stackplot_F11_sub} shows that while differences over much of the line are modest relative to the h \& k lines, there are locations were a few $\times10$~\% differences are present, and that, similar to the resonance lines, certain features only appear in the PRD solution. 

\subsection{Line Formation}\label{sec:formcomps}

\begin{figure*}
	\centering 
	\hbox{
	\hspace{0.4in}
        \subfloat{\includegraphics[width = 0.45\textwidth, clip = true, trim = 0.cm 0.cm 0cm 0cm]{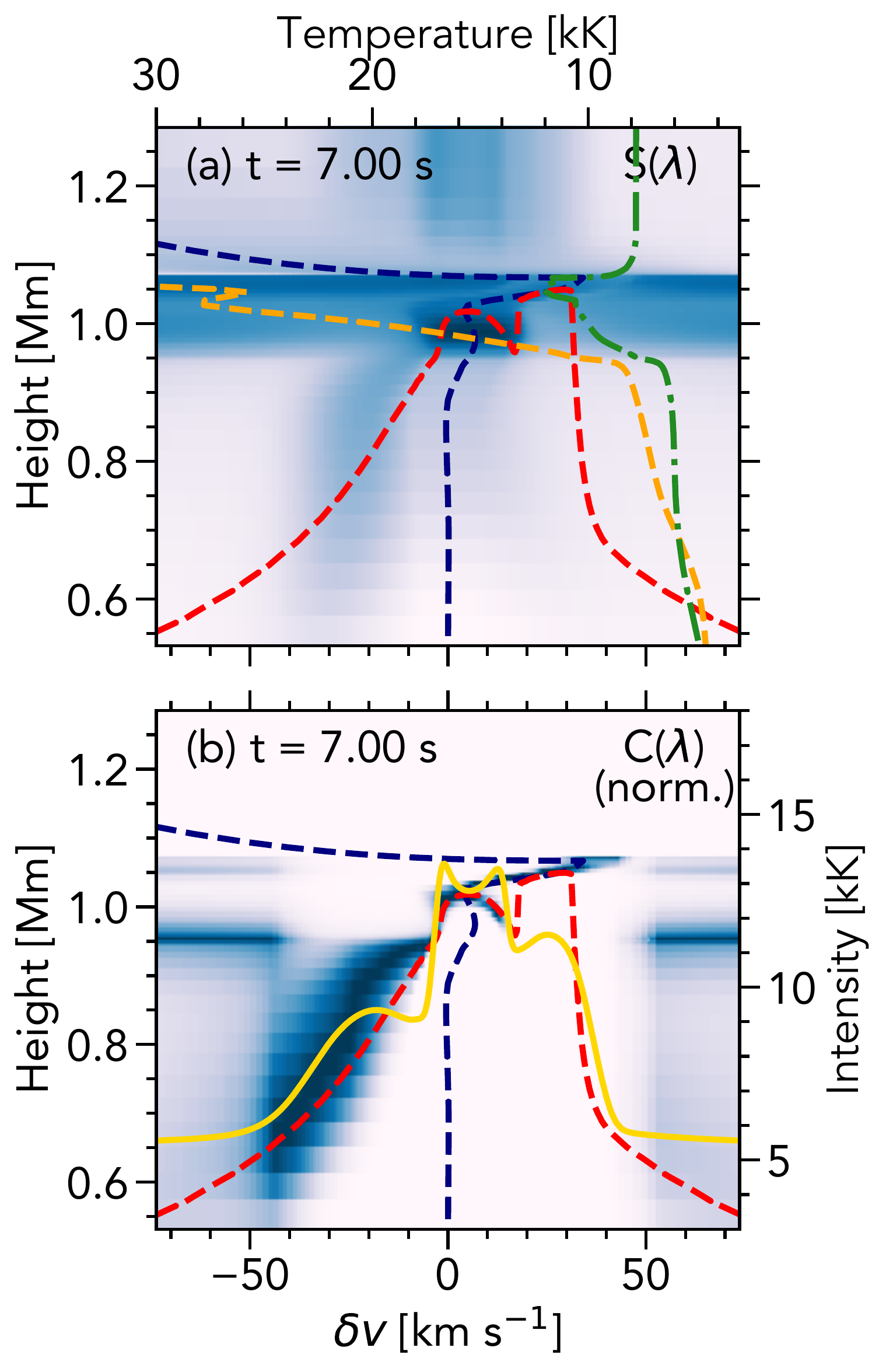}}	
        \subfloat{\includegraphics[width = 0.45\textwidth, clip = true, trim = 0.cm 0.cm 0cm 0cm]{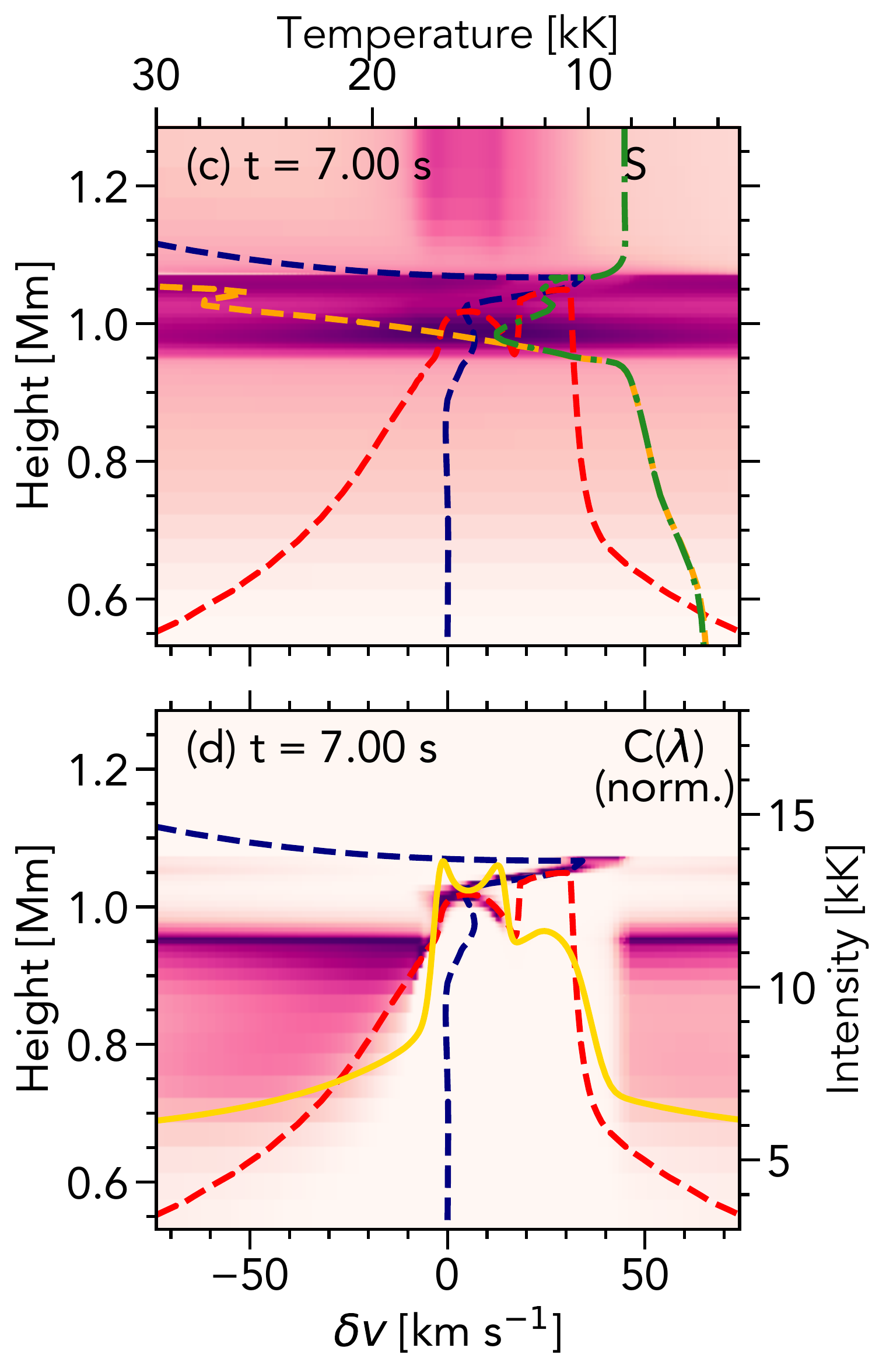}}	
        }
	\caption{\textsl{Illustrating Mg~\textsc{ii} k line formation in the PRD (a,b) and CRD (c,d) scenarios. The background images in panels (a) \& (c) are the source function $S_{\lambda}(z)$ as a function of height, and wavelength (shown as a Doppler shift scale, centred on $\delta v = 0$~km~s$^{-1}$, corresponding to $\lambda_{\mathrm{rest,k}}$).  The background image in panels (b) \& (d) is the contribution function to the emergent intensity, $C_{\lambda}(z)$. Also shown are: the $\tau_\lambda = 1$ layer (red dashed line), the atmospheric velocity (upflows are negative; blue dashed line), the gas temperature as a function of height (lefthand and upper axis, orange  dashed line), the source function of the line core (green dot-dashed line) and the emergent profile, in units of radiation temperature is overlaid (righthand axis; yellow line).}}
	\label{fig:Cfn_prd_vs_crd}
\end{figure*}

To understand the formation of line features we can use the contribution function to the emergent intensity, $C_{\lambda\mu}(z)$, which effectively describes where in the atmosphere the radiation originates \citep[e.g.][]{1986A&A...163..135M,1998LNP...507..163C}. Integrating through a chosen depth scale yields the emergent intensity $I_{\lambda\mu}$. Defining on a height scale $z$, we can write the specific intensity in a plane-parallel semi-infinite atmosphere as:

\begin{equation}\label{eq:contfneqn}
	I_{\lambda\mu} = \int \frac{1}{\mu}S_{\lambda}(z)e^{-\tau_{\lambda}\nu}\chi_{\lambda}\mathrm{d}z = \int C_{\lambda,\mu}(z) \mathrm{d}z,
\end{equation}

\noindent where $\mu = \cos\theta$ describes the viewing angle ($\theta$ being the angle between the line of sight and the normal), $S_\lambda(z)$ is the source function (with the line source function being independent of $\lambda$ in the CRD case), $\tau_{\lambda}$ is the optical depth, and $\chi_{\lambda}$ is the monochromatic opacity. Hereafter the $\mu$ subscript is dropped and we focus on near-disk centre results. 

\begin{figure*}
	\centering 
	{\includegraphics[width = .95\textwidth, clip = true, trim = 0.cm 0.cm 0.cm 0.cm]{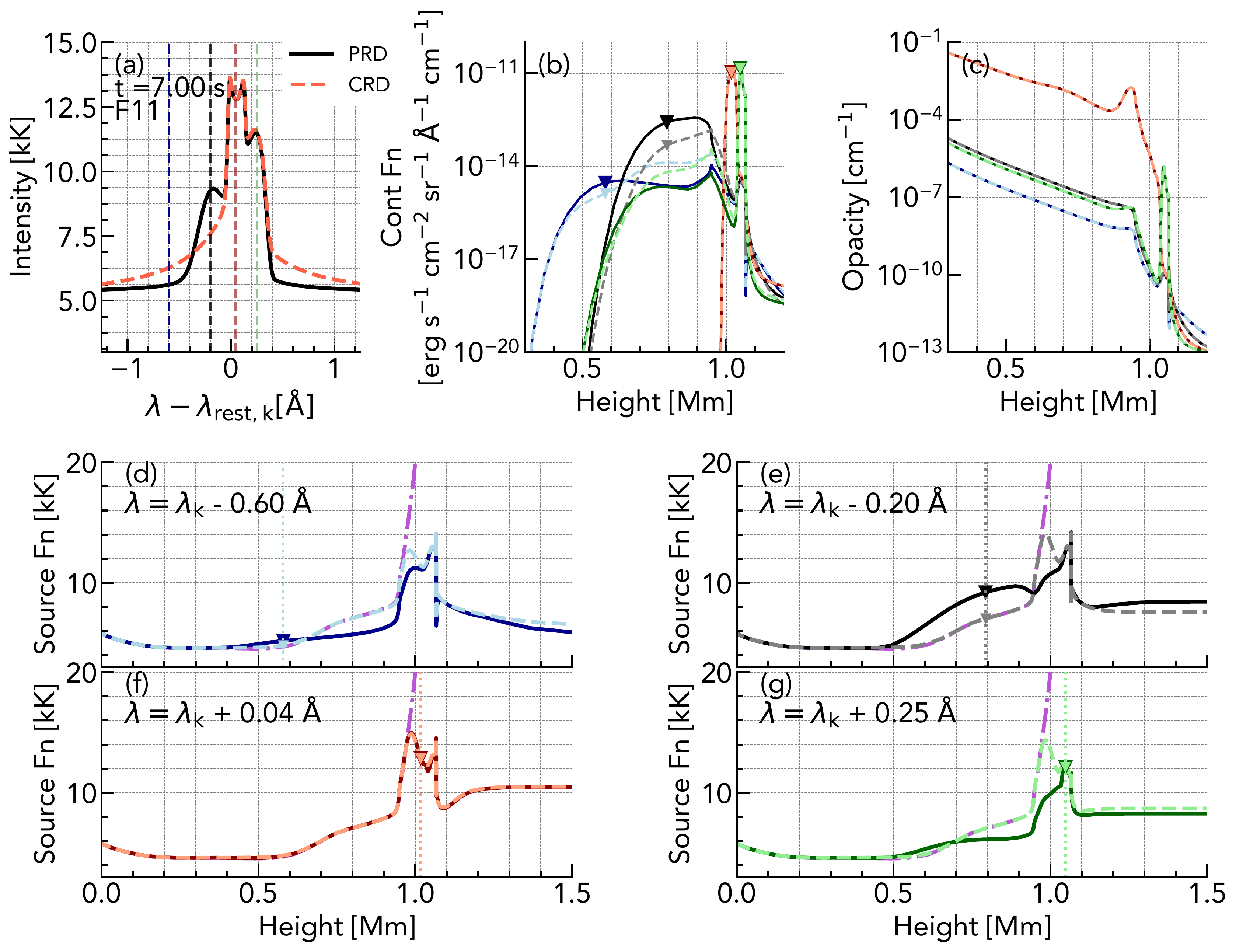}}	
	\caption{\textsl{Detailed line formation at selected wavelengths. Panel (a) shows the Mg~\textsc{ii} k line profile computed in PRD (black) \& CRD (dashed red), at $t=7$~s in the F11 simulation. The dashed vertical lines are the wavelengths for which the contribution functions, opacities, and source functions are shown in panels (b-g). In all the other panels the darker coloured solid lines are the PRD solutions, the lighter coloured dashed lines is the CRD solutions: blue is $\lambda_{\mathrm{rest}} - 0.60$~\AA, black is $\lambda_{\mathrm{rest}} - 0.20$~\AA, red is $\lambda_{\mathrm{rest}} + 0.04$~\AA, and green $\lambda_{\mathrm{rest}} + 0.25$~\AA. Panel (b) shows the contribution functions, with symbols showing the height at which $\tau_\lambda = 1$ for each wavelength. Panel (c) shows the opacity. Panels (d-g) show the source functions, with symbols and dotted lines indicating the height at which $\tau_\lambda = 1$. Also shown in purple is the atmospheric temperature (i.e. we here compare the source functions to the Planck function).}}
	\label{fig:Cfn_prd_vs_crd_detailed}
\end{figure*}

Figure~\ref{fig:Cfn_prd_vs_crd} illustrates the differences between the PRD and CRD line formation for the Mg~\textsc{ii} k line in the F11 simulation at $t=7$~s. In that figure the background image in the top panels (a,c) is the source function. The frequency dependence of the source function in the PRD solution is clear, with a larger source function in the core compared to the wings, since it is dependent on the redistribution function and the radiation field. In the CRD case the line source function is uniform in frequency, with redistribution from the wings to core. The source function follows the local temperature at all wavelengths, decoupling from the Planck function at the same height in the chromosphere. In the PRD case, however, different frequencies decouple from the Planck function at different heights due to coherency. Here we define the line core as the wavelength at which the height of $\tau_{\lambda} = 1$ is maximum (i.e. the part of the optically thick component that forms highest in the atmosphere), and at this time the core is located in the redshifted component ($\sim\lambda_{\mathrm{rest,k}}+25$~\AA. Even though in PRD the source function has decoupled from the Planck function much lower in the atmosphere compared to the CRD case which remains coupled until $\sim0.9$~Mm (the orange dashed line shows the Planck function), the source functions at the emission height are identical, and have increased in response to the flare. 

Perhaps the largest difference is that there is an increase in the source function in the blue wing, extending over a few hundred km, present in the PRD solution but absent in the CRD solution. This grows in strength with time, along with the bump in the redwing that is shared in both the PRD and CRD solutions. 

\begin{figure*}
	\centering 
	\hbox{
	\hspace{0.4in}
        \subfloat{\includegraphics[width = 0.45\textwidth, clip = true, trim = 0.cm 0.cm 0cm 0cm]{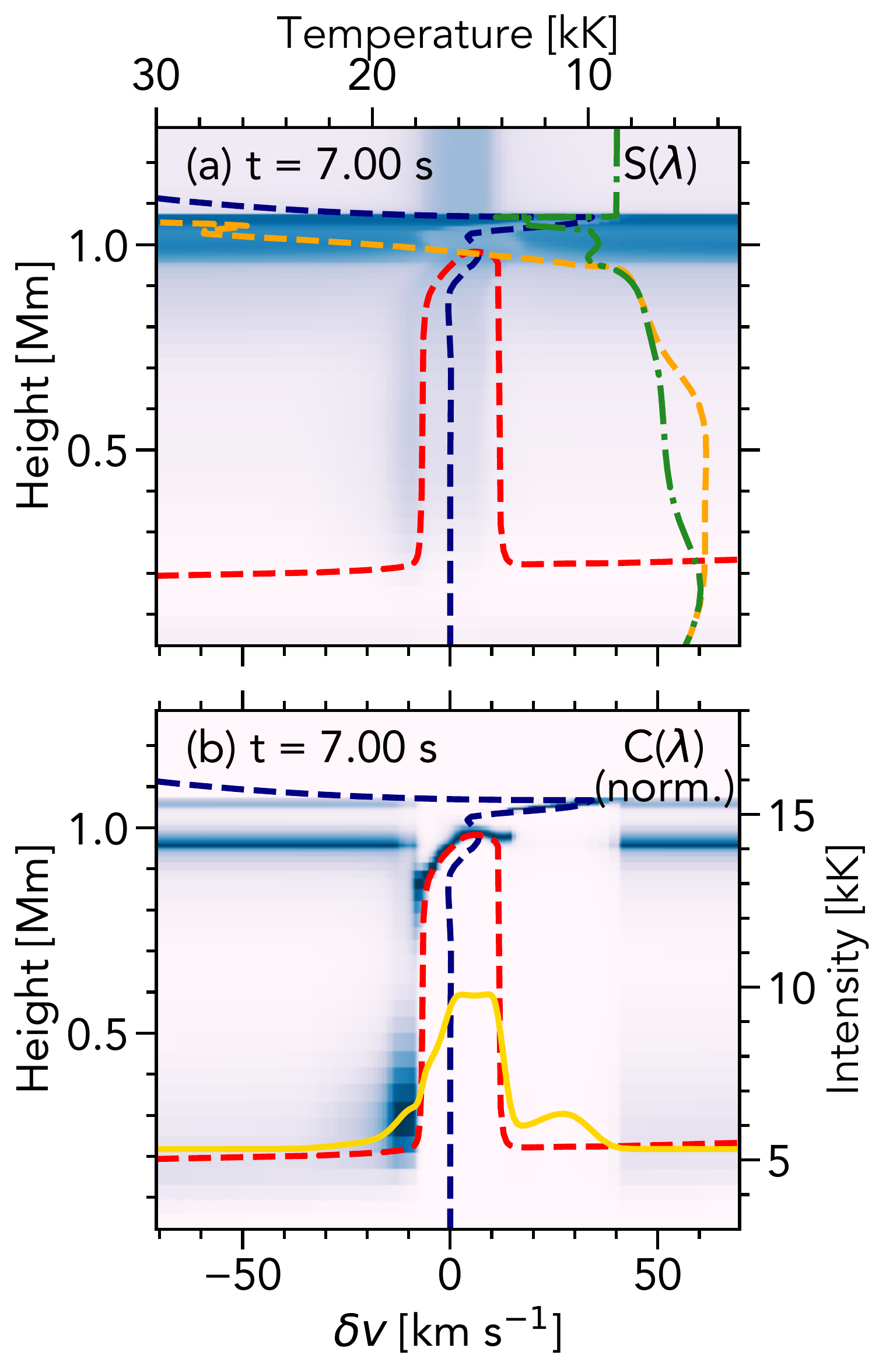}}	
        \subfloat{\includegraphics[width = 0.45\textwidth, clip = true, trim = 0.cm 0.cm 0cm 0cm]{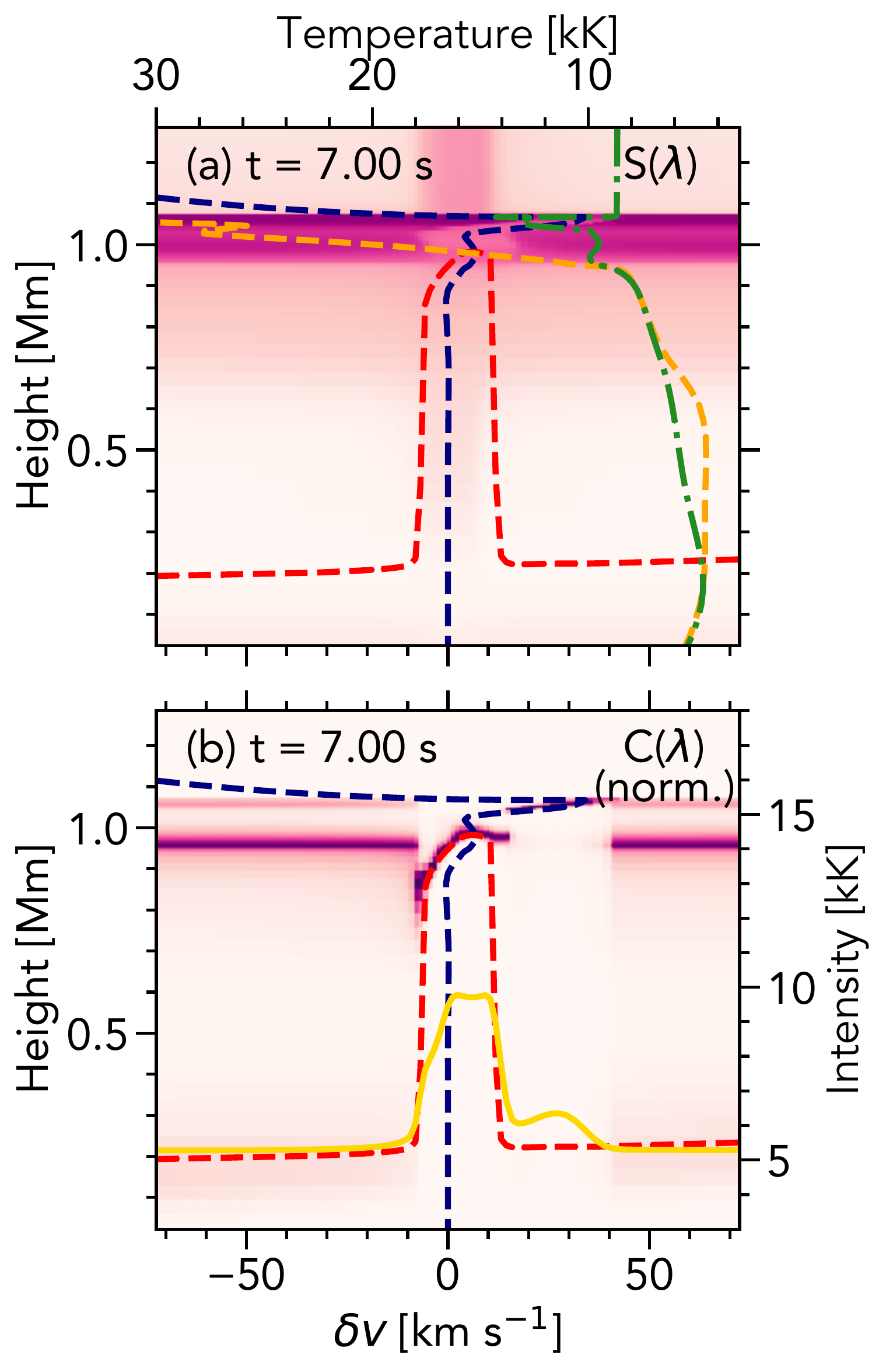}}	
        }
	\caption{\textsl{Illustrating Mg~\textsc{ii} 2791~\AA\ line formation in the PRD (a,b) and CRD (c,d) scenarios. Images and lines are as described in Figure~\ref{fig:Cfn_prd_vs_crd}}}
	\label{fig:Cfn_prd_vs_crd_subline}
\end{figure*}
The cause of these two bumps in the wings is different, as shown in the bottom panels Figure~\ref{fig:Cfn_prd_vs_crd}(b,d), which show the contribution function. Emission is considered optically thick if it forms close to the $\tau_{\lambda} = 1$ layer, and optically thin if it forms far above that layer. In both CRD and PRD there are two optically thick line components, one nearly stationary, and the other redshifted by approximately $15-30$~km~s$^{-1}$. This redshifted component is caused by a dense condensation that is propagating deeper with time, emitting Doppler shifted Mg~\textsc{ii} photons. At this time ($t=7$~s) it has reached a depth sufficient for the opacity in the red wing of Mg~\textsc{ii} to increase. Previously the red wing feature was weaker and optically thin, still caused by emission from the condensation but from a greater altitude so that the Mg~\textsc{ii} red wing opacity was not so increased. However, the blue wing feature in the PRD solution is not caused by an upflowing source. Instead it is a PRD effect. The absorption profile has shifted to the red so that there is a preference for absorbing redder wavelengths. In PRD, bluer photons are re-emitted coherently without scattering to other wavelengths, where they would be trapped by the much larger opacity. This feature grows in strength because the opacity in the core and the red wing increases when the condensation propagates deeper. Note also that this component is a mix of optically thick and thin emission. 

Figure~\ref{fig:Cfn_prd_vs_crd_detailed} shows the contribution function, opacity, and source functions for selected wavelengths in the k line at $t=7$~s in the F11 simulation, showing in a more quantitative sense the differences between the PRD and CRD solutions. At all wavelengths the PRD and CRD source functions converge between $z=1-1.1$~Mm. The larger temperatures and densities drive more collisions reducing the coherency fraction there, so that the CRD and PRD solutions are similar. 

The Mg~\textsc{ii} 2791~\AA\ subordinate line formation is shown for comparison in Figure~\ref{fig:Cfn_prd_vs_crd_subline}. Similar features to the resonance lines are present, with the same difference between the PRD and CRD solution. The redshifted component caused by the condensation is optically thin at this time because the subordinate line forms lower in the atmosphere, so that the condensation has to propagate deeper to sufficiently increase the opacity structure for this line. Despite this, the increased source function and contribution function in the blue wing can be seen. Note here that the subordinate lines form in the upper chromosphere only a couple of hundred km deeper than the resonance lines. This is quite different from the quiet Sun scenario, where they form in the lower atmosphere \citep{2015ApJ...806...14P}. This means that quiet Sun diagnostics cannot be applied in the flaring scenario \citep[see also Figure 11 in][]{2019ApJ...871...23K}.

A similar effect was found in the Ca~\textsc{ii} K line, comparing the CRD and PRD solutions for that line revealed a blue wing bump that grew in strength with an optically thick red wing component. As with the Mg~\textsc{ii} subordinate line this happened somewhat later in the Ca~\textsc{ii} K line since it forms lower in the chromosphere, so the condensation took slightly longer ($\sim1.5$~s) to reach heights where the effect became noticeable. 

This highlights the ambiguity resulting from interpreting asymmetries and Doppler motions in optically thick lines, with the blue wing feature being a result of a condensation not evaporation \citep[see also][]{2015ApJ...813..125K,2016ApJ...827..101K,2018ApJ...862...59B}. 

It is important to note here that these results used a value of microturbulence of $2$~km~s$^{-1}$, the nominal value used in \texttt{RADYN}. Defining the magnitude of microturbulence present is ambiguous, particularly in flares. Increasing microturbulence will increase the width of the Doppler core, having a consequential effect on the amount of redistribution. Some authors have attempted to use increased microturbulence to explain the narrower-than-observed line widths, though this alone is unlikely to be the main cause \citep{2017ApJ...842...82R,2019arXiv190412285Z}. However, as an experiment we increased microturbulence to a value of $10$~km~s$^{-1}$. This smeared the stationary and redshifted optically thick components, and smoothed out the blue-wing bump so that the profile shapes appeared more similar in CRD and PRD. The CRD solution still overestimated the wing intensity. Thus, our conclusion that PRD is necessary stands, regardless of the level of microturbulence chosen.

\section{Treatment of Partial Frequency Redistribution}\label{sec:adprd}
\begin{figure*}
	\centering 
	{\includegraphics[width = .75\textwidth, clip = true, trim = 0.cm 0.cm 0.cm 0.cm]{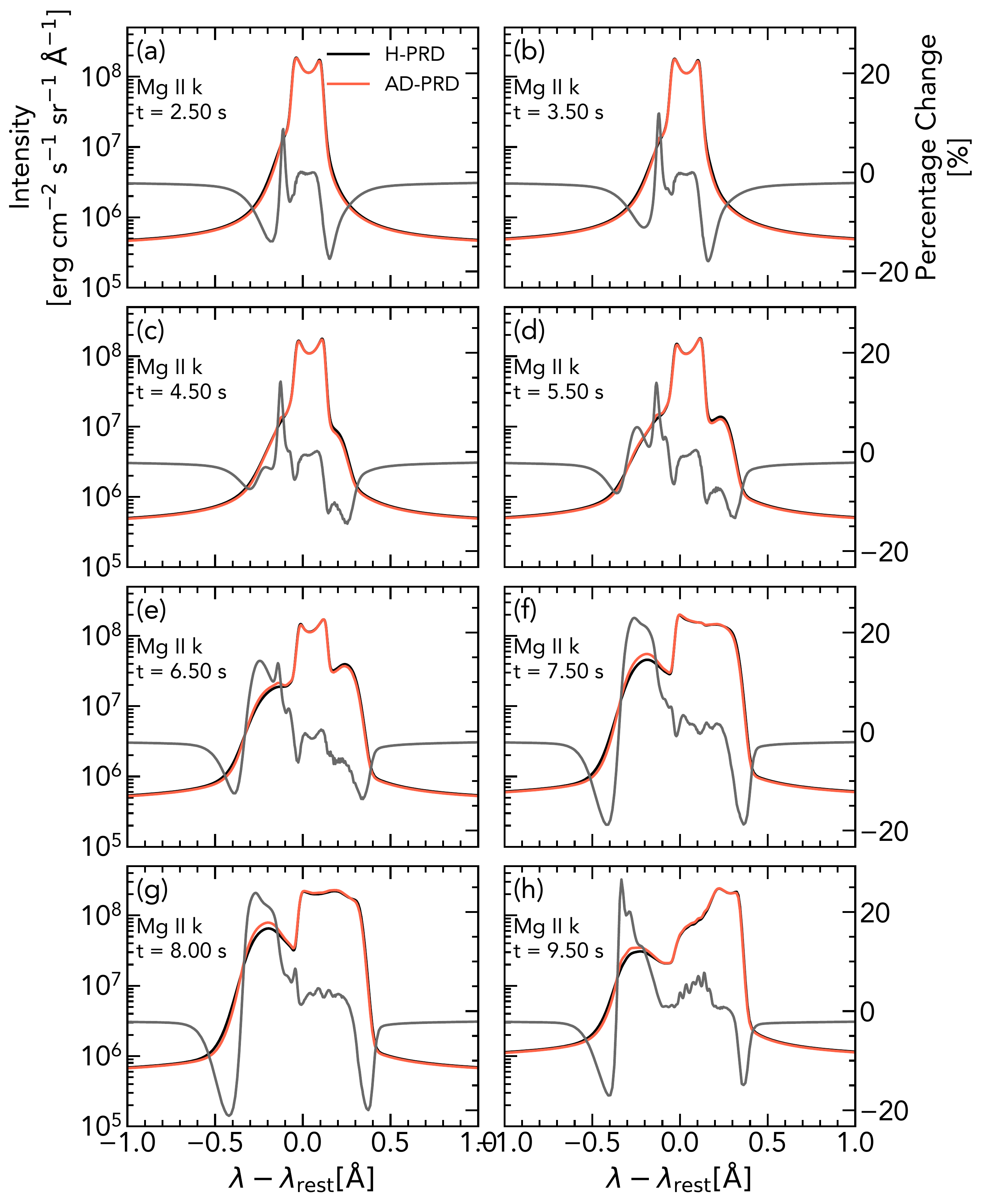}}	
	\caption{\textsl{Comparing full angle-dependent PRD (AD-PRD) to the Hybrid PRD (H-PRD) scheme of \cite{2012A&A...543A.109L}, for various snapshots in the F11 simulation. The black lines are the H-PRD solution, the red lines are the AD-PRD solution, and the grey lines show the percentage change.}}
	\label{fig:prdtreatment}
\end{figure*}
Previously we adopted the fast angle-approximated PRD treatment (H-PRD) of \cite{2012A&A...543A.109L}, which saves considerably on computational time and was shown to be a sufficiently accurate approximation to full angle-dependent PRD (AD-PRD). The H-PRD scheme is typically employed when simulating Mg~\textsc{ii} in flares using \texttt{RH} \citep[e.g.][]{2016ApJ...827..101K,2017ApJ...842...82R,2019ApJ...871...18R, 2019arXiv190412285Z}. 

This H-PRD scheme was tested by \cite{2012A&A...543A.109L} in the scenario of relatively small velocity fields compared to those present in flares. The mass motions driven during solar flares can be significant, with downflows of several tens of km~s$^{-1}$, and upflows of a few hundred km~s$^{-1}$. Therefore, this approximated scheme may not be applicable during flares, a prospect that we test in this section.

Using the F11 simulation we ran several snapshots with the Mg~\textsc{ii} h \& k and subordinate lines computed using AD-PRD. The snapshots chosen were $t = [2.5, 3.5, 4.5, 5.5, 6.5, 7.5, 8, 9.5]$~s, which covered several points during the development of mass flows in the flare, sampling a range of velocity gradients. To save somewhat on computational time we used the Mg~\textsc{ii} atomic level populations from the previously computed H-PRD simulation as the starting solution so that the initial guess was closer to the final result. Even with this effort to speed up convergence, the difference in the time taken to complete these simulations was substantial in comparison to the hybrid scheme. Using AD-PRD the time to complete each snapshot was 3-4 \textsl{days}, compared to only a few minutes for the H-PRD case. The AD-PRD solution took approximately $400$ times longer to complete than the H-PRD solution. Lowering the main \texttt{RH} convergence threshold to $1\times10^{-2}$ in the AD-PRD experiments reduced convergence time to around one day, with negligible difference in the AD-PRD emergent Mg~\textsc{ii} profiles (at most a $\sim0.5$~\% change). Performing high cadence studies flare simulations using AD-PRD will obviously be a computationally expensive and time-consuming endeavour, particularly if multiple flare simulations are involved. 

Figure~\ref{fig:prdtreatment} shows the Mg~\textsc{ii} k line computed for the F11 simulation using H-PRD, and AD-PRD. There is excellent agreement in line shape, with all features reproduced. There are, however, intensity differences. This can vary from only a few percent to up to $15-20$~\% in narrow wavelength regions across the line where redistribution effects are maximised (recall the blue wing feature discussed previously). Balancing the computational cost to the accuracy of the results, we believe that, for most purposes, H-PRD is acceptable for use in flare simulations, despite the localised intensity differences that may result. It is, of course, up to an individual to decide if these intensity differences are tolerable for their specific study. 

\section{Using a Simpler Model Atom}\label{sec:smalleratom}
\begin{figure*}
	\centering 
	{\includegraphics[width = .75\textwidth, clip = true, trim = 0.cm 0.cm 0.cm 0.cm]{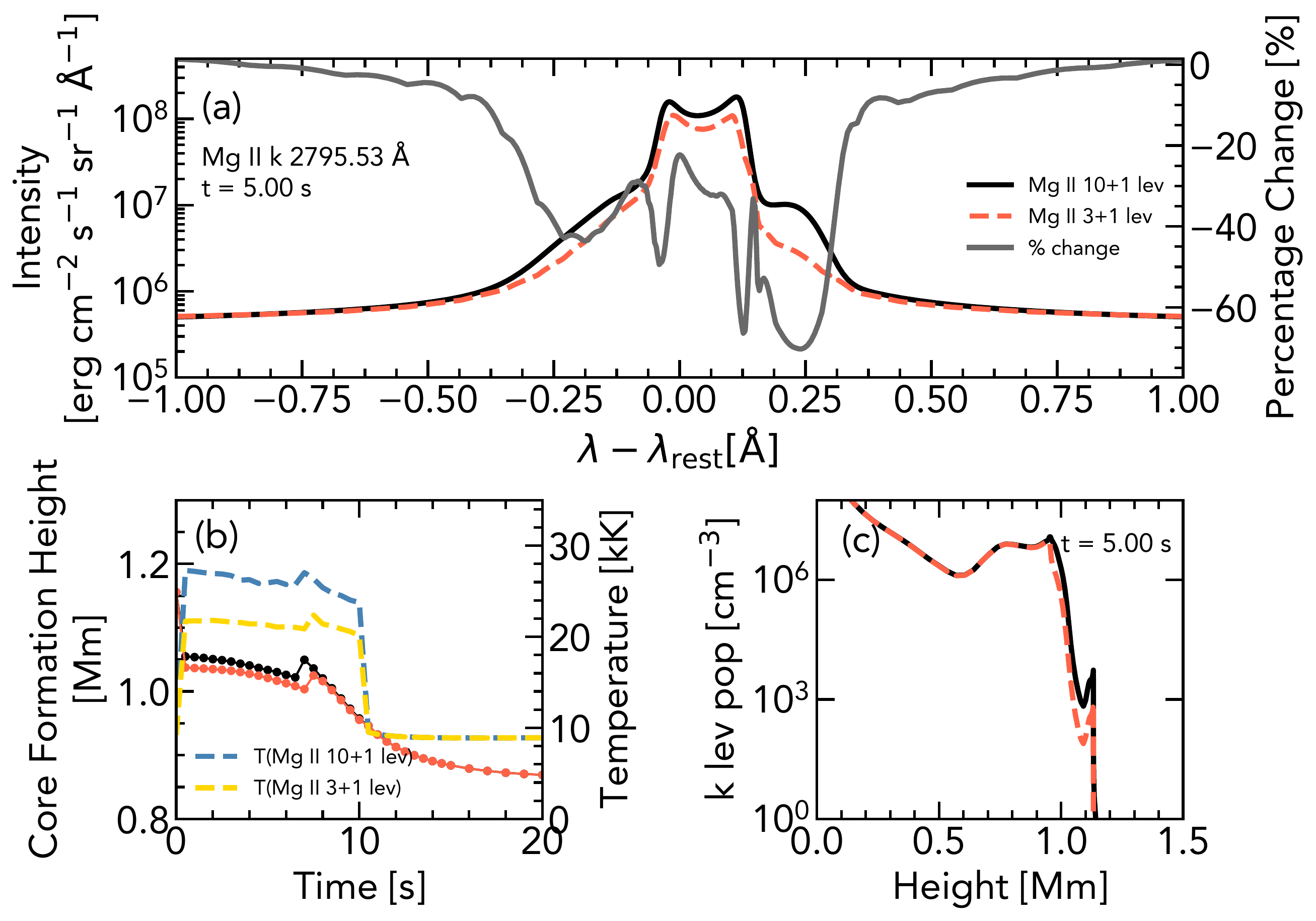}}	
	\caption{\textsl{Illustrating the impact of the number of levels in model Mg~\textsc{ii} atom. In all panels the black lines are the 10-level-plus-continuum model atom, and the red are the 3-level-plus-continuum model atom. Panel (a) shows the Mg~\textsc{ii} k line profile at $t=5$~s in the F11 simulation. The grey line shows the percentage difference. Panel (b) shows the formation height (height at which $\tau_{\nu} = 1$ is maximum) for the Mg~\textsc{ii} k line as a function of time. The temperature at the formation heights are shown also (blue line is 10+1 level, and yellow is 3+1 level). Panel (c) shows the population of the k line upper level}}
	\label{fig:small_atom_kline}
\end{figure*}
\begin{figure}
	\centering 
        {\includegraphics[width = 0.45\textwidth, clip = true, trim = 0.cm 0.cm 0cm 0cm]{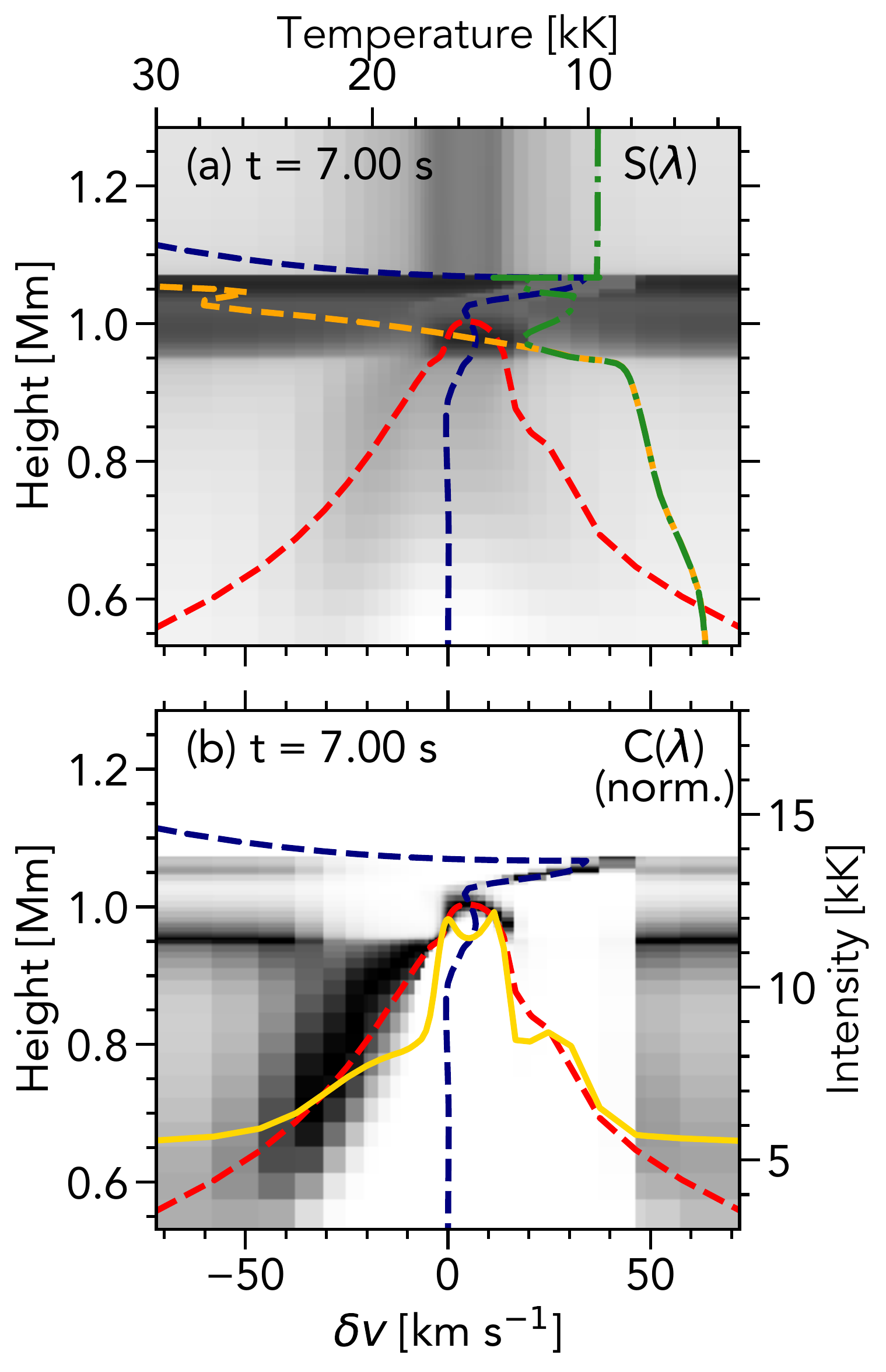}}	
	\caption{\textsl{Source functions (a) and contribution functions (b) for the Mg~\textsc{ii} k line at $t=7$~s in the F11 simulation using the 3-level-plus-continuum model atom. Images and lines are as described in Figure~\ref{fig:Cfn_prd_vs_crd}}}
	\label{fig:small_atom_Cfn}
\end{figure}

It was demonstrated by \cite{2013ApJ...772...89L} that to adequately model Mg~\textsc{ii} in quiescent atmospheres, an atom that included excited upper states was required, in order to support an ionisation/recombination loop that populated the h \& k upper levels via recombinations to higher lying excited states, followed by cascades through to the h \& k upper levels. To show the impact of using too simple a model atom in the flaring scenario, where atmospheric extremes could potentially lead to even more spurious results if too simple an atomic model is used, we show here the results from a three-level-with-continuum Mg~\textsc{ii} atom, consisting of the Mg~\textsc{ii} ground state, the h \& k line upper levels, and the ground state of Mg~\textsc{iii}. 

As shown in Figure~\ref{fig:small_atom_kline}(a) there are non-negligible differences in both line intensity and in the shape of the profiles. This is a direct result of smaller population densities of the resonance lines upper levels, due to the lack of recombinations to higher lying excited states and subsequent cascades to the $3p\,^2\!P^{\rm o}_{3/2,1/2}$ levels. The lack of these recombinations changes the Mg~\textsc{ii} ion fraction and opacity stratification, resulting in a somewhat lower $\tau_{\lambda} = 1$ height, compared to the larger model atom. The lower formation height means that the source function is smaller, due to both the lower temperature at that height and the smaller upper level populations. Figure~\ref{fig:small_atom_kline}(b) shows the formation heights of the lines, along with the temperatures at those heights, and Figure~\ref{fig:small_atom_kline}(c) shows the lower population density when using the 3+1 level model atom. 

Impacts of velocity gradients consequently appear at different times, with different magnitudes, so that the line profiles do not always share common features during the flare simulations. This is demonstrated in Figure~\ref{fig:small_atom_Cfn}, which shows the source and contribution functions for the smaller model atom (compare to the same in Figure~\ref{fig:Cfn_prd_vs_crd}(a,b) for the standard 10+1 level atom). For the larger model atom, the fraction of Mg~\textsc{ii} at the height of the condensation is sufficient to produce a substantially larger opacity compared to the smaller model atom, resulting in a redshifted optically thick component to the line. At this same time for the smaller model atom, the condensation results in a weaker, optically thin, feature to appear in the red wing. 

We recommend that the model atom of \cite{2013ApJ...772...89L} be used, or at least a model atom capable of supporting cascades through higher lying excited levels, as described by \cite{2013ApJ...772...89L}. 

\section{Including Mg~\textsc{I}}\label{sec:biggeratom}
It has been shown that for modelling quiet Sun h \& k profiles, the inclusion of Mg~\textsc{i} mainly affected the very far wings with no effect on the line cores \citep{2013ApJ...772...89L}, and consequently if one is interested primarily in the h, k, or subordinate lines then omitting Mg~\textsc{i} is safe to do.  We assess here if this result also applies in the more extreme environment of flares. A 75-level-with-continuum model of Mg was used, with 65 levels of Mg~\textsc{i} added to the 10 level model Mg~\textsc{ii} atom plus the ground state of Mg~\textsc{iii}. 

\begin{figure*}
	\centering 
	{\includegraphics[width = .75\textwidth, clip = true, trim = 0.cm 0.cm 0.cm 0.cm]{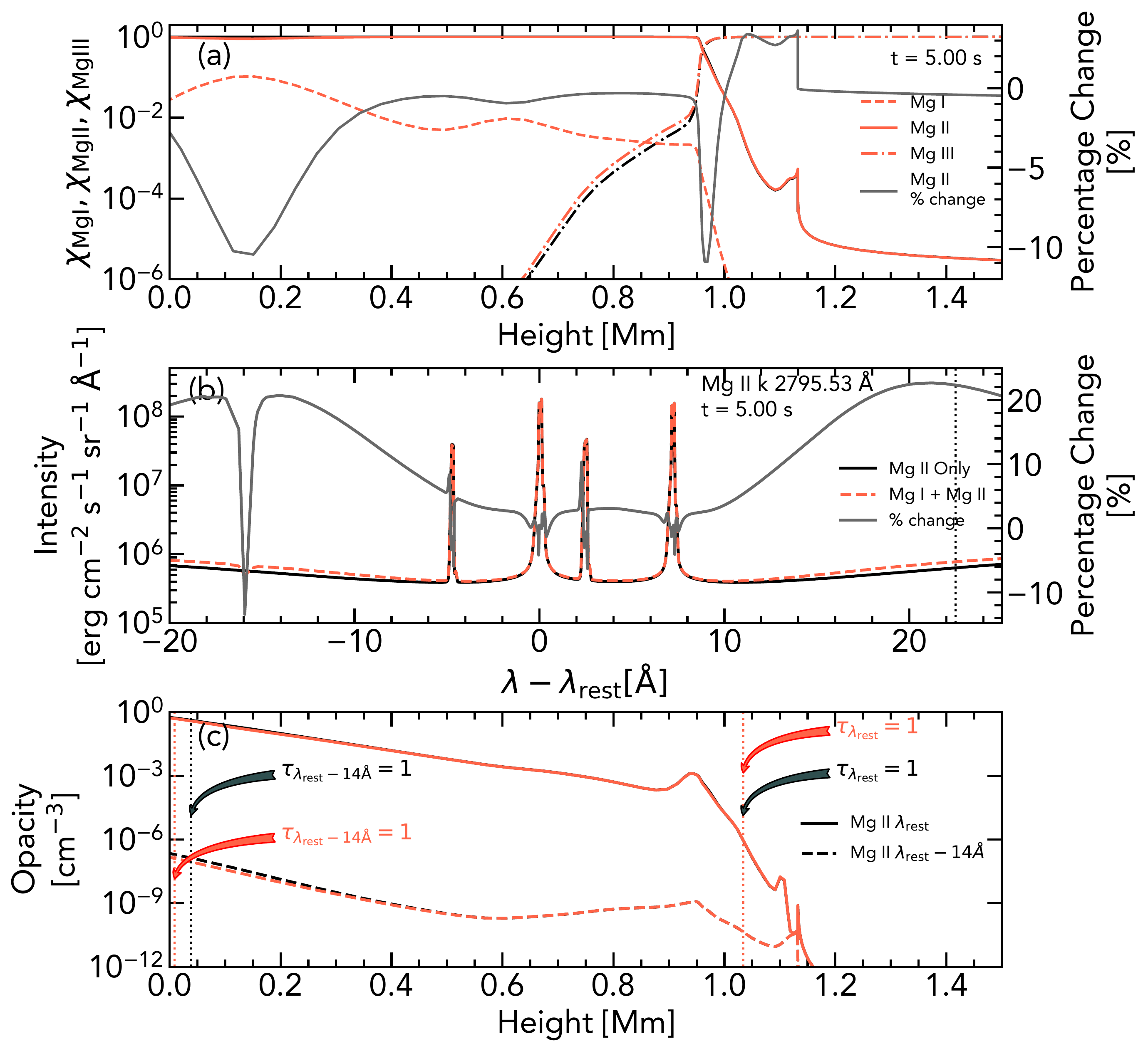}}	
	\caption{\textsl{Illustrating the impact of including Mg~\textsc{i} in the model atom. In all panels black shows the `standard' case, and red shows the solution when Mg~\textsc{i} is included, at $t=5$~s in the F11 simulation. Panel (a) shows the ionisation fraction stratification for Mg~\textsc{i} (dashed line), Mg~\textsc{ii} (solid line), and Mg~\textsc{iii} (dot-dashed line). The percentage change when of the Mg~\textsc{ii} fraction when Mg~\textsc{i} is included is the grey solid line. Panel (b) shows the emergent NUV spectrum, where again the grey line shows the percentage change due to the inclusion of Mg~\textsc{i}. Panel (c) shows the opacity at $\lambda_{\mathrm{rest}}$ (solid lines), and $\lambda_{\mathrm{rest}-14\AA}$ (dashed lines). On that panel the height of optical depth unity is indicated.}}
	\label{fig:Mgifraction}
\end{figure*}

Even with the inclusion of Mg~\textsc{i} in our model atom, the majority of chromospheric/photospheric Mg exists as Mg~\textsc{ii} (Figure~\ref{fig:Mgifraction}(a)). Through the temperature minimum region $X_{Mg\textsc{i}}\sim10$~\%, dropping through the chromosphere to  $X_{Mg\textsc{i}}\sim1$~\% or smaller. Figure~\ref{fig:Mgifraction}(a) shows the Mg ion fractions at $t=5$~s into the F11 flare simulation. In that figure we also show the percentage change of $X_{Mg\textsc{ii}}$ resulting from including Mg~\textsc{i}. Despite this small fraction, the amount of Mg~\textsc{ii} is still somewhat reduced, which changes the opacity stratification, and the intensity of the Mg~\textsc{ii} lines. This has only a small impact on the core of the line, resulting in a tolerable intensity difference in the line cores (on the order of a few percent). Further into the line wings, however, the effect becomes noticeable with a $15-20$~\% difference as shown in Figure~\ref{fig:Mgifraction}(b). The reduced fraction of Mg~\textsc{ii} in the lower atmosphere reduces the opacity, so that the wings form somewhat deeper (Figure~\ref{fig:Mgifraction}(c))

Including Mg~\textsc{i} in our experiments increased the time to reach convergence by a factor of $\sim2.5$. Given the minor intensity differences we feel it is safe to ignore Mg~\textsc{i} if one is only interested in the line cores. 

IRIS also observes the (\textsl{quasi}-) continuum near $\lambda = [2810-2834]$~\AA, the intensity of which is affected by the opacity of the h \& k absorption profiles \citep{2017ApJ...836...12K}. If one is concerned with modelling the h \& k line far wings or the continuum in that region Mg~\textsc{i} should probably be included.


\section{Including Other Species}\label{sec:otherspecies}
\begin{figure*}
	\centering 
	{\includegraphics[width = .75\textwidth, clip = true, trim = 0.cm 0.cm 0.cm 0.cm]{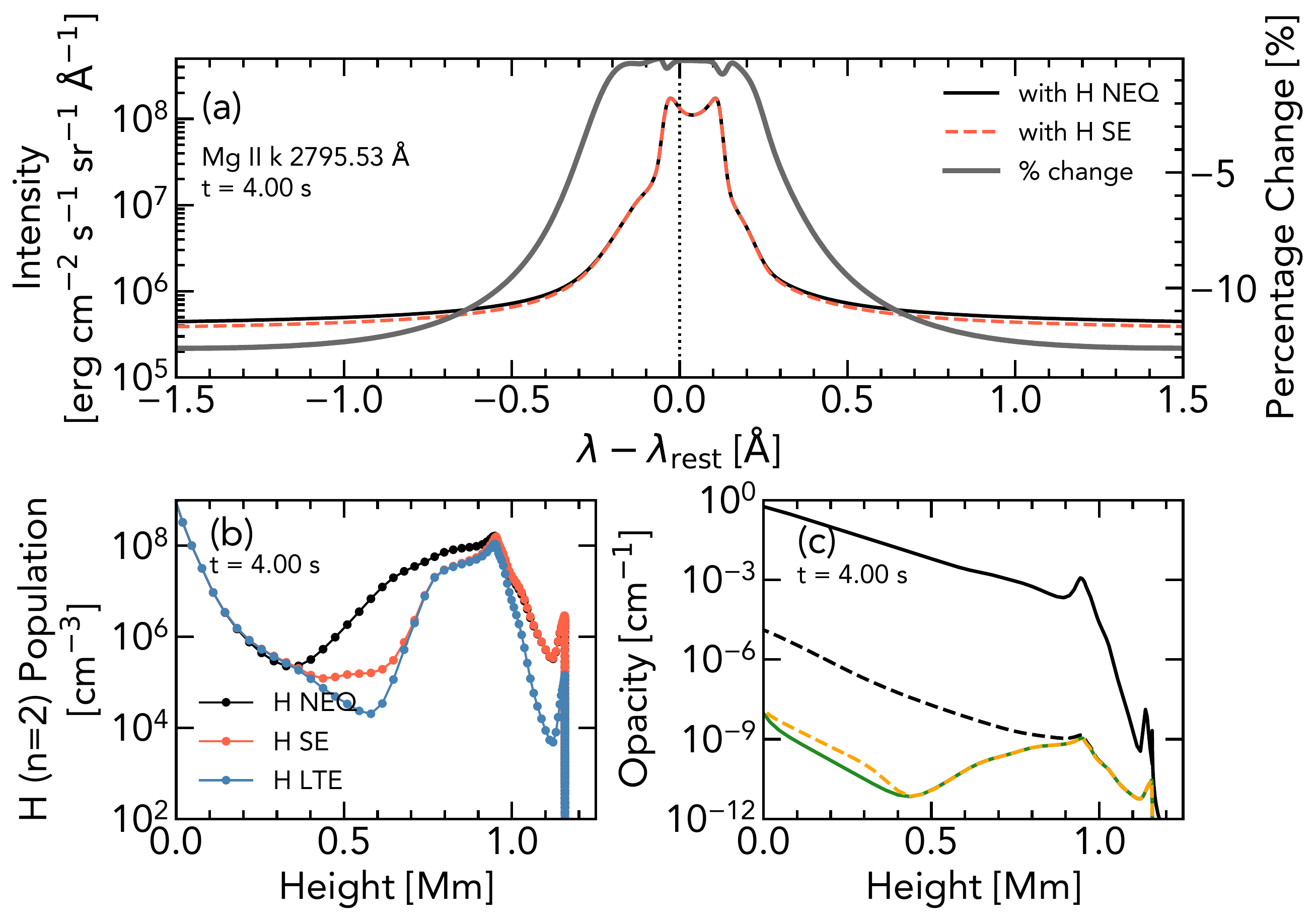}}	
	\caption{\textsl{Illustrating the impact of the treatment of hydrogen. (a) the Mg~\textsc{ii} k line profile at $t=4$~s in the F11 simulation, computed with NEQ \& non-thermal hydrogen populations (black line), and SE hydrogen populations (red line). The grey line shows the percentage difference. (b) Hydrogen level 2 populations from the F11 simulation at $t=3$~s, where the black line shows the NEQ \& non-thermal solution from \texttt{RADYN} and the red line is the SE solution from \texttt{RH}. The blue line shows the LTE solution. (c) sources of opacity. Black solid line is the total opacity at line center (Mg~\textsc{ii}, hydrogen + other species), the black dashed line is the same for $\lambda_{\mathrm{rest}} - 2$\AA, the green line is the hydrogen contribution, and the gold dashed line shows the hydrogen + other species (with no Mg~\textsc{ii} contribution).}}
	\label{fig:hydtreatment}
\end{figure*}

Hydrogen is an important source of opacity in the solar atmosphere, dominating the background and continuum at NUV and optical wavelengths.  In our benchmark, and in the other experiments, we used the hydrogen atomic level populations obtained directly from the \texttt{RADYN} simulations, which account for NEQ ionisation and non-thermal effects. These populations may not always be available to a user wishing to simulate Mg~\textsc{ii} so we tested the impact on Mg~\textsc{ii} of allowing \texttt{RH} to solve the SE equations for hydrogen. 

The F11 \texttt{RADYN} simulation was re-processed with hydrogen (and calcium, though this would not really affect the Mg~\textsc{ii} result) solved in SE. The hydrogen populations can differ substantially between the solutions, shown in Figure~\ref{fig:hydtreatment}, which has a resulting effect on the NUV/optical opacity. With a larger amount of hydrogen in the n = 2 state, the atmosphere is more opaque to Balmer wavelength photons. The LTE populations are also shown for comparison.

The Mg~\textsc{ii} lines were not greatly affected by the hydrogen solution. Differences in the core intensity were negligible, though with increasing $\Delta \lambda$ from the line core into the line wings, discrepancies began to increase. Figure~\ref{fig:hydtreatment} shows the k line computed with hydrogen in SE and with the NEQ, non-thermal hydrogen populations, at $t=4$~s in the F11 simulation (the 2791\AA\ subordinate line shows similar differences). As can be seen in that figure, differences range from almost zero to a few percent in the core, increasing to $>10$~\% into the wings. This is to be expected as the Mg~\textsc{ii} line opacity dominates the core, but in the wings the underlying hydrogen Balmer continuum becomes more important. 

Figure~\ref{fig:hydtreatment}(c) shows various sources of opacity in the Mg~\textsc{ii} k line. The solid black line shows the full opacity (Mg~\textsc{ii}, H NEQ plus the other NLTE and LTE species included as described previously) at $\lambda_{\mathrm{rest,k}}$, and the dashed black line shows $\lambda_{\mathrm{rest,k}} - 2$~\AA. The background H NEQ opacity (green line) is many orders of magnitude smaller than the core Mg~\textsc{ii} opacity, though does contribute more to the wing opacity. The gold dashed line shows the opacity of NEQ H plus the other NLTE and LTE species excluding the Mg opacity. There is an increase in the temperature minimum region opacity compared to the NEQ H-only case, but the core of the Mg~\textsc{ii} line is still not affected by the inclusion of these other species. Mg~\textsc{ii} and hydrogen are the main contributors. 

Given the relative un-importance of other species as sources of opacity at these wavelengths, we conclude that it is not necessary to include species other than hydrogen when forward modelling the Mg~\textsc{ii} NUV lines observed by IRIS. Our tests showed no impact on the emergent profiles if the other species were included or not.  Ideally one would use the hydrogen NEQ populations from \texttt{RADYN}, or a similar code, but these effects are confined to the wings and so the hydrogen SE populations are probably satisfactory for most purposes. However, the computational expense of including the other species was not prohibitive and it is useful to have lines from multiple species with which to compare against observations. 

Another reason one may wish to include the more accurate NEQ hydrogen populations is their role in processes such as charge exchange. We did not study the importance of charge exchange with neutral hydrogen or protons on Mg~\textsc{ii}, but it is known to be important for O~\textsc{i} \citep{2015ApJ...813...34L} and for Si~\textsc{iv} \citep{2019ApJ...871...23K}. 


\section{Photoionization from coronal and transition region radiation}\label{sec:coronalradiation}
\begin{figure*}
	\centering 
	\hbox{
	\hspace{0.75in}
	\subfloat{\includegraphics[width = .35\textwidth, clip = true, trim = 0.cm 0.cm 0.cm 0.cm]{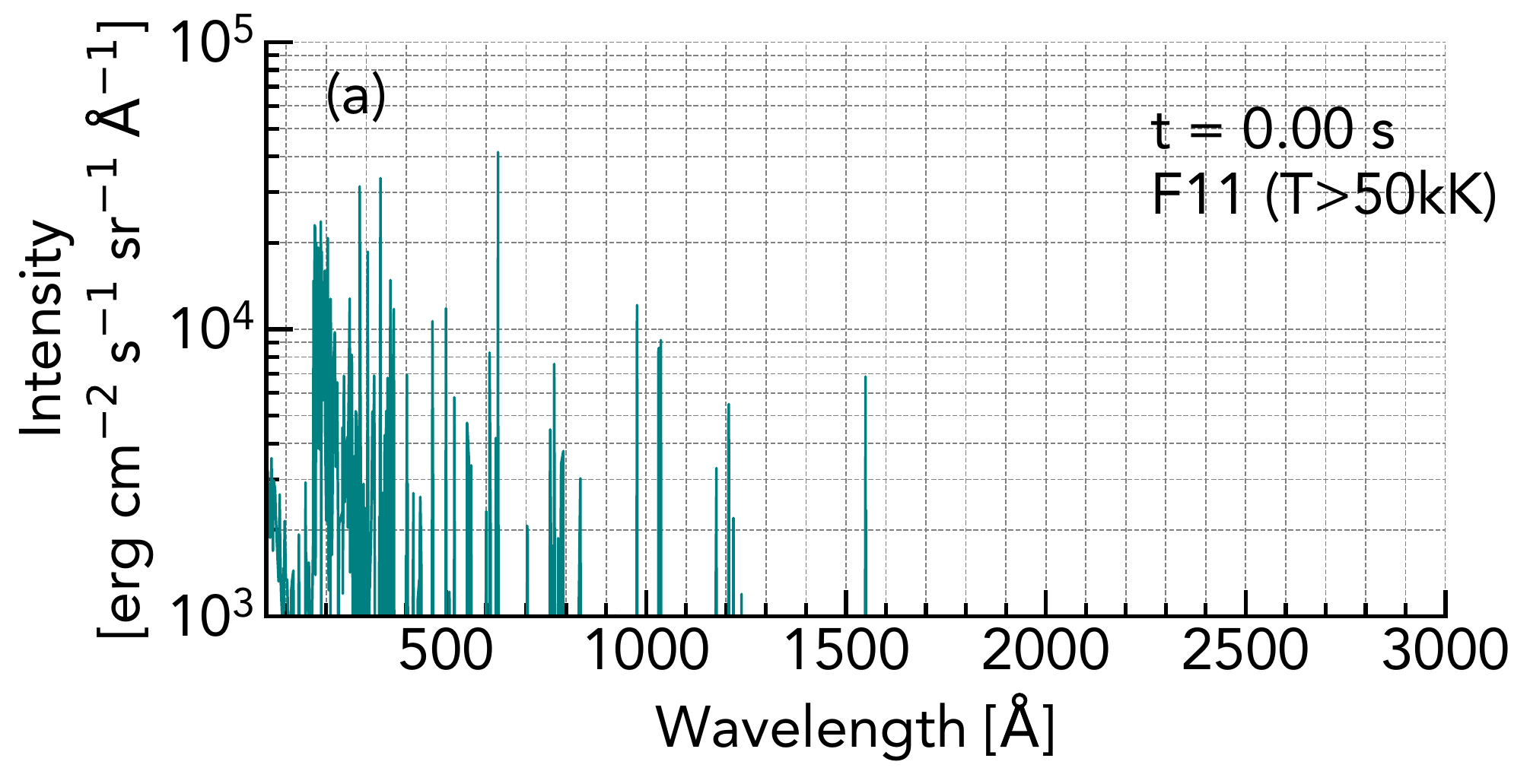}}
	\subfloat{\includegraphics[width = .35\textwidth, clip = true, trim = 0.cm 0.cm 0.cm 0.cm]{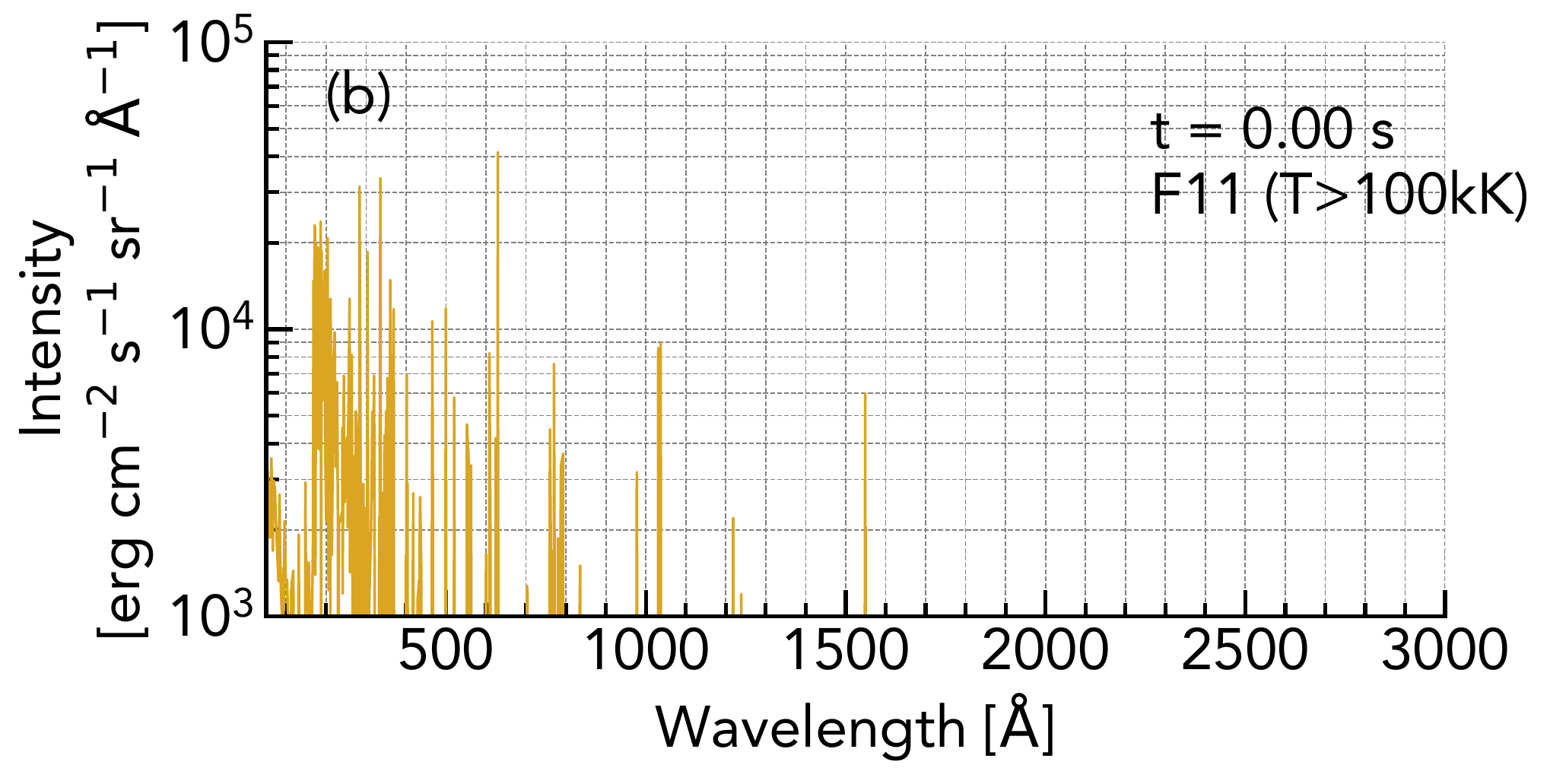}}
	}
	\hbox{
	\hspace{0.75in}
	\subfloat{\includegraphics[width = .35\textwidth, clip = true, trim = 0.cm 0.cm 0.cm 0.cm]{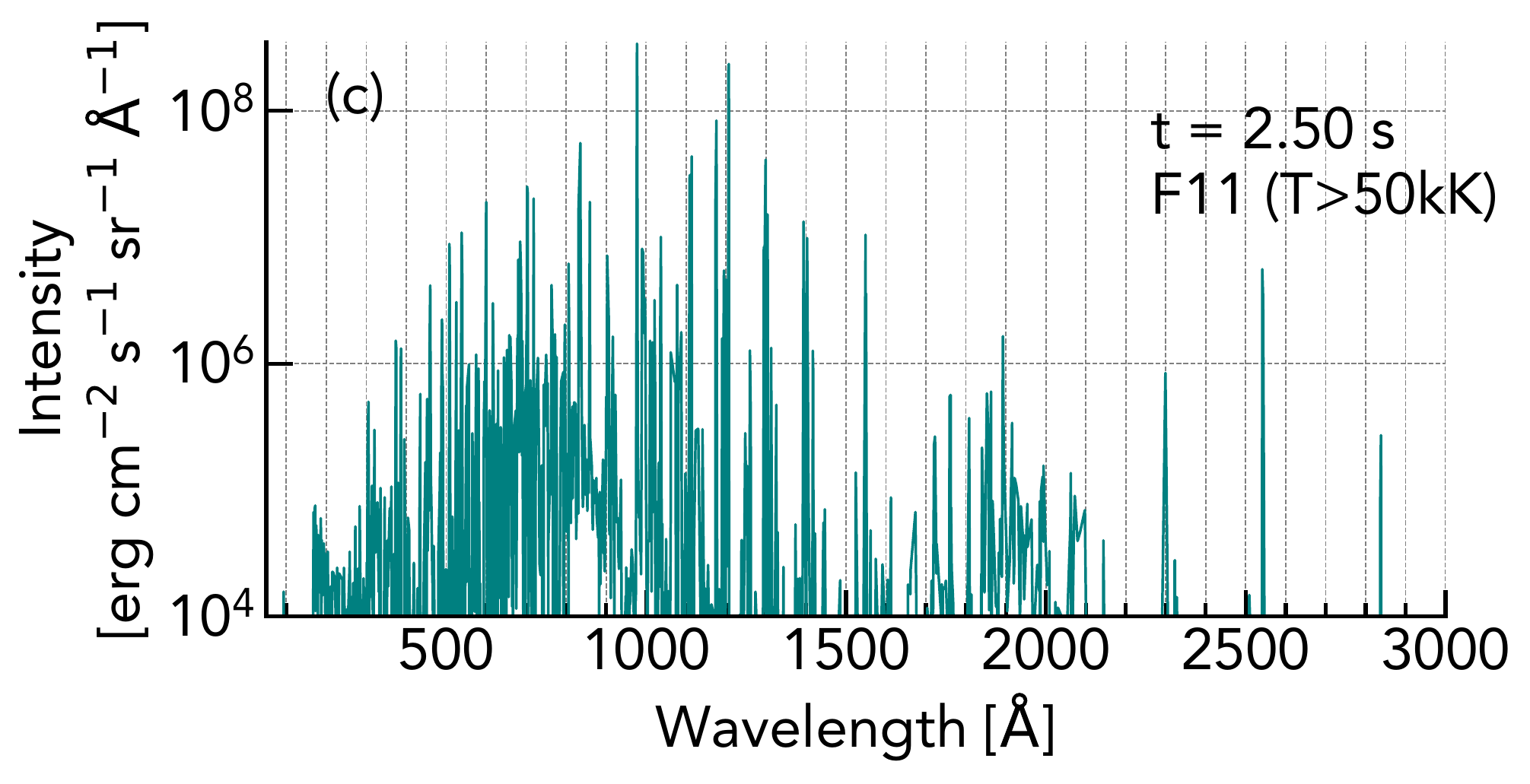}}
	\subfloat{\includegraphics[width = .35\textwidth, clip = true, trim = 0.cm 0.cm 0.cm 0.cm]{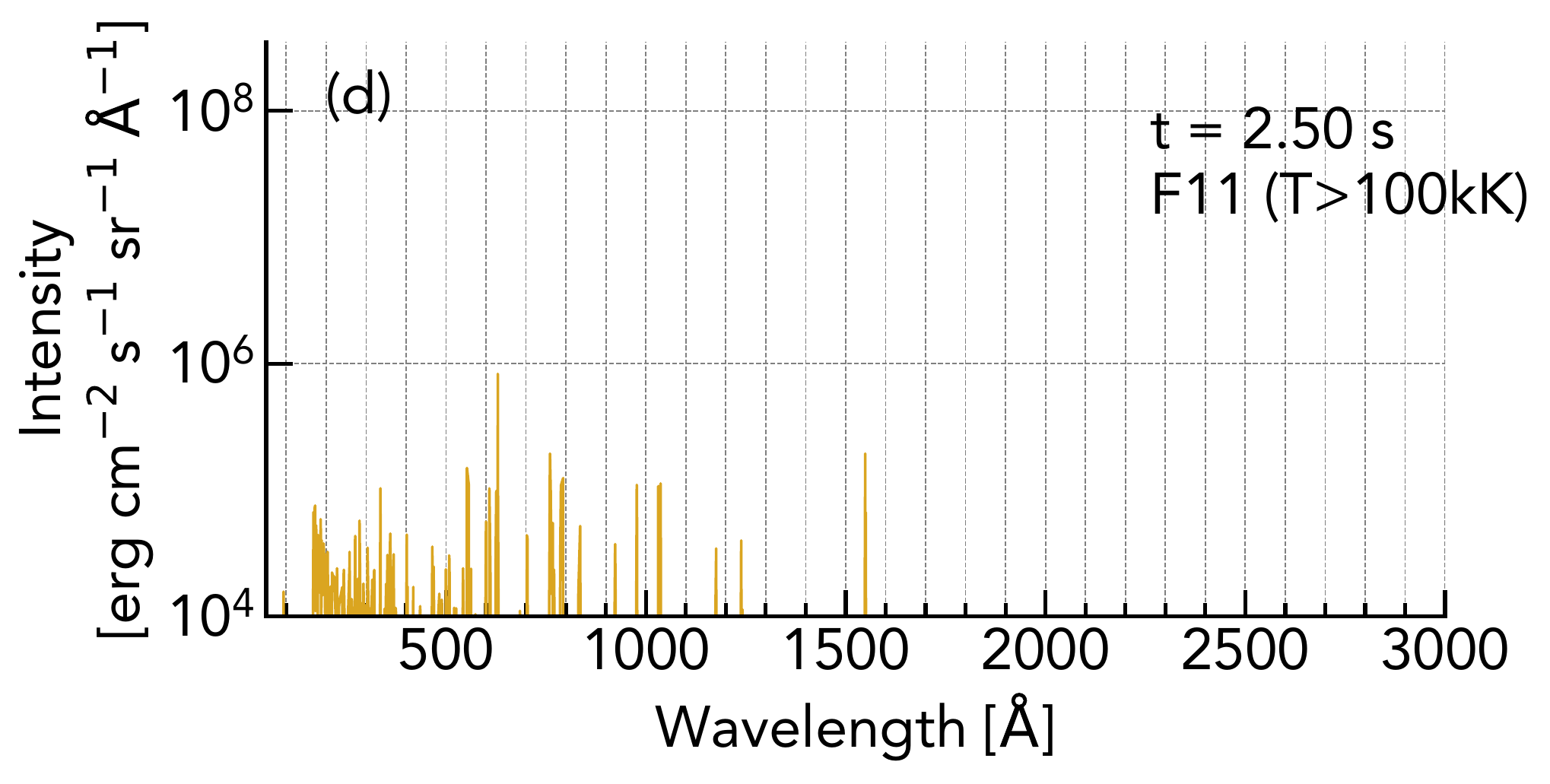}}
	}
	\hbox{
	\hspace{0.75in}
	\subfloat{\includegraphics[width = .35\textwidth, clip = true, trim = 0.cm 0.cm 0.cm 0.cm]{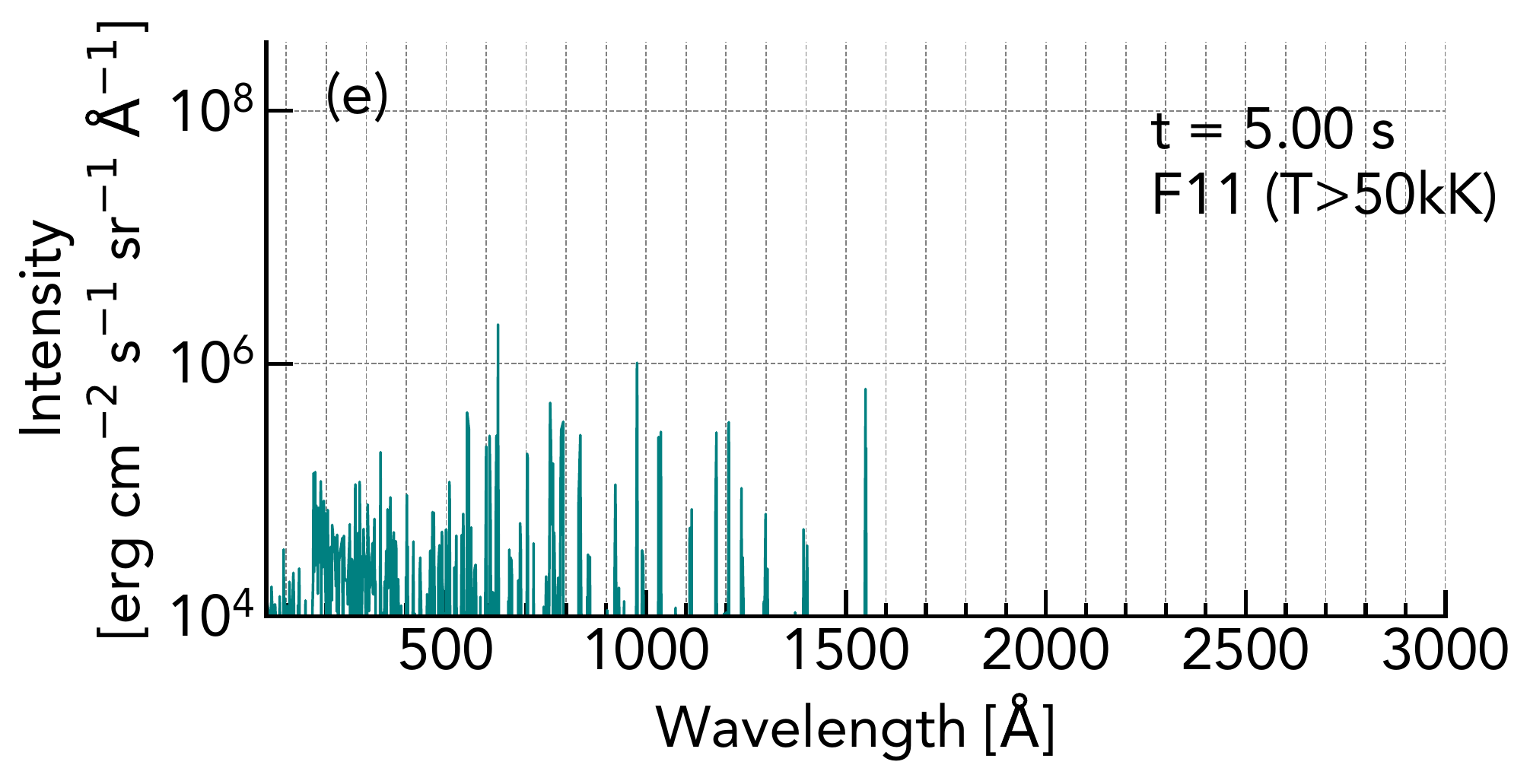}}
	\subfloat{\includegraphics[width = .35\textwidth, clip = true, trim = 0.cm 0.cm 0.cm 0.cm]{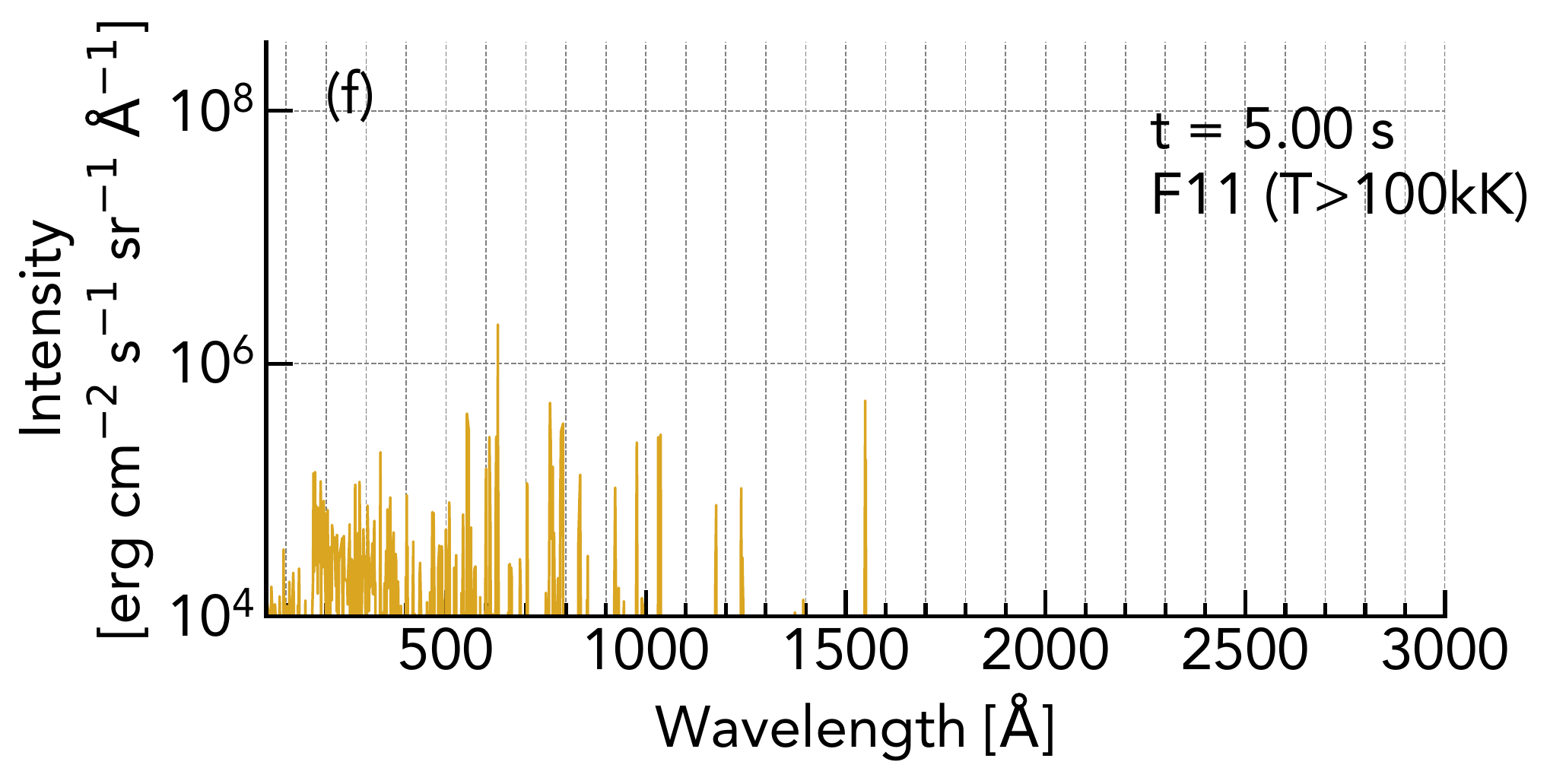}}
	}
         \hbox{
	\hspace{0.75in}
	\subfloat{\includegraphics[width = .35\textwidth, clip = true, trim = 0.cm 0.cm 0.cm 0.cm]{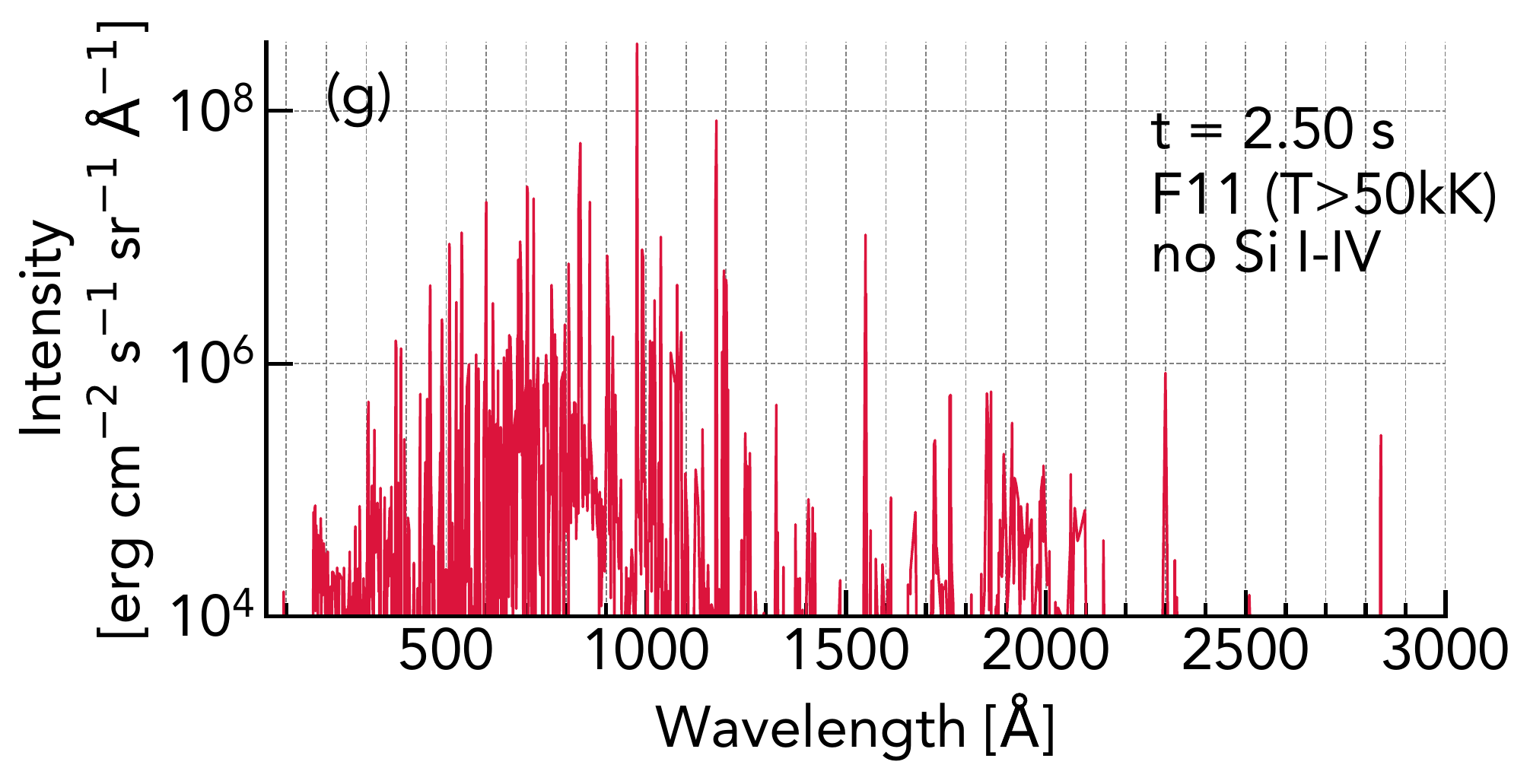}}
	\subfloat{\includegraphics[width = .35\textwidth, clip = true, trim = 0.cm 0.cm 0.cm 0.cm]{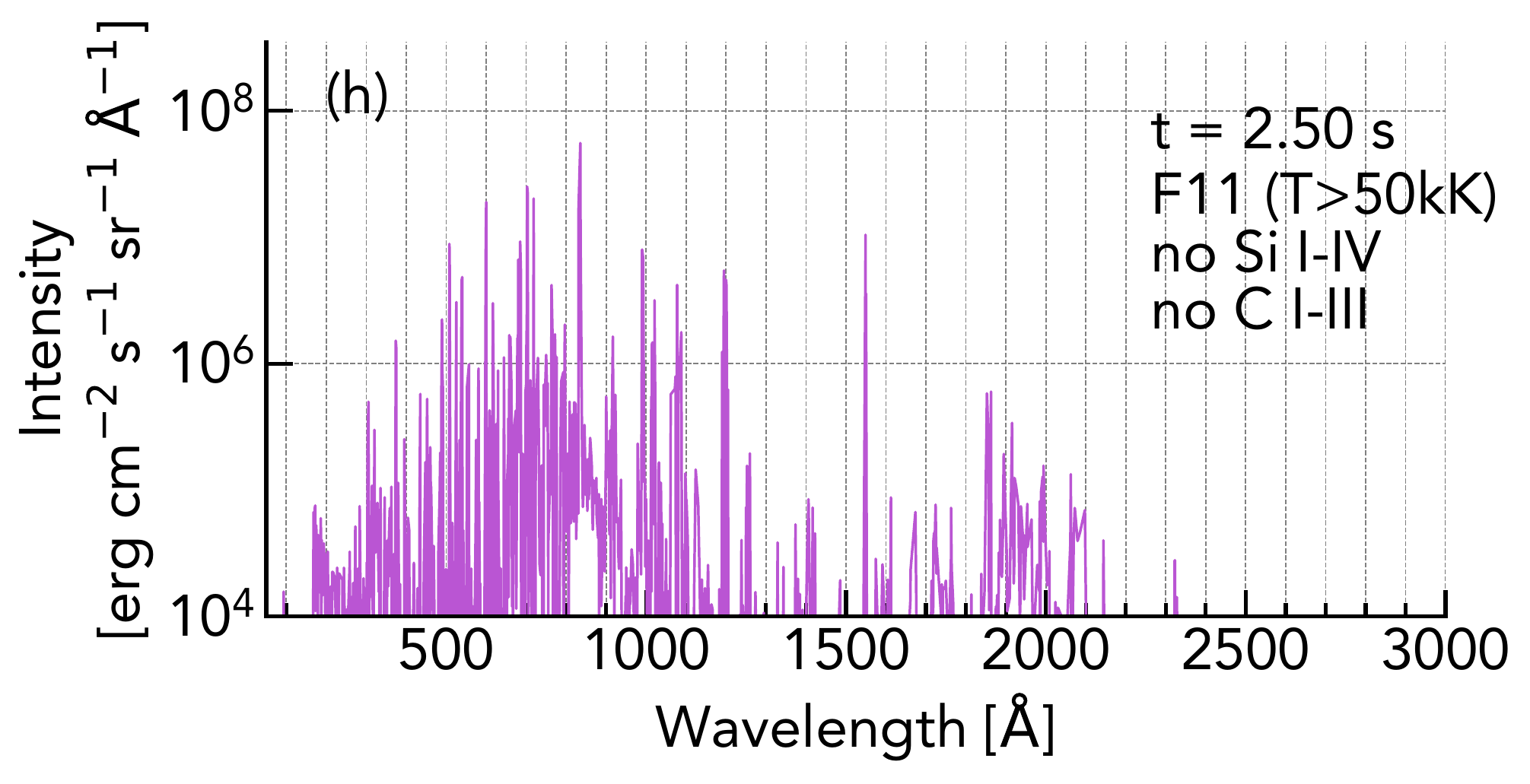}}
	}
	\caption{\textsl{A portion of the irradiating spectrum in the F11 simulation at different times. Panels (a,c,e) show the spectrum where emissivities originating from cells with $T>50$~kK were summed to obtain the intensity. Panels (b,d,f) show the spectrum where a higher threshold of $T>100$~kK was considered. Note the different scale at $t=0$~s (a,b). The higher spectrum in the former scenario illustrates that the irradiation from the lower-mid TR dominates in this flare. Removing Si~\textsc{i}-\textsc{iv} (g) and additionally C~\textsc{i}-\textsc{iii} (h) results in the removal of strong lines, notably at $\sim1206$~\AA\ and $\sim977$\AA. }}
	\label{fig:irradiation}
\end{figure*}

\begin{figure*}
	\hbox{
	\hspace{1.5in}
	\subfloat{\includegraphics[width = .5\textwidth, clip = true, trim = 0.cm 0.cm 0.cm 0.cm]{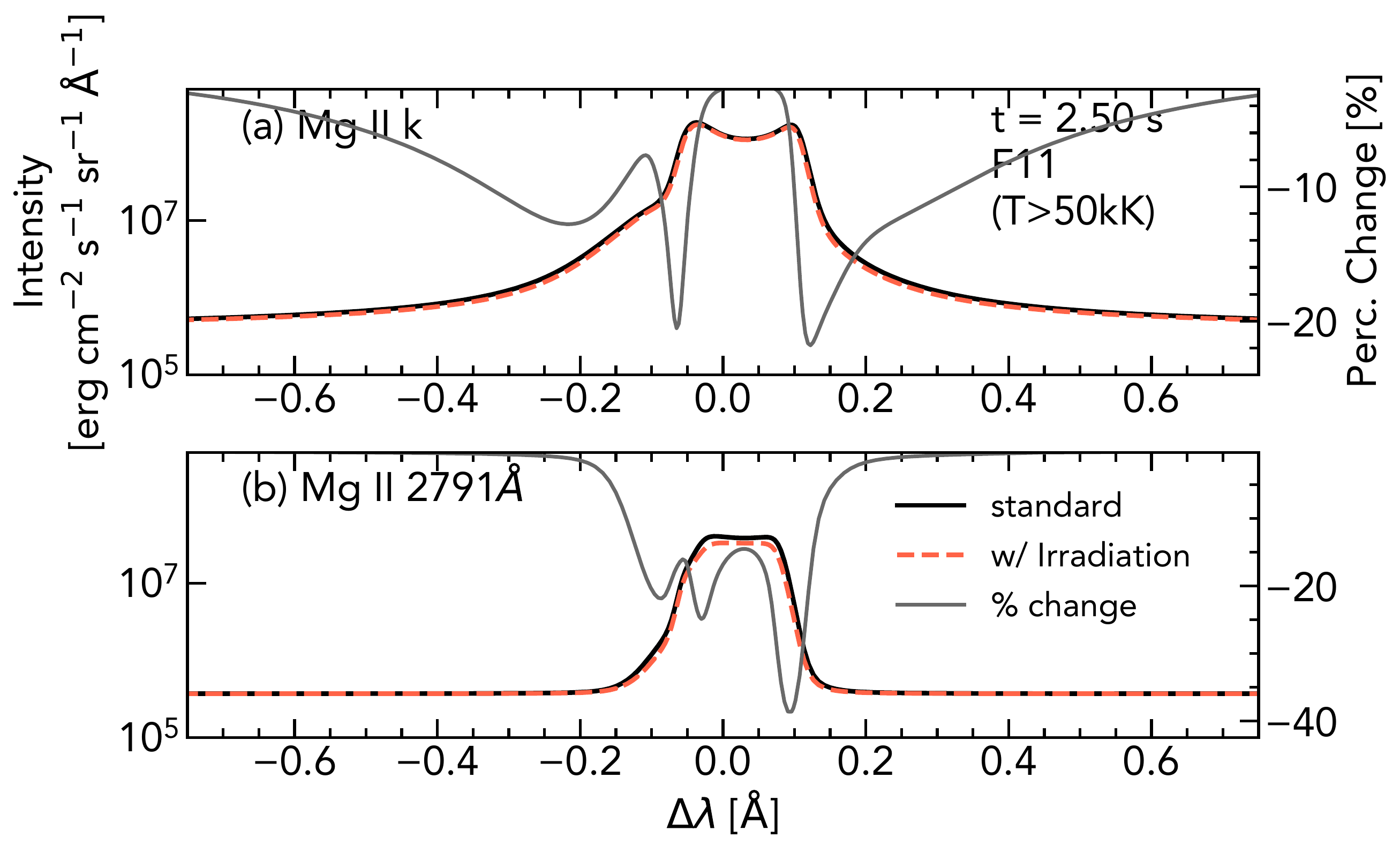}}
	}
	\hbox{
	\hspace{1.5in}
	\subfloat{\includegraphics[width = .5\textwidth, clip = true, trim = 0.cm 0.cm 0.cm 0.cm]{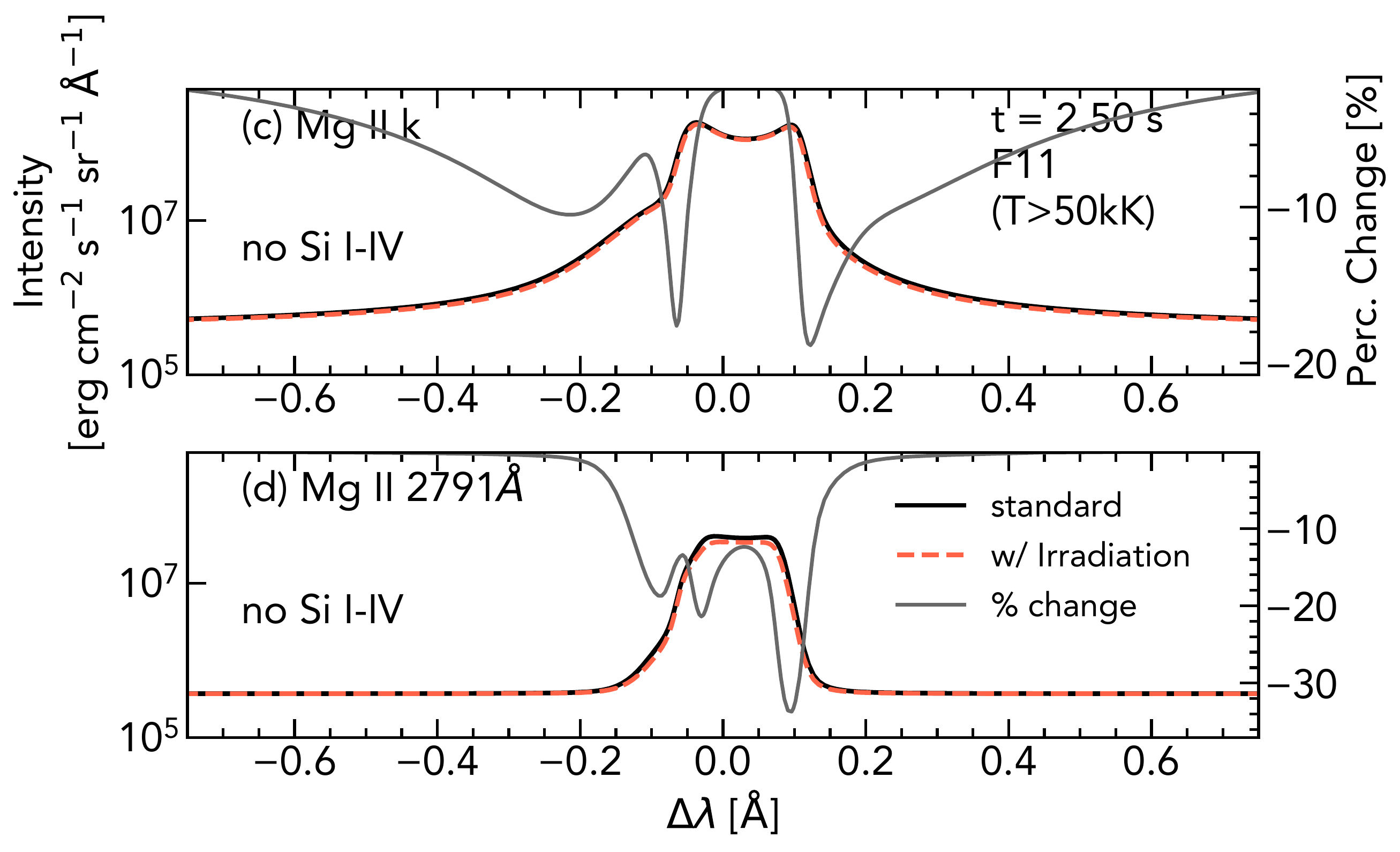}}
	}
	\hbox{
	\hspace{1.5in}
	\subfloat{\includegraphics[width = .5\textwidth, clip = true, trim = 0.cm 0.cm 0.cm 0.cm]{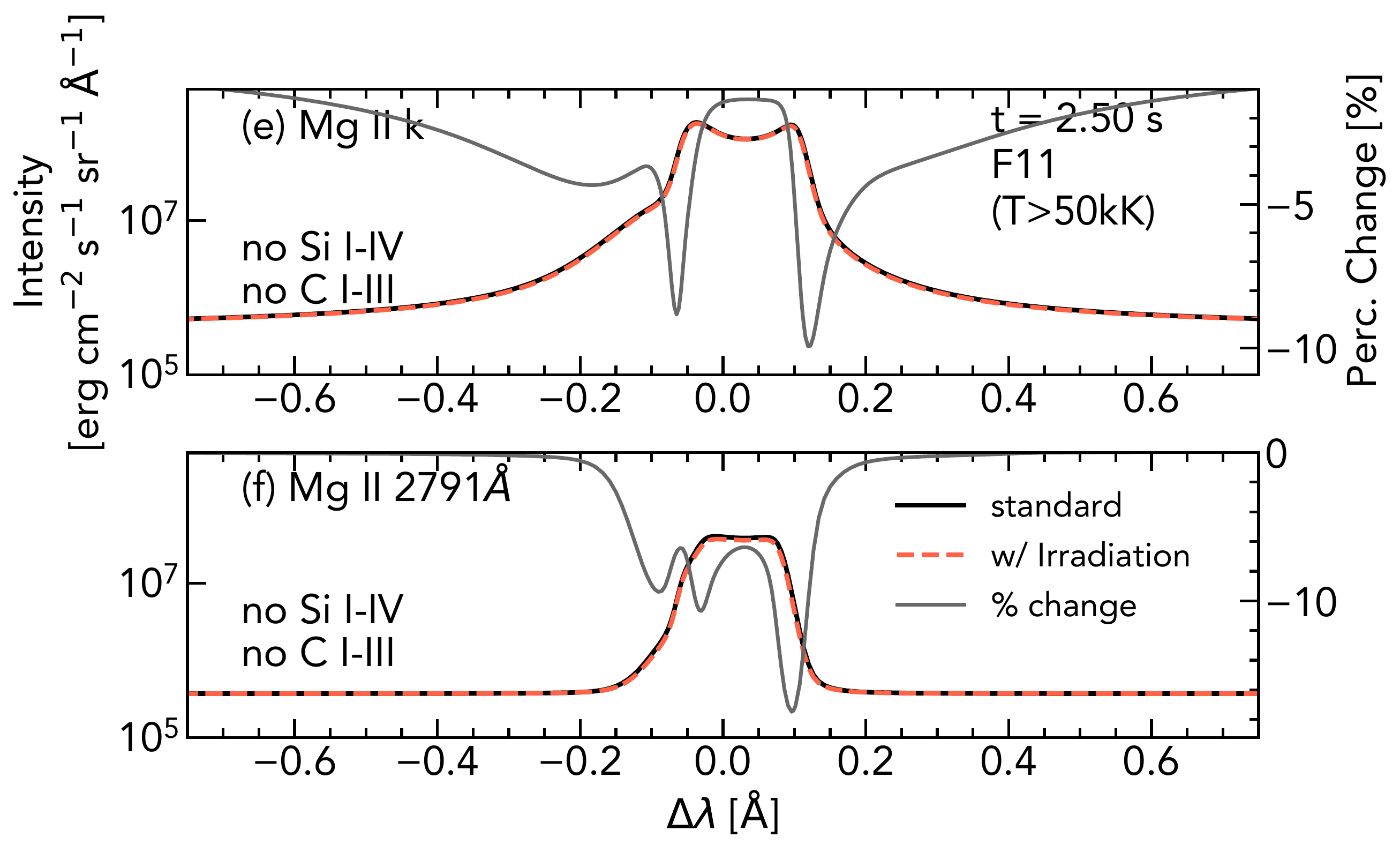}}
	}
	\caption{\textsl{The Mg~\textsc{ii} k line (a,c,e) and 2791~\AA\ subordinate line (b,d,f) in the standard setup (black line) and when TR/coronal irradiation (from $T>50$~kK) is included (red dashed lines) at $=2.5$~s in the F11 simulation. The grey lines show the percentage change. The irradiance in panels (a,b) was with every transition in CHIANTI with the detailed transitions from \texttt{RH} and helium removed. Panels (c,d) also removes Si~\textsc{i}-\textsc{iv}. Panels (e,f) then additionally removes C~\textsc{i}-\textsc{iii}}}
	\label{fig:f11_irradiance_profiles}
\end{figure*}

The corona in flares can reach temperatures in excess of $10$~MK, with ablation of chromospheric material increasing the density sufficiently so that there is a substantial increase in the flux of X-ray, extreme-ultraviolet, and ultraviolet emission (together: XEUV). At the same time the TR density increases, strongly enhancing radiation produced there. This optically thin emission is directed both outward to the observer, and downward towards the lower solar atmosphere, where it can heat the plasma and photoionise. For example, the emission of the He~\textsc{i} $10830$~\AA\ line is strongly influenced by coronal radiation, with photoionisation of helium and subsequent recombination to the $2s~^{3}S_{1}$ level \citep[e.g][]{2014ApJ...784...30G}. The plasma heating effects of XEUV backwarming is included in the \texttt{RADYN} simulation (as are photoionisations) so that the atmospheric structure being solved by \texttt{RH} also includes this source of additional heating. However, the potential impacts of an increased photoionisation rate is not considered, as downward directed coronal radiation is not typically included in post-processing of flare simulations with radiation transport codes. An exception to this was an investigation into the origin of H$_{2}$ emission, including using a semi-empirical flare atmosphere. The fluorescence from strong transition region lines was included as a downward incident radiation that resulted in enhanced molecular transitions \citep{2018ApJ...855..134J}. 

There is already functionality in \texttt{RH} to introduce a coronal source of radiation, but this requires a bespoke input file for each snapshot, with specific intensity of the injected radiation defined as a function of wavelength, covering all wavelengths that \texttt{RH} models. Producing a bespoke input file for hundreds of snapshots is a laborious task, especially if the atomic models are changed (which changes the wavelengths on which the injected intensity must be defined). It is preferable to provide a single common input file, and have \texttt{RH} internally compute the intensity to inject based on the atmospheric stratification. Using a similar method to that employed by \texttt{RADYN} we modified \texttt{RH} to permit a straightforward inclusion of coronal irradiance when post-processing flare simulations.

Contribution functions, $G({n_{e},T,\lambda})$ from all transitions in the CHIANTI atomic database were tabulated on a grid that spanned $\lambda = [1,40500]$~\AA\ with $\Delta\lambda = 1$~\AA\ bins, $\mathrm{log}T=[4,9]$ with $\Delta\mathrm{log}T = 0.05$ bins, and $\mathrm{log}N_{e} = [6,16]$ with $\Delta\mathrm{log}N_{e} = 0.5$ bins. Bound-bound transitions already modelled by \texttt{RH} were omitted, as was Helium since those transitions are optically thick. The emissivity at each temperature and electron density in the grid was computed by $j_{\lambda} = A_{b}G(\lambda, ne, T)n_{e}n_{H}$, where $A_{b}$ is the elemental abundance, and $n_{H}$ is the hydrogen density. The abundance table \texttt{sun$\_$coronal$\_$2012$\_$schmelz$\_$ext.abund} from the CHIANTI database was used where low FIP elements are enhanced , while the high FIP elements are not \citep{2012ApJ...755...33S,2011SoPh..268..255C,2009LanB...4B..712L}. For $n_{H}$ we use the typical assumption for a hot optically thin plasma that $n_{H} = 0.83n_{e}$. When computing $j_{\lambda}$ we modelled the lines as Gaussians with thermal broadening according to the temperature in each cell of our grid, and a bin size of $1$~\AA. The units of this input grid were then erg~cm$^{-3}$~s$^{-1}$~sr$^{-1}$~\AA$^{-1}$. 

This grid of emissivities was input to \texttt{RH}, and the resulting emissivity in each grid cell of the atmospheric model was obtained by performing a trilinear interpolation to the local values of temperature and electron density, for each wavelength bin required by the \texttt{RH} active set of atoms and molecules. The wavelength bins are irregular in \texttt{RH} with some exceeding the bin size of the input grid. When that occurred, we created a new temporary grid that had finer resolution than the input grid. The input emissivities were interpolated at each point on this new wavelength grid, which was then integrated through wavelength to obtain the total emission within the original \texttt{RH} bin (erg~cm$^{-3}$~s$^{-1}$~sr$^{-1}$). This was divided by the \texttt{RH} bin size to return the averaged emissivity in that bin (erg~cm$^{-3}$~s$^{-1}$~sr$^{-1}$~\AA$^{-1}$). In this manner the emission was conserved through the interpolation, and the potential of a transition falling through a missing wavelength bin was removed. Finally, the emissivity was converted to SI units (W~m$^{-3}$~sr$^{-1}$~Hz$^{-1}$) and integrated through height, providing the specific intensity spectrum (W~m$^{-2}$~sr$^{-1}$~Hz$^{-1}$) that was injected at the apex of the loop. 

Several snapshots of the injected spectrum from the F11 simulations are shown in Figure~\ref{fig:irradiation} showing how the irradiation varies with time and with atmospheric state. Panels (a,c,e) show the case where emission from grid cells with $T>50$~kK, and (b,d,f) from cells with $T>100$~kK, illustrating that the lower-mid TR dominates the irradiating spectrum. This assumes optically thin emission, though it is likely that opacity effects will be present for certain species. We have already removed Helium for this reason, and future work will investigate in detail the opacity of certain species that are likely to show opacity effects, such as C~\textsc{iii}, refining the emissivity grid accordingly.  For this present work we performed two additional experiments. In the first we removed transitions of Si~\textsc{i} - \textsc{iv} based on the finding in \cite{2019ApJ...871...23K} that Si~\textsc{iv} lines can become optically thick in flares. In the second we also removed transitions of C~\textsc{i} - \textsc{iii}. Figure~\ref{fig:irradiation}(g,h) show the injected spectra for these two scenarios (both used a $50$~kK temperature threshold.
 
 Emission from grid cells with $T>50$~kK only was initially considered for the irradiation.  Coronal/TR irradiaton did affect the Mg~\textsc{ii} formation at certain times during the simulation, the significance of which depended on the transitions included in the irradiating spectrum. Increased photoionisations at these times lowered the Mg~\textsc{ii} fraction, decreasing the upper level populations of the resonance and subordinate lines, and decreasing the line intensity. These effects were only present at $t<5$~s, when the coronal irradiation was maximum. After $t=5$~s the injected spectrum became much weaker, due to the compression of the mid-lower transition region, the layer which contributed most to the irradiation. Both the resonance and subordinate line intensities were affected, though the profile shapes were preserved. Subordinate line cores were depressed by $\sim20$~\%, and the inner wings by $\sim40$~\%. The resonance line inner cores were little affected but moving outward from the inner cores the lines were depressed by up to $\sim20$~\% into the inner wings. Figure~\ref{fig:f11_irradiance_profiles}(a,b) shows the Mg~\textsc{ii} k line and $2791$~\AA\ line profiles at a time when irradiation had non-negligible effects.  Removing Si~\textsc{i}-\textsc{iv} transitions and then additionally C~\textsc{i}-\textsc{iii} reduced the magnitude of the intensity decrease, suggesting that these transitions contributed a large amount of Mg~\textsc{ii} photoionisations. 
 
If the temperature threshold at which TR/coronal emissivity begins to be considered is increased to $T>100$~kK then the irradiating intensity drops significantly and there is no impact on the Mg~\textsc{ii} formation. This tells us that it is the enhanced radiation originating in the lower-mid TR that has the potential of photoionising Mg~\textsc{ii}. 

These experiment were repeated for the F10 simulation, which showed no impact on the Mg~\textsc{ii} formation even with the lower $50$~kK threshold. In that simulation there is not sufficient heating at higher densities to enhance radiation from the mid-lower TR to the point that it irradiates the chromosphere to the same degree as found for the F11 simulation.

An in-depth investigation of the appropriate temperatures and species to consider for irradiation will be the focus of future work, as this process is also of interest in the formation of the He~\textsc{i} $10830$~\AA\ and He~\textsc{i} $\mathrm{D}_{3}$ lines, which have recently shown interesting dimming effects at the leading edge of flare ribbons \citep[e.g.][]{2016ApJ...819...89X,2019A&A...621A..35L}. For now, we note that lower-mid TR and coronal irradiation does have the potential to impact the formation of Mg~\textsc{ii}, should ideally be included, but that further research is required to determine the most accurate means to do so.  Those modelling Mg~\textsc{ii} in flares should be aware of this additional source of ionisation.


\section{Summary \& Conclusions}\label{sec:conclusions}
We have compared various approaches of modelling the Mg~\textsc{ii} NUV spectra during solar flares, using three radiation hydrodynamic flare simulations from the \texttt{RADYN} code, that were post-processed through the \texttt{RH} radiation transport code. This is an effort to ensure that we are including sufficient physics to properly model these lines, which have been routinely observed in flares by the IRIS spacecraft, and offer intriguing diagnostic potential of the flare chromosphere and a crucial observable with which to critically attack flare models. These model approaches either added or removed complexity, with the resulting impact on the Mg~\textsc{ii} lines discussed both conceptually and quantitatively. 

In the course of experimenting with the physics included in the models we made two modifications to the \texttt{RH} code. Those were:

\begin{enumerate}[label=(\Alph*)]
\item Fixing input populations for specified atomic, or molecular, species, so that the radiation transport problem is solved using populations obtained from another source. We used this to include the \texttt{RADYN} NEQ hydrogen populations with non-thermal collisional excitation and ionisations also considered. With this method the NEQ hydrogen opacity was used by \textsc{RH} \textsl{and} the resulting radiation was output (as opposed to being treated as background only);
\item Inclusion of TR and coronal irradiation. \texttt{RH} can now accept a grid of emissivities as a function of temperature, electron density, and wavelength. In each grid cell above a defined temperature threshold ($50$~kK in our case) the local temperature and density is used to interpolate the emissivity at each wavelength modelled by \texttt{RH}'s atomic models. This is is then integrated through depth to obtain the intensity under the assumption that it is optically thin, and is subsequently injected as a downward-directed source of radiation from the apex of the loop. We provided every transition in the CHIANTI atomic database, with the transitions already solved by \texttt{RH} and certain other species removed, but the module is written such that a user can provide any grid they wish.
\end{enumerate}

In summary, our experiments showed that: 
\begin{enumerate}[label=(\roman*)]
\item PRD is required for modelling Mg~\textsc{ii} in flares. The enhanced density in stronger flares does reduce, somewhat, the magnitude of the differences between CRD and PRD, but the wing intensities are still significantly overestimated with CRD. Features in the line wings due to redistribution effects also result in differences to the profile shapes that may be misinterpreted as being due to mass flows. 
\item Angle-dependent PRD is more accurate than the hybrid-PRD treatment when velocities are large. The profile shapes are unchanged, just the intensity in localised parts of the line. Given the substantial increase in computational time, we conclude that H-PRD is appropriate for most flare studies, so long as the caveats are understood. 
\item It is not necessary to model Mg~\textsc{i} if one is only interested in the h \& k lines, and subordinate lines, but there are intensity and formation differences in the far wings if Mg~\textsc{i} is omitted. Therefore, if the (\textsl{quasi}-) continuum observed by IRIS is of interest, then Mg~\textsc{i} should be included with Mg~\textsc{ii}.
\item The dominant sources of opacity at the Mg~\textsc{ii} resonance line cores and near wings are are Mg~\textsc{ii} and hydrogen, with hydrogen only really impacting the line wings. Only those species need be included in NLTE, with other sources safely treated in LTE. If the far wings or \textsl{quasi-} continuum are of interest then other species (e.g. Mg~\textsc{i}, Si~\textsc{i}) should be included. 
\item Using the NEQ hydrogen populations did result in a more accurate modelling of the opacity in the Mg~\textsc{ii} line wings, so that if possible those populations should be used. However, the differences were not very large. If it is not possible to include NEQ hydrogen populations for some reason then the results will not be very different.
\item Using a three-level-with-continuum atom was found to be insufficient. A model atom containing excited states capable of supporting recombinations with cascades down to the h \& k upper levels must be used. 
\item Lower-mid transition region irradiance does have an impact on the Mg~\textsc{ii} formation, albeit with the effect seemingly very dependent on the transitions and temperatures considered. Irradiation from locations $T>50$~kK resulted in Mg~\textsc{ii} line intensity differences of $\sim10-30$~\%. However, if irradiation only from locations $T>100$~kK is included then there is negligible impact on Mg~\textsc{ii}. Removing Si~\textsc{i}-\textsc{iv} and C~\textsc{i}-\textsc{iii} from the irradiating spectrum reduces the magnitude of the differences. So, irradiation from the lower-mid TR should potentially be included, but this requires further investigation.
\end{enumerate}

When modelling Mg~\textsc{ii} in flares the above findings should be considered, and a suitable setup used. These experiments all assumed SE for the formation of Mg~\textsc{ii}. Paper II in this series will discuss NEQ effects, which we found do have an impact during the initial heating and cooling phase of the flares \citep{NEQ_inprep}. \\

\textsc{Acknowledgments:} \small{GSK was funded by an appointment to the NASA Postdoctoral Program at Goddard Space Flight Center, administered by USRA through a contract with NASA. JCA acknowledges funding support from the Heliophysics Supporting Research and Heliophysics Innovation Fund programs. This research was supported by the Research Council of Norway through its Centres of Excellence scheme, project number 262622, and through grants of computing time from the Programme for Supercomputing. We thank Dr. Han Uitenbroek for making the \texttt{RH} code publicly available, and Dr. Adrian Daw and Dr. Don Schmit for useful discussions regarding coronal irradiation.}

\bibliographystyle{aasjournal}
\bibliography{Kerr_etal_prdMgII}

\appendix

\section{Ly~$\alpha$ PRD}\label{sec:HydPRD}
Here we show Ly~$\alpha$ line profiles and hydrogen populations, for \texttt{RADYN} flare snapshots forward modelled using \texttt{RH}, comparing the PRD and CRD solutions. This was done for all three flares (F9, F10, and F11). In the weaker flare PRD effects are still clear, with CRD overestimating the line wing intensity and underestimating the emission peaks. The atomic level populations and proton density also vary somewhat. For the moderate and strong flares, where the electron density was significantly more enhanced, the PRD solution approaches the CRD solution, and atomic level populations show only modest changes. 
\begin{figure*}
	\centering 
	\hbox{
	\subfloat{\includegraphics[width = .5\textwidth, clip = true, trim = 0.cm 0.cm 0.cm 0.cm]{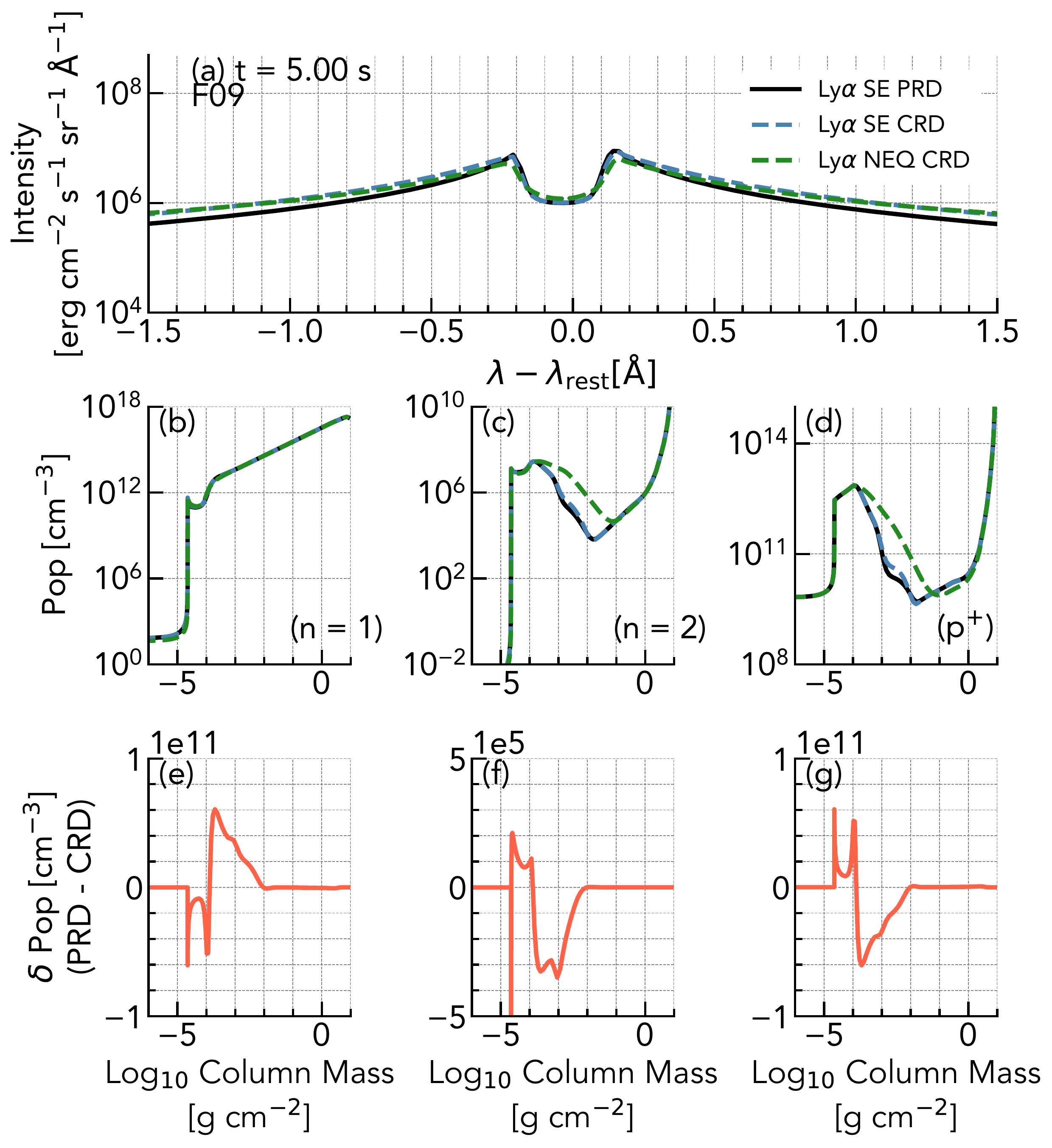}}
	\subfloat{\includegraphics[width = .5\textwidth, clip = true, trim = 0.cm 0.cm 0.cm 0.cm]{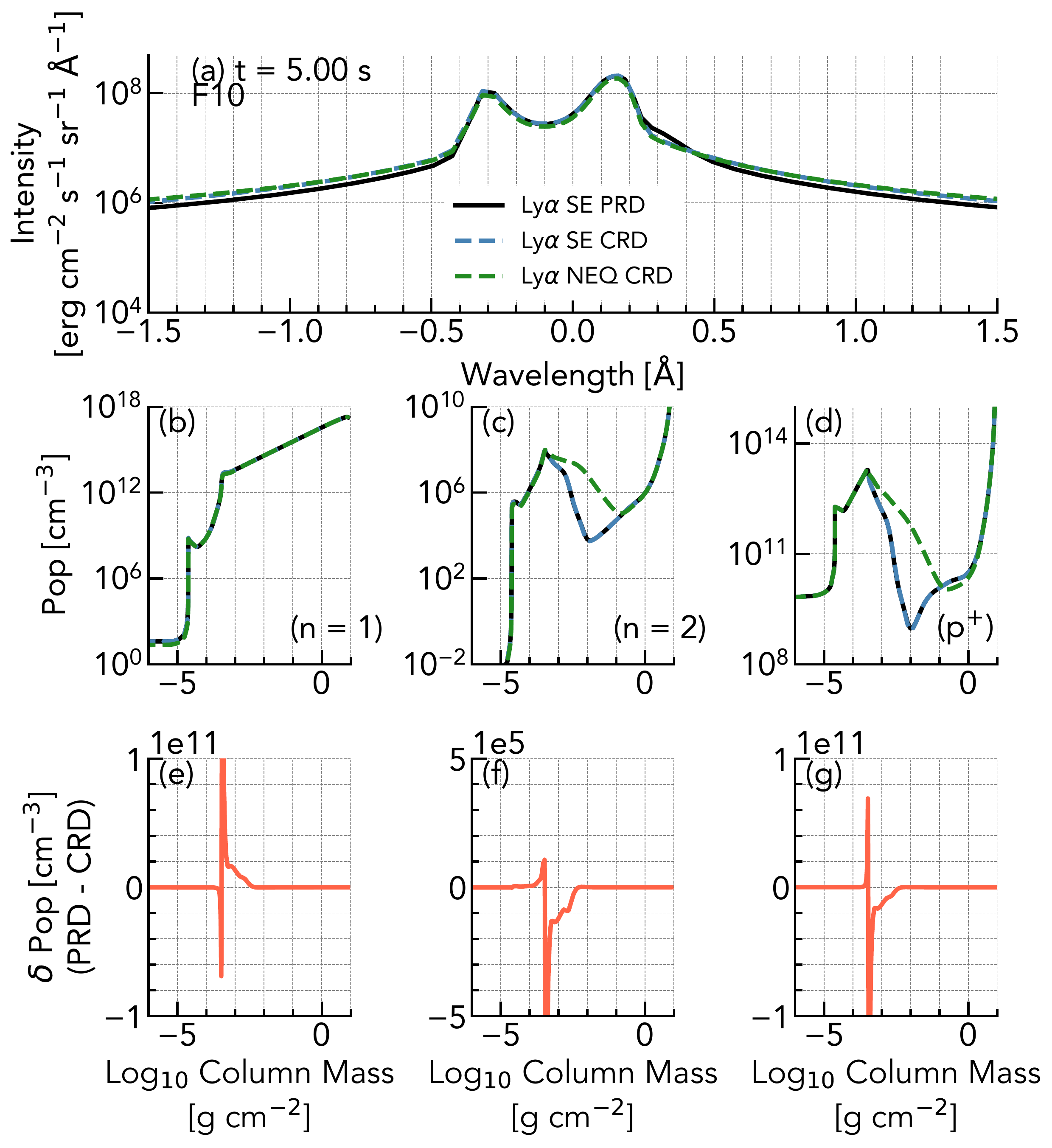}}
	}
	\hbox{
	\hspace{1.5in}
	\subfloat{\includegraphics[width = .5\textwidth, clip = true, trim = 0.cm 0.cm 0.cm 0.cm]{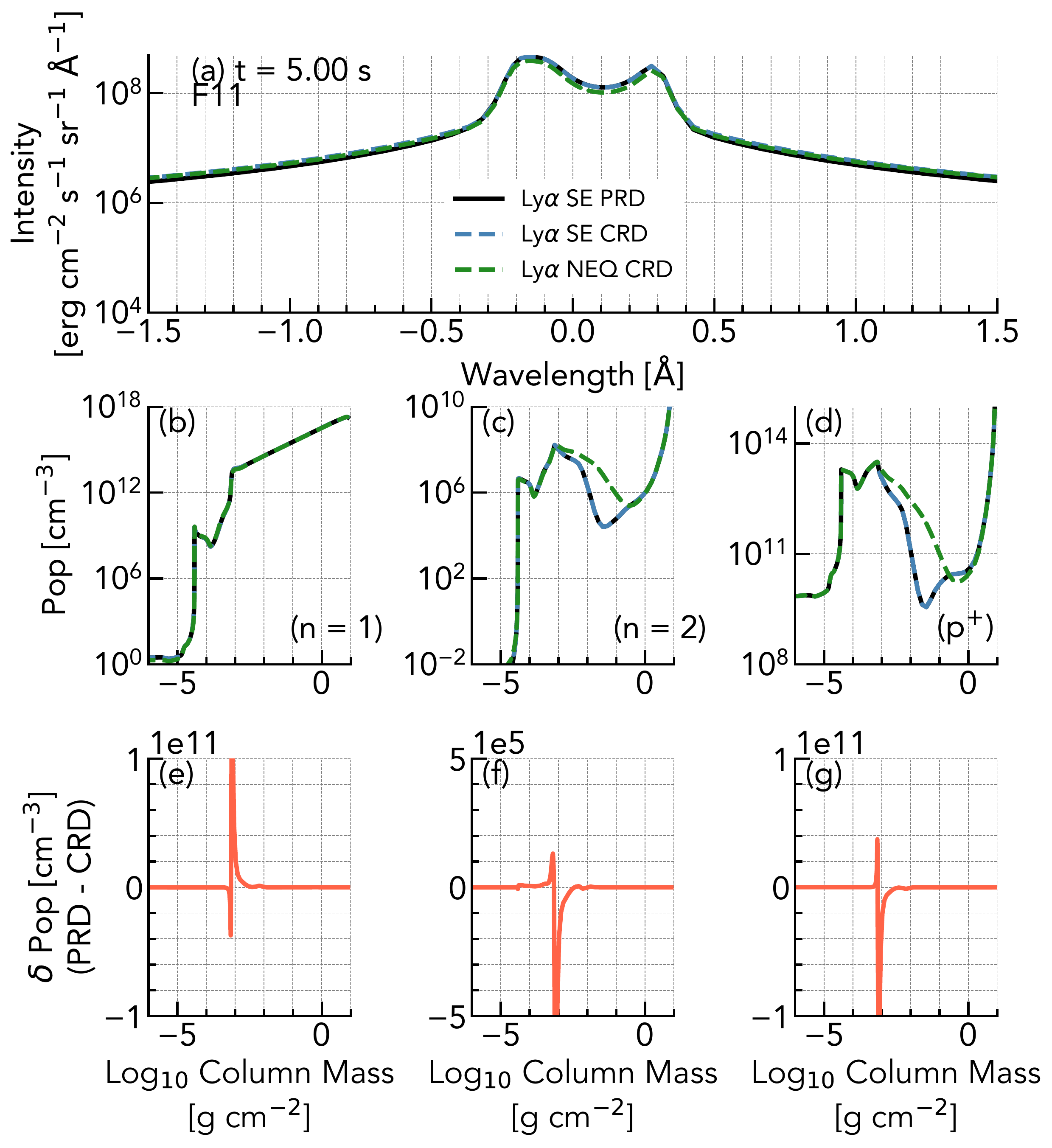}}
	}
	\caption{\textsl{Comparing the Ly~$\alpha$ line in CRD and PRD. Black solid lines show the SE PRD solution, the blue dashed lines show the SE CRD solution, and the green dashed lines show the NEQ CRD solution (populations directly from \texttt{RADYN}. The upper left block show the F9 flare, the upper right the F10 flare, and the bottom the F11 flare. Panel (a) show the Ly~$\alpha$ profiles, panels (b-d) the $n=1, n = 2$, \& proton densities, and panels (e-g) the difference in those populations (SE PRD - SE CRD).}}
	\label{fig:H_PRD_or_CRD_F09}
\end{figure*}

\end{document}